\documentclass[10pt,journal,twocolumn]{IEEEtran}

\ifCLASSOPTIONcompsoc
  \usepackage[nocompress]{cite}
  \else
  \usepackage{cite}
  \fi
  
  \ifCLASSINFOpdf
  
  \else
\fi
\usepackage{multirow} 
\usepackage{algpseudocode}
\usepackage{algorithm}
\usepackage{rotating}
\usepackage{kantlipsum} 
\usepackage{commath}
\allowdisplaybreaks
\usepackage{mathtools}  
\usepackage{verbatim}
\usepackage{xr-hyper} 

\usepackage{epstopdf}
\usepackage{wrapfig}
\usepackage{latexsym}
\usepackage{amssymb}
\usepackage{amsthm}
\usepackage{amsfonts}
\usepackage{amsmath} 
\usepackage{graphicx}
\usepackage{latexsym}
\usepackage{booktabs}
\usepackage{subcaption} 
\usepackage{breqn}

\newtheorem{thm7}{Lemma}
\algnewcommand\algorithmicinput{\textbf{INPUT:}}
\algnewcommand\INPUT{\item[\algorithmicinput]}
\algnewcommand\algorithmicoutput{\textbf{OUTPUT:}}
\algnewcommand\OUTPUT{\item[\algorithmicoutput]}
\usepackage[table]{xcolor}

\begin{document}

\title{\huge Joint Energy Efficient and QoS-aware Path Allocation and VNF Placement for Service Function Chaining}


\author{Mohammad M. Tajiki, Stefano Salsano,~\IEEEmembership{Senior Member,~IEEE,} Luca Chiaraviglio,~\IEEEmembership{Senior Member,~IEEE,}  Mohammad Shojafar,~\IEEEmembership{Senior Member,~IEEE,} Behzad Akbari,~\IEEEmembership{Senior Member,~IEEE}
\IEEEcompsocitemizethanks{\protect
\IEEEcompsocthanksitem S. Salsano and L. Chiaraviglio are with the Department of Electronic Engineering at the University of Rome – Tor Vergata – Rome, Italy E-mail: \{stefano.salsano, luca.chiaraviglio\}@uniroma2.it\protect
\IEEEcompsocthanksitem M. Shojafar is with the Consorzio Nazionale Interuniversitario per le Telecomunicazioni (CNIT) – Rome, Italy E-mail: mohammad.shojafar@cnit.it \protect
\IEEEcompsocthanksitem MM. Tajiki and B. Akbari are with the ECE Department, University of Tarbiat Modares, Tehran, Iran - Email: \{mahdi.tajiki,b.akbari\}@modares.ac.ir
}}

\markboth{Extended version of submitted paper - v7 - July 2018}%
{Tajiki \MakeLowercase{\textit{et al.}}: Joint Energy Efficient and QoS-aware Path Allocation and VNF Placement for Service Function Chaining}


\IEEEtitleabstractindextext{
\begin{abstract}
Service Function Chaining (SFC) allows the forwarding of a traffic flow along a chain of Virtual Network Functions (VNFs, e.g., IDS, firewall, and NAT). Software Defined Networking (SDN) solutions can be used to support SFC reducing the management complexity and the operational costs. One of the most critical issues for the service and network providers is the reduction of energy consumption, which should be achieved without impact to the quality of services. In this paper, we propose a novel resource allocation architecture which enables energy-aware SFC for SDN-based networks. To this end, we model the problems of VNF placement, allocation of VNFs to flows, and flow routing as optimization problems. Additionally, we model the problem of flow rerouting to reduce the impact of resource fragmentation on the network utilization.
Thereafter, heuristic algorithms are proposed for the different optimization problems, in order to find near-optimal solutions in acceptable times. The performances of the proposed algorithms are numerically evaluated over a real-world topology and various network traffic patterns. The results confirm that the proposed heuristic algorithms provide near-optimal solutions while their execution time is applicable for real-life networks.
\end{abstract}
\begin{IEEEkeywords} 
Software Defined Network (SDN), Service Function Chaining (SFC), Quality of Service (QoS), Energy Consumption, VNF Placement; 
\end{IEEEkeywords}
}
\maketitle
\IEEEdisplaynontitleabstractindextext
\IEEEpeerreviewmaketitle
\section{Introduction}\label{sez:1}
    \IEEEPARstart{T}{raffic} flows in packet networks may need to pass through different hardware middle-boxes (e.g., IDS, proxy, and firewall) in their end-to-end paths. Network Function Virtualization (NFV) replaces hardware middle-boxes with flexible and innovative software applications known as \textit{Virtual Network Functions (VNFs)} to reduce the capital and operational expenditures and increase the flexibility of providing the services \cite{fischer2013virtual}. On the other hand, the Software Defined Networking (SDN) paradigm offers the possibility to control the forwarding of packets from a logically centralized point of view, easing the introduction of efficient and flexible algorithm to optimize the utilization of network and processing resources. 
    
    This is commonly referred to as Service Function Chaining (SFC) \cite{halpern2015service}.
    	    
    VNF chains or simply chains are required to process large volumes of traffic within a very short period of time to facilitate real-time streaming applications that comprise majority of traffic in today's networks. Failure to provide the desired throughput of a chain may lead to violating of the service level agreements (SLAs) incurring high penalties. Hence, achieving high throughput of VNFs is of paramount importance. Consequently, several works focus on providing SFC in SDNs. An SFC taxonomy that considers architecture and performance dimensions as the basis for the subsequent state-of-the-art analysis is introduced in \cite{medhat2017service}.
            
    On the other hand, the high energy consumption by computer networks incurs high costs for providers. Furthermore, some countries have set carbon taxes on the emitted CO\textsubscript{2} to enforce the environmental sustainability  \cite{hepburn2006regulating}. Therefore, considering the amount of energy consumed as a parameter of resource allocation algorithm helps service providers to reduce the energy and carbon cost as a major sector of their total cost \cite{khosravi2017dynamic}. In this context, different natural questions arise, such as: Is it possible to propose resource allocation architecture to cover SFC over SDN switches? How to optimally trade-off between energy consumption of servers and network allocation side effects on SDN switches? How to properly implement SFC without affecting the Quality of Service (QoS) parameters in real-world network? The goal of this paper is the shed light on these issues. More in detail, we propose a resource allocation architecture and a set of novel energy-aware SFC algorithms for SDN networks. To this end, we propose a resource allocation architecture based on the SDN principles, mathematically formulate the problems of resource allocation and propose several heuristic algorithms. In a nutshell, our main contributions are as follows:
        \begin{itemize}
        \item we propose a resource allocation architecture consisting of five components (i.e., Single Flow Resource Allocation, Global Resource Reallocation, Network Monitoring, Server/Network Configurator, and Congestion Detector/Predictor).
        \item we model the Service Function Chaining problem (energy-aware VNF placement and path allocation) to assign the resources to the flows (path allocation) and switch the VMs between three different states (OFF, ON-IDLE, ON-ACTIVE).
        \item we mathematically formulate the problems of resource allocation for the proposed architecture. We consider both the initial resource allocation to a flow and the global reallocation. The corresponding problems are cross-layer optimization problems considering energy, service function chaining, and QoS constraints and belong to the class of Integer Non Linear Programming (INLP) in our first natural formulation
        \item we linearized the non linear constraints in aforementioned optimization problems, obtaining ILP problems that can be solved using ILP solvers for small-sized network scenarios. 
        \item in order to address the optimization problems in larger scale network scenarios, we propose near-optimal heuristics. We show that the heuristic approaches are applicable in real-world networks. 
        \item in order to evaluate the proposed algorithms, we compare the heuristic algorithms with the optimal ones from several aspects: i) energy consumption, ii) path length, iii) network side-effect iv) average/maximum link utilization, v) average/maximum node utilization, and vi) computational complexity.
        \item finally, we implemented a demand/resource generator to create network resources and flows with different requirements (e.g., VNF chaining, tolerable delay, flow size, servers' functionalities, etc.) based on mathematical distribution.
        \end{itemize}
            
    Considering the ETSI NFV MANO architecture \cite{etsi-mano}, the proposed Application Layer modules would be need to be mapped in the NFVO orchestrator and VNF Manager functional entities, while the Control Layer modules should be mapped into an SDN controller. Anyway, the precise mapping into the NFV MANO architecture is left for further study.
            
    We believe that this work can be the first step towards the deployment of algorithms tailored to the energy-aware management of network traffic with SFC considerations. Future work will be dedicated to making the algorithms robust against burst traffics. Another field of future interest would be considering the queuing delay of the switches and the processing delay of the server to improve the quality of users experiment. 
        
    The remainder of the paper is organized as follows: In Section~\ref{relatedWork}, the related work is discussed. Section \ref{problemDefinition} states the proposed architecture and an outline of the proposed schemes. Section~\ref{problemForumlation} discusses the system model, parameters, objective function and constraints. Additionally, the proposed VNF placement and routing approaches are discussed in Section~\ref{problemForumlation}. The heuristics are discussed in Section~\ref{sec:heuristic}. The performance analysis of the proposed schemes are presented in Section~\ref{resultAnalysis}. Finally, Section~\ref{conclusion} concludes the paper and presents future directions.
            
    
\section{Related Work}\label{relatedWork}
    This section is divided into three different categories: VNF/VM placement problems \cite{mills2011comparing,filiposka2016balancing,bhamare2017optimal,bari2015orchestrating,even2016approximation,xu2015effective,rocha2015network,zhao2015joint,calcavecchia2012vm,meng2010improving,eramo2017migration}, routing and service function chaining \cite{zhang2016co,abdelsalam2017implementation,khoshbakht2016sdte,kulkarni2017neo,soares2015toward,tajiki2017optimal,reddy2016robust,ghaznavi2016service,jiang2012joint,wang2016joint,Akbari2016qrtp}, and green service function chaining \cite{marotta2017energy,shojafar2016energy,khosravi2017dynamic,tang2015hybrid,gu2015joint,eramo2017approach}. From now on we refer to service function as \textit{VNF (Virtual Network Function)}. In the following, we will briefly discuss each category. 
    
\subsection{VNF/VM Placement}
    In this category, the problem is to optimize the placement of VMs in a way that an objective is gained, e.g., maximizing network throughput or minimizing energy consumption. A survey of the recent works on VNF resource allocation that tackle chain composition and SFC embedding is presented in~\cite{herrera2016resource}.
    Some works focus on proposing measurements to compare different VM placement algorithms, e.g., \cite{mills2011comparing} and \cite{filiposka2016balancing}. More in detail, authors in \cite{filiposka2016balancing} compare several algorithms to evaluate their capabilities on balancing resource usages versus total number of used physical machines. They presents the relation between \textit{the data center characteristics}, \textit{size of the cloud services} and \textit{their diversity} and the performance of the heuristics.
        
    Focusing on VM placement algorithms, authors of \cite{bhamare2017optimal} study the problem of deploying SFCs over NFV architecture. Specifically, they investigate virtual network function placement problem for the optimal SFC formation across geographically distributed clouds. Moreover, they set up the problem of minimizing inter-cloud traffic and response time in a multi-cloud scenario as an Integer Linear Programming (ILP) optimization problem, along with some other constraints such as total deployment costs and SLAs.
        
	Besides, authors of \cite{bari2015orchestrating} use ILP to determine the required number and placement of VNFs that optimize network operational costs and utilization without violating SLAs. In \cite{even2016approximation} an approximation algorithm for path computation and function placement in SDNs is proposed. Similar to \cite{bari2015orchestrating}, they propose a randomized approximation algorithm for path computation and functions placement. In addition, the paper \cite{rost2016service} considers offline batch embedding of multiple VNF chains. They adopt the objectives of maximizing the profit by embedding an optimal subset of requests or minimizing the costs when all requests need to be embedded.
	    
	There are lots of VM placement algorithms for data centers \cite{xu2015effective,rocha2015network,zhao2015joint,calcavecchia2012vm}. The authors of \cite{xu2015effective} state a VM placement algorithm for data center networks to make a trade-off between the energy consumption of the network and the SLA performance. Similarly, in \cite{rocha2015network} a Mixed ILP (MILP) traffic-aware VM placement for data center Networks is presented. The authors of \cite{zhao2015joint} propose an approach which jointly optimize both topology design and VM placement in order to make an scalable solution. To this end, they exploit MILP to formulate the problem. Additionally, they propose a heuristic based on Lagrange relaxation decomposition.
	    
	The authors of \cite{meng2010improving} formulate the VM placement as an optimization problem and propose a two-tier approximate heuristic to solve the problem. Furthermore, they compare the impact of the network architectures and the traffic patterns on the performance of their approach. 
	    
	These methods are interesting in structure and SDN problem formulation, however, none of them considers the problem of service function chaining with respect to the energy consumption of the VMs in SDN.
	    
\subsection{Routing and Service Function Chaining}
    In this category, the problem is to find a routing in which SFC requirements are met, i.e., all required VNFs deliver to the flows. To this end, it is possible to define some extra objectives such as minimizing the network congestion. There are lots of works in this category e.g., \cite{zhang2016co,abdelsalam2017implementation,kulkarni2017neo,soares2015toward,tajiki2017optimal,reddy2016robust,ghaznavi2016service,jiang2012joint,wang2016joint}.
        
    The authors of \cite{zhang2016co} propose a heuristic algorithm to find out a solution for VNF chaining. It employs two-step flow selection when a SFC with multiple network functions needs to scale out. Furthermore, authors in \cite{abdelsalam2017implementation} introduce a VNF chaining which is implemented through segment routing in a linux-based infrastructure. To this end, they exploit IPv6 segment routing network programming model to support SFC in a NFV scenario. Moreover, authors of \cite{kulkarni2017neo} propose a scheme which provides flexibility, ease of configuration and adaptability to relocate the VNFs with minimal control plane overhead. Reference \cite{soares2015toward} focuses on enabling telco infrastructures to orchestrate and manage SFC toward cloud infrastructures.
    
    Moreover, in \cite{reddy2016robust} a optimization model based on the concept of $\Gamma$-robustness is proposed. They focus on dealing with the uncertainty of the traffic demand. In \cite{ghaznavi2016service} an optimization model to deploy a chain in a distributed manner is developed. Their proposed model abstracts heterogeneity of VNF instances and allows them to deploy a chain with custom throughput without concerning about individual VNF's throughput. 
        
    On the other hand, the paper \cite{jiang2012joint} solves a joint route selection and VM placement problem. They design an offline algorithm to solve a static VM placement problem and an online solution traffic routing. More in detail, they expand the Markov approximation to gain their objectives. In \cite{wang2016joint} a joint resource allocation and service function chaining is proposed. The authors use a cost model to make a trade-off between service performance and network costs. They exploit MILP to model the problem and propose a heuristic to solve it. 
        
    Although the aforementioned solutions in this category are interesting, non of the them considers the problem of service function chaining with respect to the energy consumption of the VMs.

\subsection{Green Service Function Chaining}
    In this category, the problem of routing or VM placement in SFC is considered, while the objective is to assign resources such that energy consumption of the network is minimized. In detail, the authors in \cite{marotta2017energy} introduce an optimization model to minimize the energy consumption while considering a set of VNF chains. The model explicitly provides robustness to unknown or imprecisely formulated resource demand variations. To gain this, it powers down unused routers, switch ports and servers, and calculates the energy optimal VNF placement. The most challenging part of this work is that they do not consider ordering for the VNFs. In other words, it is impossible to guarantee that the flow meets VNF $x$ before VNF $y$.
        
    On the other hand, authors of \cite{shojafar2016energy} present an energy-aware VM placement algorithm to maximize the admitted traffic delivered to mobile clients and minimize the jointly computation and communications energy consumption in cloud data centers. Although the approach is interesting, they do not consider the impact of traffic pattern in their algorithm. In \cite{khosravi2017dynamic} an energy-aware VM placement method for geographically distributed cloud data centers is proposed. The authors investigate parameters that have impact on carbon footprint and energy cost and formulate the total energy cost as a function of the energy consumed by servers plus overhead energy.
         
    Similar to \cite{khosravi2017dynamic,shojafar2016energy}, authors of \cite{tang2015hybrid} and \cite{gu2015joint} propose two algorithms for the VM placement problem in data centers which optimize the energy consumption. To be specific, in \cite{tang2015hybrid}, authors propose a hybrid genetic algorithm for VM placement problem considering the energy consumption in both communication network and physical machines. Similarly, authors of \cite{gu2015joint} mathematically formulate the VM placement problem and propose a heuristic to solve it. Both \cite{khosravi2017dynamic} and \cite{tang2015hybrid} focus on VM placement and ignore real-time routing of the traffic flows.
        
    The authors of \cite{eramo2017approach} propose a scheme that uses three different algorithms to do the VNF placement, SFC routing, and VNF migration in response to changing workload. Their objective is to minimize the rejection of SFC bandwidth and reduce the energy consumption. Although their work is pretty interesting, there are several weak points in their approach. First of all, their approach is applicable for networks with predicable traffic, i.e., they suppose that the traffic pattern is repeated in a time interval. Moreover, they assume the amount of network traffic demands for all slots of the time interval and based on this knowledge, they turn on or off servers. Finally, the authors considers all of the possible physical path as an input to their algorithm which is not an applicable assumption for medium and big networks. 
    
    {This work is the extended version of our previous work \cite{HajMeity}. In this paper, we propose a new routing architecture where five different sub-problems raise: initial assignment of resources, long-term/short-term rerouting, and online/offline algorithms while in our previous conference paper we have a simple routing architecture where we only focus on online short-term rerouting problem. Consequently, we have five heuristic algorithms in this paper which are completely different from the heuristic algorithm which is proposed in \cite{HajMeity} (one heuristic algorithm is proposed in our previous work). In \cite{HajMeity}, we do not have ordering constraints for the VNFs so we cannot enforce sequences of VNFs. Besides, delay constraint is not considered in \cite{HajMeity}.}
            
	         
\section{Reference Scenario}\label{problemDefinition}
	In this section, the proposed resource allocation architecture is discussed by illustrating the architectural components and providing an outline of the main considered procedures. In general, we assume that a Network Operator is managing an SDN/NFV enabled network and it is allocating resources (e.g. processing capacity in nodes, bandwidth capacity on network links) to a set of \textit{flows} that needs to be processed by the network. We consider here ``macro-flows'' or ``aggregated'' flows, i.e. flows that have a relatively long life. To give examples, these flows may correspond to VPN connections or to classes of user traffic that need to be processed in the same way. We also refer to these ``aggregated'' flows as \textit{traffic demands}. 
	In our model, each flow/traffic demand is associated with: an ingress point and an egress point in the operator network; a set of VNFs that needs to be executed on the flow, this is typically a chain of VNFs, that is an ordered list; a required capacity (flow rate), this can be known in advance or unknown (and in the latter case it can be estimated when the flow is active); QoS requirements (e.g. maximum tolerable delay).
	According to the NFV principles, the VNFs (processing services) are executed by Virtual Machines (VMs) that are running on physical servers. The operator needs to map the chain of VNFs associated with the flow to a set of VM instances, which need to be located on physical servers. Moreover, thanks to the SDN approach, we assume that the Network Operator can explicitly configure the routing of the flows in its network, i.e. the path that the packets of a flow take to go from the ingress point to the first VNF, then from a VNF to the next VNF in the chain and finally from the last VNF of the chain to the egress point.
	    
	In general, when allocating the resources to the flows, the Network Operator tries to optimize one or more objective functions, like for example: reducing costs due to energy consumption, maximizing the QoS perceived by the flows, balancing load on the links, minimizing the maximum link utilization. In particular, in our model the Network Operator can use an objective function which is a combination of energy consumption and network reconfiguration cost. We consider as constraints the maximum link utilization, the maximum utilization of the processing modules, the maximum end-to-end sum of fixed link delays. In this work we are focusing on the minimization of energy consumption so in our experiments we will not consider the network reconfiguration cost in the objective functions. As for the energy consumption, in the proposed model the physical servers on which the VMs can be run can be turned ON and OFF, and this is supposed to be a relatively "long term" operation as it requires some time for a server to boot/shutdown. Moreover, a server which is in the ON state can be either ACTIVE (i.e. fully operational, the VNFs are running on it) or IDLE (i.e. no VNFs are running, but since the VM is in ON mode VNF can be instantly reactivated when there is a requirement).
	    
	We consider different resource allocation procedures. \textit{Single Flow Resource Allocation} is used when a single new flow needs to be supported by the Network Operator on top of a running configuration. In this case, we do not consider any reallocation of the active flows and the optimization problem only consider a single flow. Then we consider different types of \textit{Global Resource Reallocation} procedures that can potentially reallocate the running flows. In this case the optimization is performed globally on the set of flows. We consider two types of global reallocation procedures: in one type we can turn on and off the servers (long-term energy management), in the second type we can only change the status of ON servers from ACTIVE to IDLE and vice-versa (short-term energy management). Moreover, depending on the timescale in which we can run the reallocation procedure, we consider \textit{online} and \textit{offline} algorithms. In particular, when the network should be reconfigured in a real-time manner, e.g. following a congestion event, the \textit{online} algorithms will be invoked. On the other hand, the \textit{offline} algorithms can be used to periodically reconfigure the network based on the traffic history and predictions. These can be used for example to plan different resource allocation for different times of the day (e.g. at the midnight the demands always reduces, therefore, an offline algorithm can specify state of the processing nodes for midnight hours).
	    
\subsection{Architecture}
\begin{figure*}
    	\begin{center}
    		\includegraphics[width=\textwidth]{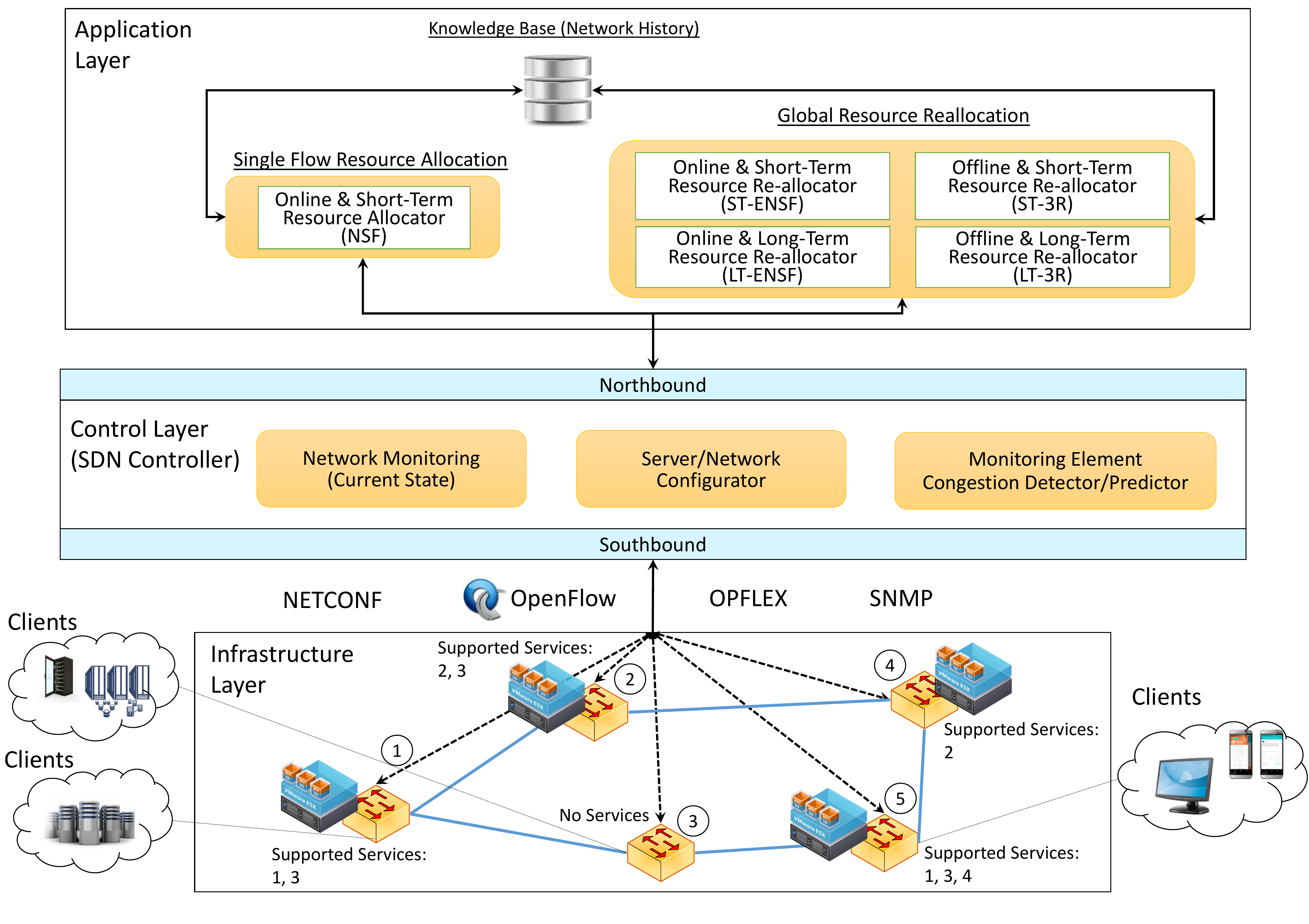}
    		\caption{System Architecture.}
    		\label{fig:Architecture}
    	\end{center}
        \end{figure*}

    In this section, the proposed architecture and its components is presented. The architecture is conceptually aligned with the SDN layering discussed in \cite{haleplidis2015software}. With reference to Fig.~\ref{fig:Architecture}, we consider three different layers. The \textit{Infrastructure} layer consists of networking and processing devices and corresponds to the Forwarding Plane and Operational Plane presented in \cite{haleplidis2015software}. The \textit{Control} layer interacts with the networking devices in order to program their behavior from a logically centralized perspective. It corresponds to the Control Plane and Management Plane in \cite{haleplidis2015software}. Finally, the \textit{Application} layer includes the applications and services that define the overall network behavior. 
        	
    Looking at Fig.~\ref{fig:Architecture}, the Infrastructure Layer includes the networking devices and the servers on which we can run the VMs. We refer to the networking devices as \textit{switches}, following an SDN-based terminology, but these devices could also act as routers. The servers are connected to a switch (multiple servers can be connected to the same switch). The switches can also operate as networking nodes only, without any connected server. We refer to the pair (switch, server) as a \textit{node}. 
    	
    We assume that there is a logically centralized SDN controller in the Control layer, connected to the set of SDN switches via the Southbound protocols (examples of Southbound protocols are shown in Fig.~\ref{fig:Architecture}). The main role of the SDN controller is to setup the forwarding tables of the single SDN switches in order to properly configure the packet forwarding. The SDN controller interacts with the switches and gathers information on the topology and on the network traffic. The SDN controller can include additional functionality, that we represent as additional modules in Fig.~\ref{fig:Architecture}. In particular, in our architecture we consider two monitoring modules. These modules are considered as a part of the controller to speed up the process and reduce the overhead of gathering information from the switches. 
        	
    The Control layer offers a set of functionality to the Application layer, through the Northbound API. The application and services in the Application layer use this API in order to implement the desired behavior. In our architecture, the proposed resource allocation/reallocation modules are included in the Application layer. These modules takes the decisions about the mapping of VNF chains into servers/VMs and the path selection. Then they request the Control layer to enforce them through the Northbound API. In the other direction, the Control layer provides the Application layer with the information about topology and network traffic. 
        	
    As can be seen in Fig. \ref{fig:Architecture}, the proposed architecture has the following modules: 
    \begin{itemize}
        \item \textit{Single Flow Resource allocator}: this module belongs to the application layer. When a new flow enters to the network, the Single Flow Resource allocator assigns the required resources to the flow. This module does not reroute existing flows and only focus on newly arrived ones. 
        \item \textit{Global Resource reallocator}: this module belongs to the application layer. It is capable to perform a reallocation of some flows as a reaction to some event (like congestion) or periodically to optimize the use of resource or the user perceived QoS.
        \item \textit{Knowledge Base}: this module is used to gather information about the previous states of the network and to predict the future state of the network\footnote{Note that monitoring modules and traffic prediction algorithms are out of the scope of this paper.}.
        \item \textit{Server/Network configurator}: this module in the Control Layer enforces the processing nodes and networking equipment to act based on the decision made in the Application layer. It will be mapped in an SDN controller.
        \item \textit{Congestion detector/predictor}: this module generates a reconfiguration request in two cases: i. when there is a congestion in the network or the traffic load on links goes above a predefined link utilization, and ii. when based on the network history and traffic pattern it predicts that the network will face congestion in future. This module is a part of the Control Layer, and can be also seen as a module of the SDN controller.
        \item \textit{Network monitoring}: this module periodically monitors the links status, updates the ``knowledge base'', and provides the traffic matrices for other modules. We assume that the switches perform some traffic measurements along with the forwarding of the packets and forward these measurements to the controller. Similar to the congestion detector/predictor, this module is a part of control layer and can be considered as a module of the SDN controller.
        
    {The Network Function Virtualization Management and Orchestration (NFV MANO) is an architectural framework for managing and orchestrating virtualized network functions (VNFs) and other software components. The European Telecommunications Standards Institute (ETSI) Industry Specification Group (ISG NFV) defined the MANO architecture to facilitate the deployment and connection of services as they are decoupled from dedicated physical devices and moved to virtual machines (VMs). Our proposed architecture along with the corresponding algorithms are considered as a higher layer which exploits this standard to apply the required management in VNFs. Therefore, our algorithm decides on the actions (where to put the VNFs) and the NFV MANO decides on how to apply them.}
        
    \end{itemize}

\subsection{Outline of the resource management procedures and algorithms}\label{sec:outline}
	In this subsection, a brief overview of the total process of allocation and reallocation of the resources is provided. As shown in Fig. \ref{fig:Architecture}, different algorithms can be used inside the Single Flow Resource Allocator and Global Resource Reallocator modules. In this context, the long-term algorithms specify whether a server is ON or OFF and the the placement of VNFs on the servers, by considering the network history information (called \textit{knowledge base}). On the other hand, short-term algorithms specify whether a VM is ACTIVE or IDLE on a server (but cannot turn on or off a server), by using  current status of the network (i.e., traffic load and link utilization).
    	
	Single Flow Resource Allocation is an online and short-term algorithm in which resources are assigned to the newly arrived flows in an instant manner. We mathematically formulate this problem as ``SFRA'' in section \ref{sect:SFRA}.
	On the other hand, the Global Resource Reallocation module reassign resources to the flows in a way that the energy consumption is optimized while the other constraints are considered so that the utilization of links and server is kept under control and the flow QoS is guaranteed. We mathematically formulate this problem as ``GRR'' in section \ref{sec:GRR}.
    	
    Since the optimization problems that we have defined (both SFRA and GRR) are NP-hard problems, several heuristics and relaxed versions are proposed for the different variants of resource allocation and reallocation procedures that we have considered. In the following, the proposed algorithms for each module is presented.
    	
    \begin{itemize}
    	\item  {\textit{Single Flow Resource Allocator}: this module is mathematically formulated in section~\ref{sect:SFRA}. The input of Single Flow Resource Allocation is one flow as a pair of (source, destination), the average rate of flows until now, and the current state of the network. The output is a set of links and servers assigned to that flow. When a new flow enters to the network, this algorithm should assign the required resources to the flow instantly, hence, in section \ref{sect:NSF} we propose a fast heuristic called ``Nearest Service Function first (NSF)'', which focuses on the second problem. Note that \textit{SFRA} and \textit{NSF} consider one flow at each time.}
        \item \textit{Online Long-Term Resource Re-allocator}: if some links are congested or congestion is predicted to happen soon, then an online long-term reconfiguration algorithm will be invoked. It should be mentioned that in this case, there is an estimation of the rates of the flows based on the current status of the network. ``Long Term Energy-aware NSF (LT-ENSF)'' is the fast VNF placement and energy-aware reallocation algorithm which performs this long-term reallocation. LT-ENSF considers all flows simultaneously (i.e., it can reconfigure the network for all flows at once). This algorithm does not only specifies the routing of the network, but also changes the state of servers from ON to OFF or vice-versa. This algorithm is described precisely in section \ref{section:LT-ENSF}. The input of this algorithm is similar to the Online Short-Term Resource Re-allocation, however, the output is the selected path for each flow, the state of all servers (ON or OFF) and the placement of VNFs.  
        \item \textit{Online Short-Term Resource Re-allocator}: when the network needs to be reconfigured in an online manner, an heuristic algorithm called ``Short Term Energy-aware NSF (ST-ENSF)'' is invoked (section \ref{sec:ST-ENSF}). ST-ENSF reroutes flows and changes the ACTIVE/IDLE state of servers that are in ON state. ST-ENSF is invoked periodically, update the configuration of the networks in order to optimize the energy consumption. This algorithm considers all flows, simultaneously (like LT-ENSF). It is notable that ST-ENSF does not only specify the new routing of existing flows but it also changes the ACTIVE/IDLE state of some servers. The input of this algorithm is the current state of network (the measured rate of active flows, the sources and destinations of active flows, the state of servers: ON (Active, IDLE) or OFF, the VNFs that can be supported on the servers, and the network topology). The output is the reassignment of resources to the flows: selected path for each flow and the state of currently ON servers (IDLE or ACTIVE). 
        \item \textit{Offline Long-Term Resource Reallocator}: in order to propose an offline Global Resource Reallocator, we relaxed the mathematical formulation proposed in section \ref{sec:GRR} defining a new optimization problem called ``Relaxed Resource Reallocator (3R)'' (section \ref{sec:3R}). We can solve this problem with a standard ILP solver and we refer to this procedure as 3R algorithm. If the 3R algorithm is invoked with the network topology and the rate of flows (measured or predicted), then it will perform an offline long-term resource reallocation. In this case, the 3R algorithm is considered as a long-term algorithm since it specifies whether a server is in ON or OFF state and it decides which VNFs are deployed in which server. 
        \item \textit{Offline Short-Term Resource Reallocator}: If the 3R algorithm (described in section \ref{sec:3R}) is invoked with the current state of network (the network topology, the measured rate of flows, the state of servers: ON (Active, IDLE) or OFF, and supported VNFs per servers), it performs an offline short-term reconfiguration. In this case, the 3R algorithm is considered as a short-term algorithm since it specifies whether an ON server be in IDLE or ACTIVE state. In brief, if we specify whether the servers are in ON or OFF state and the placement of the VNFs on the servers, then 3R works as a short-term algorithm, on the other hand, if this information is not specified then the 3R output is a long-term reconfiguration.
    \end{itemize}
    	    
    Algorithm \ref{alg:outline} presents an outline of the proposed management workflow, which combines the above described procedures. There is a timer called \textit{Long\_Term\_timer} that calls 3R algorithm which turns \textit{on} or \textit{off} the servers based on the predicted rate of flows (lines 2-4 of Algorithm \ref{alg:outline}). We call it Long\_Term\_timer because the impact of invoking 3R algorithm has a long term impact on the energy consumption of the networks. There is another event that may trigger the offline resource reallocation algorithm which is congestion prediction (lines 5-7). In this way, if a link utilization crosses a predefined threshold then the 3R algorithm is invoked with current state of the network as the input parameter. The current state of the network consists of the network topology, the measured rate of active flows, the state of servers: ON (Active, IDLE) or OFF, and supported VNFs per servers. 
    	    
    On the other hand, three different events may lead to perform an online reaction: 1) the arrival of a flow (line 9-11, NSF algorithm is invoked), 2) congestion detection/prediction (line 12-14, LT-ENSF algorithm is invoked), and 3) a predefined timer for network reconfiguration (line 15-18, ST-ENSF algorithm is invoked). It is notable that the knowledge base is updated periodically (lines 20-22).
        
    	\begin{algorithm}
        	\caption{Outline}
        	\label{alg:outline}
        	\normalsize
        	\allowdisplaybreaks
        	\begin{algorithmic}[1]
            	\break
            	\While{true}
            	\If{Long\_Term\_timer elapses}
            	    \State{\textit{$3R$(Knowledge Base)}}
            	    \State{Reset Long\_Term\_timer}
            	\ElsIf{Congestion\_Avoidance\_Alarm elapses}
            	    \State{\textit{$3R$(Current state)}}
            	    \State{Reset Long\_Term\_timer}
                \Else
                	\If{new flow arrives}
                	    \State{\textit{NSF()}}
                	\EndIf
                	\If{congestion is detected or predicted}
                	    \State{\textit{LT-ENSF()}}
                	    \State{Reset GRR\_timer}
                	\ElsIf{GRR\_timer elapses}
                	    \State{\textit{ST-ENSF()}}
                	    \State{Reset GRR\_timer}
                	\EndIf
                \EndIf
                \If{Update\_timer elapses}
            	    \State{\textit{Update\_Knowledge\_Base()}}
            	    \State{Reset Update\_timer}
            	\EndIf
            	\EndWhile
        	\end{algorithmic}
    	\end{algorithm}       	
            
\section{Problem Formulation}\label{problemForumlation}
\begin{figure}[!htbp]
    	\begin{center}
    		\includegraphics[width=1\columnwidth]{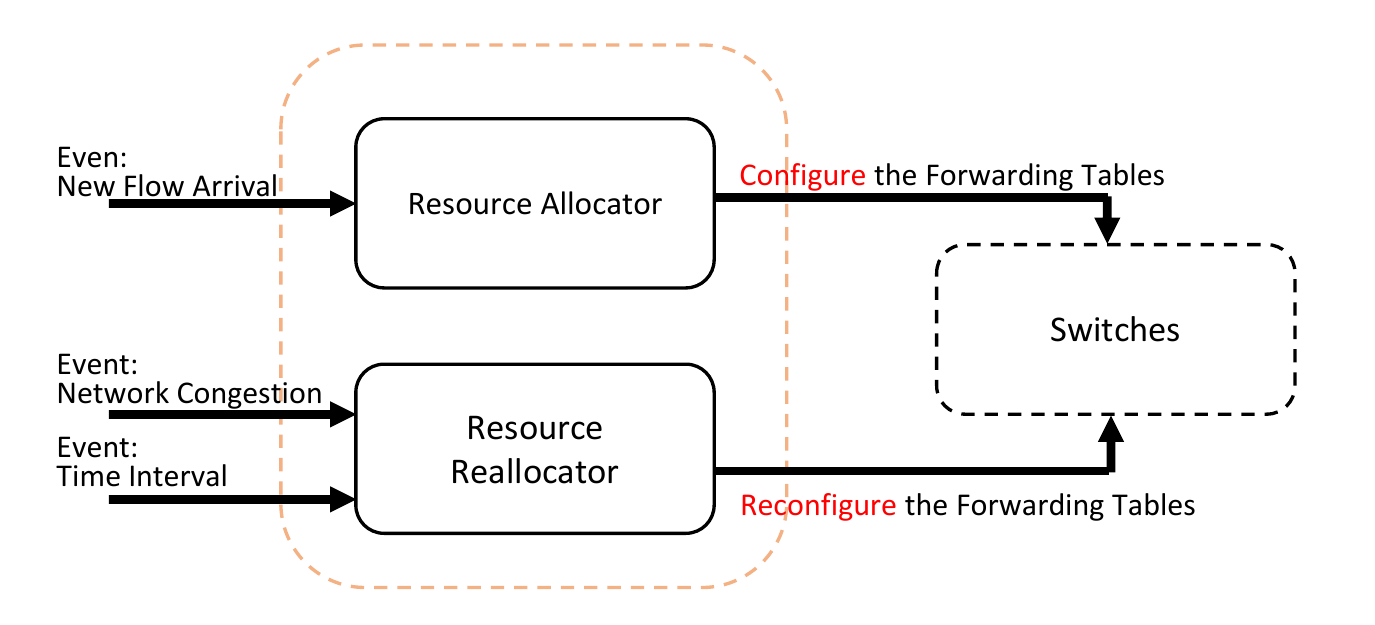}
    		\caption{Software Architecture.}
    		\label{fig:software_architecture}
    	\end{center}
        \end{figure}

	In this section, the components of the proposed architecture are discussed in detail. As it can be seen in Fig. \ref{fig:software_architecture}, two types of algorithms are used in the proposed architecture: 1) Single Flow Resource Allocation (SFRA), and 2) Global Resource Reallocation (GRR). The Single Flow Resource Allocation algorithms considers one flow at a time. The output of these algorithms provide the new entries to be inserted into the forwarding tables of the switches to route the new flow. On the other hand, the Global Resource Reallocation algorithms consider all flows, simultaneously, i.e., they consider the impact of flows on each other. GRR algorithms updates the forwarding tables of switches to reroute existing paths. When there is no information about the rate of a new flow, SFRA algorithms are invoked to allocate the resources to this new flow. On the other hand, GRR algorithms are used to reallocate resources based on the estimated rates of the existing flows. In algorithm~\ref{alg:outline}, NSF is a Single Flow Resource Allocation algorithm while ST-ENSF, LT-ENSF and 3R are Global Resource Reallocation algorithms.
	    
    Briefly, in Fig. \ref{fig:software_architecture}, if a new flow arrives at a switch, the controller uses an SFRA algorithm to route the flow. Thereafter, the controller configures the switches that on the selected path by sending controlling messages to them. On the other hand, if the network gets congested (or, if a predefined time intervals elapsed), a global reconfiguration can help to reduce the congestion and/or the energy consumption, improve the throughput, and ensure that the requirements of the flows are fulfilled. The configuration algorithms are executed in the Application layer. To this end, the controller collects the information and detects/predicts the congestion and informs the application layer. Thereafter, the controller will receive the instructions from the Application Layer and enforce these decisions by communicating with the switches.
        
\subsection{System Model and Assumptions}
    We assume an SDN-based network using a southbound protocol to program the switches dynamically. Each flow requires to pass through a sequence of VNFs in its path from the source switch to the destination switch (\textit{SFC requirements} of a flow). As for \textit{QoS constraints}: i) each flow has a maximum tolerable propagation delay that should be guaranteed; ii) the utilization of each link has not to exceed a given threshold; and iii) the utilization of each server (i.e., its processing load) has to be lower than a given threshold. The servers have \textit{three} different modes: OFF, ON-IDLE (in short IDLE), ON-ACTIVE (in short ACTIVE). In the OFF mode, the energy consumption of the server is negligible since it is in a sleep state, which may be exploited to remotely turn on the server via software if needed. For example, the network card of the server may be required to be powered on, even it the server is off, to detect Wake on Lan (WoL) packets (see e.g., \cite{chiaraviglio2010polisave}). On the other hand, when the ACTIVE mode is set, the server works on its full performance, and the energy consumption is the same as the full rate working energy consumption. {Finally, in the IDLE mode, the energy consumption is a fraction $\delta$ of the energy consumption in ACTIVE mode.This parameter depends on the server and its features in terms of power management. In particular, the $\delta$ parameter can be set from the power requirements reported on the server data sheets or from real measurements (see e.g., \cite{chiaraviglio2017measurement}).} Our goal is to configure the network in a way that optimizes the \textit{energy consumption} and the \textit{number of flow table entries} (or \textit{flow table changes}), while simultaneously meeting both the SFC requirements and the QoS constraints. Minimizing the table entries will result in minimizing the average number of hops. Consider a selected path $p$ for a flow $f$. If $p$ has $l$ hops then it should cross $l$ switches in its network path. This results in $l$ table entries into the switches. Hence, the total number of table entries is proportional to the average number of hops for the flows. Minimizing the flow table changes reduces the reconfiguration overhead of network, i.e. the number of control messages that the controller will send to the switches. Consequently it can reduce the reconfiguration time and the control load on the network devices.
    	    
        \begin{table}[t]
        	\caption{Main Notation.}\label{tab:notation}
        	\centering
        	\resizebox{1\columnwidth}{!}{%
        	\begin{tabular}{|c|l|l|} 
        		\hline
        		& \textbf{Symbol} & \textbf{Description}\\
        		\hline
        		& $\mathcal{N}$ & Set of nodes (switches and servers)\\
        		& $\mathcal{F}$ & Set of flows\\
        		& $\mathcal{X}$ & Set of VNFs\\
        		& $N$ & Number of switches: $\abs{\mathcal{N}}\triangleq N$ \\
        		& $F$ & Number of of flows: $\abs{\mathcal{F}}\triangleq F$  \\
        		& $X$ & Number of different VNFs : $\abs{\mathcal{X}}\triangleq X$  \\
        		& $E$ & Number of links  \\
        		& $i,\:j$ & indexes of switches\\
        		& $x$ & indexes of VNFs\\
        	    & $f$ & indexes of flows\\
        	    & $\Psi$ & Maximum number of requested VNFs\\
        	    & $K^f_{\psi}$ & Sequence of requested VNFs\\
        		& $B^{\max}_{(i,j)}$ & Matrix of link capacity\\
        		\multirow{6}{0.1cm}{\begin{sideways}\textbf{Parameters}\end{sideways}} 
        		& $L_{(i,j)}$& Current traffic load\\
        		& $D_{(i,j)}$ & Links propagation delay\\
        		& $T^f$ &  Vector of traffic rate\\
        	    & $D^{\max}_{f}$ & Maximum tolerable propagation \\ 
        		& $s^{f}$ & Vector of source switch\\
        		& $d^{f}$ & Vector of destination switch\\
        		& $M^f_{(i,j)}$ & Current routing matrix: 1 means flow $f$\\&&~~~~~~~~~ crosses link $i\rightarrow j$ \\
        		& ${\mu}_L$ & Maximum link utilization ratio \\
        		& ${\mu}_S$ & Maximum server utilization ratio \\ 
        		& $P_{x}$ & Required processing of $x^{th}$ VNF \\
        		& ${C}^{\max}_{i}$ & Servers processing capacity\\
        		& $\rho_{(i,x)}$ & Current processing load:\\&&~~~~~~~~~ per $x^{th}$ VNF on $i^{th}$server \\ 
        		& $S_{(i,x)}$ & Matrix of supported VNFs: 1 means VNF\\&&~~~~~~~~~ $x$ associated with server $i$\\ 
        		& $V^f_{x}$ & Requested VNFs\\
        		& $\mathcal{E}_{i}$ & Energy consumption of servers \\
        		& $\mathcal{E}'_{i}$ & {Context based energy consumption of servers} \\
        		& $\delta$ & Coefficient of energy consumption in\\&&~~~~~~~~~  IDLE mode\\
        		& $O^{t-1}_{i}$ & Previous status of servers (ACTIVE/IDLE \\&&~~~~~~~~~ or OFF/ON) \\ \hline
        		\multirow{4}{0.1cm}{\begin{sideways}\textbf{Variable}\end{sideways}}
        		& $R_{(i,j)}^f$ & rerouting matrix \\ 
        		& $Q^f_{(i,j)}$& rerouting matrix with nodes order \\ 
        		& $U_{(i,x)}^f$ & 1 means Flow $f$ receives service from\\&&~~~~~~~~~ VNF $x$ in server $i$\\ 
        		& $O^t_{i}$ & Next modes of servers (ACTIVE/IDLE or\\&&~~~~~~~~~ ON/OFF)  \\
            \hline
            	\end{tabular}
            	}
            \end{table}

    Table \ref{tab:notation} summarizes the notation used in the paper. Consider a network with $N$ SDN-enabled switches, we represent the network topology with a matrix $B^{\max}_{N\times N}$ where $b_{(i,j)}$ denotes the capacity of the link from the switch $i$ to the switch $j$. Similarly, the propagation delay of links is modeled via matrix $D_{N\times N}$ where $d_{(i,j)}$ denotes the propagation delay of the link from the switch $i$ to the switch $j$. The current traffic load on each link is expressed using $L_{N\times N}$. Let $F$ be the number of flows in the network. In order to simplify the understanding of the notation, a sample for the each element of the proposed notation is presented. Therefore, consider the topology illustrated in Fig. \ref{fig:Architecture} (the links are represented with solid blue lines), the matrix $B$ and $D$ are as following ($N=5$):
            \[B^{\max}=
              \begin{bmatrix}
                0 & \textcolor{red}{b_{(1,2)}} & \textcolor{red}{b_{(1,3)}} & 0 & 0\\
                \textcolor{red}{b_{(2,1)}} & 0 & 0 & \textcolor{red}{b_{(2,4)}} & 0\\
                \textcolor{red}{b_{(3,1)}} & 0 & 0 & 0 & \textcolor{red}{b_{(3,5)}}\\
                0 & \textcolor{red}{b_{(4,2)}} & 0 & 0 & \textcolor{red}{b_{(4,5)}}\\
                0 & 0 & \textcolor{red}{b_{(5,3)}} & \textcolor{red}{b_{(5,4)}} & 0
              \end{bmatrix}\]
            \[D=
              \begin{bmatrix}
                \infty & \textcolor{red}{d_{(1,2)}} & \textcolor{red}{d_{(1,3)}} & \infty & \infty\\
                \textcolor{red}{d_{(2,1)}} & \infty & \infty & \textcolor{red}{d_{(2,4)}} & \infty\\
                \textcolor{red}{d_{(3,1)}} & \infty & \infty & \infty & \textcolor{red}{d_{(3,5)}}\\
                \infty & \textcolor{red}{d_{(4,2)}} & \infty & \infty & \textcolor{red}{d_{(4,5)}}\\
                \infty & \infty & \textcolor{red}{d_{(5,3)}} & \textcolor{red}{d_{(5,4)}} & \infty
              \end{bmatrix}\]
            
    The vectors $s^f$ and $d^f$ determine the source and destination of flows, respectively. The matrix $M_{N\times N\times F}$ is the current routing matrix, e.g., if $M_{(i,j)}^f \in \{0,1\}$ is equal to 1 then the flow $f \in \mathcal{F}$ crosses the link $i\rightarrow j$. If we set $f=1$, $s^1=1$, and $d^1=2$, then the matrix $M^1$ is as follows:
            \[M^1=
              \begin{bmatrix}
                0 & 0 & \textcolor{red}{1} & 0 & 0\\
                0 & 0 & 0 & 0 & 0\\
                0 & 0 & 0 & 0 & \textcolor{red}{1}\\
                0 & \textcolor{red}{1} & 0 & 0 & 0\\
                0 & 0 & 0 & \textcolor{red}{1} & 0
              \end{bmatrix},\]
    where $s^1=1$ indicates that the source of the flow is the switch number 1. Therefore, we should trace the path from the first row of $M^1$. As can be seen, the third element of $M^1$ in the first row is one which means that the flow should leave the switch 1 toward the switch 3. At this point, the third row of $M^1$ should be checked. Since the $5^{th}$ column of the third row is 1, the flow will leave switch number 3 to reach the switch number 5. Thereafter, the flow will go to switch number 4 because the fourth element of row 5 in matrix $M^1$ is one. Finally, since the second column of the forth row is one, the flow will go to switch number 2. Note that we consider loop-free routing, i.e. nodes and links cannot be used twice in the routing of a flow. We will enforce this behavior with specific constraints in our formulation. The new routing matrix $M_{N\times N\times F}$ specifies the path selected for each flow
            
    On the other hand, matrix $Q_{(N,N,F)}$ is an \textit{ordering-aware} rerouting matrix, which explicitly includes the notion of order of links and nodes in the path. If flow $f$ crosses the link $i\rightarrow j$, $Q^f_{(i,j)}>0$ (in particular $Q^f_{(i,j)}$ specifies the number of previously crossed switches), otherwise $Q^f_{(i,j)}=0$. Besides, $Q^f_{(d^f,d^f)}$ is the path length for each flow (number of switches through the selected path). Accordingly, the matrix $Q^1$ that represents the same flow $f=1$ considered above will be written as:
        \begin{align}
        \label{formDec:Q}
        Q^1=\begin{bmatrix}
            0 & 0 & \textcolor{red}{1} & 0 & 0\\
            0 & \textcolor{red}{5} & 0 & 0 & 0\\
            0 & 0 & 0 & 0 & \textcolor{red}{2}\\
            0 & \textcolor{red}{4} & 0 & 0 & 0\\
            0 & 0 & 0 & \textcolor{red}{3} & 0
          \end{bmatrix},
        \end{align}
    
    In the $Q^1$ matrix, since the third column of the first row is one, the flow should leave the source switch to reach the switch number 3. Also, value 2 in the third row of the matrix specifies that the flow in the second step will leave switch number 3 to reach the switch number 5. Similarly, value three in the fifth row states that as the third step the flow will leave switch number 5 to reach to the switch number 4. Finally, the flow will go to switch number 2 which is the destination of such flow. Since the number of crossed switches in this path is five, the value of $Q^1_{2,2}$ is five.
    
    The \textit{flow rate requirement} vector $C_{F}$ specifies flows requirements. The $i^{th}$ row of the mentioned vector defines the traffic rate requirement of the $i^{th}$ flow. Similarly, vector $T_{F}$ specifies the maximum tolerable propagation delay of flows. Considering $X$ different VNFs, each flow can request at most $\Psi\leq{X}$ VNFs. 
    Matrix $R_{F\times X}$ shows the requested VNFs for each flow. Indeed, if the VNF $x$ is requested for the flow $f$, then $R_{(f,x)}$ is 1, otherwise, it is 0. The sequence of the required VNFs for all the flows is expressed by matrix $K_{F\times \Psi}$ where $K_{(f,\omega)}$ specifies the $\omega^{th}$ required VNF for flow $f$. As an example, considering $X=4$ and $\Psi=3$, taking for example flow $f=1$, the matrix $V^1$ and $K^1$ are as follows:
            \[V^1=
              \begin{bmatrix}
                0 & {\textcolor{red}{1}} & {\textcolor{red}{1}} & 0
              \end{bmatrix},\ \ \ 
            K^1=
              \begin{bmatrix}
                {\textcolor{red}{3}} & {\textcolor{red}{2}} & 0
              \end{bmatrix},\]
            
    The second and the third elements of $V^1$ are 1 which mean that VNF number 2 and 3 should deliver service to this flow. Since matrix $K^1$ specifies the ordering of the VNFs, the flow needs to receive service from the VNF 3 before VNF 2.
        
    The required processing capacity of each VNF for a unit of flow rate is expressed by the vector $P_X$, where $P_x$ specifies the required processing capacity of VNF $x\in \mathcal{X}$. Therefore, the VNF $x$ will require a processing capacity $P_x\cdot C_f$ to process the flow $f$ with rate $C_f$. The vector ${C}^{\max}_N$ identifies the processing capacities for each server. The current processing load of each VNF on a server is stated via $\rho_{N\times X}$, where $\rho_{(i,x)}$ specifies the current processing load of VNF $x$ on server $i$. The VNFs associated with each server are identified by matrix $S_{N\times X}$, therefore, $S_{(i,x)}$ specifies whether VNF $x$ is supported by server $i$ or not. We consider a server (or a cluster of servers) connected to each switch. If no server is connected to switch $i$ then $\sum_{x=1}^X(S_{(i,x)})=0$ (similarly for an OFF server). $U^F_{N\times X}$ assigns the VNFs and servers to the flows. If $U^f_{(i,x)}$ is 1, then flow $f$ receives service from VNF $x$ on server $i$. Taking $f=1$, an example matrix $U^1$ is:
        \[U^1=
          \begin{bmatrix}
            0 & 0 & {\textcolor{red}{1}} & 0\\
            0 & 0 & 0 & 0\\
            0 & 0 & 0 & 0\\
            0 & {\textcolor{red}{1}} & 0 & 0\\
            0 & 0 & 0 & 0
          \end{bmatrix}.\]
          
    The third element of $U^1$ in the first row is one, it means that the flow uses the VNF 3 in the server 1. Similarly, since the second column of the fourth row is one, the VNF 2 will be delivered to the flow in server~4.
    The vector $\mathcal{E}_N$ states the energy consumption of servers, where $\mathcal{E}_i, i \in \mathcal{N}$ specifies the energy consumption of servers $i$. {The vector $\mathcal{E}'_N$ is the context based energy consumption of servers. In long-term algorithms it is the absolute energy consumption of a node $\mathcal{E}'=\mathcal{E}$ while in short-term algorithms it is the difference of energy consumption between the IDLE and the ACTIVE mode $\mathcal{E}'=\mathcal{E}-(\delta\cdot \mathcal{E})$.} $O^{t-1}_N$ and $O^t_N$ specify the current and next state of servers respectively. Depending of the specific formulation, $O^{t-1}_N$ and $O^t_N$ will be used to represent the ON/OFF state or the IDLE/ACTIVE state of a node. Finally, $\mu_L$ and $\mu_S$ state the maximum link and server utilization, respectively.
            
\subsection{{Mathematical Formulation of constraints}}\label{section:formulation}
    In order to simplify the understanding of the constraint formulation, we divide it into three parts: 
	    \begin{itemize}
	        \item QoS constraints, \textit{unordered} SFC constraints (i.e., the order of VNFs for flows is not important) and objective function related to the number of forwarding table entries
	        \item SFC constraints with VNF Ordering
	        \item Energy consumption constraints 
	    \end{itemize}
	    
	The first part is modeled as a Binary Linear Programming (BLP) while the other two parts are in the form of Integer Quadratic Programming (IQP). In Section \ref{section:ConvertToLinear}, the nonlinear equations are converted to linear form, therefore, we will be able to use common ILP solvers for the optimization problems.
	    
\subsubsection{QoS and SFC Constraints without VNF Ordering}\label{section:formulationNoOdering}
    in the formulation reported in this section we define the constraints that force the solution to satisfy the QoS requirements of a flow (i.e., delay and traffic rate). Moreover we guarantee that the required VNFs for each flow will be crossed by the selected path for that flow. The proposed formulation is as follows:
        \begin{align}
            &\sum^N_{i=1}{U_{(i,x)}^f}= V^f_{x},\ \forall x\in \mathcal{X},\ \forall f\in \mathcal{F},\label{eq2} \\
            &\sum^N_{i=1}{R_{(i,j)}^f}\geq U^f_{(j,x)},\ \forall x\in \mathcal{X},\ \forall j\in \mathcal{N}-\{s^f\},\ \forall f\in \mathcal{F},\label{eq3} \\
            &U_{(i,x)}^f\leq S_{(i,x)},\ \forall f\in \mathcal{F},\ \forall i\in \mathcal{N},\ \forall x\in \mathcal{X},\label{eq4} \\
            &\sum^N_{i=1}{U_{(i,x)}^f}\leq 1,\ \forall f\in \mathcal{F},\:\forall x\in \mathcal{X},\label{eq5} \\
            &\sum^X_{x=1}{\left(\sum_{f=1}^F{\left(U_{(i,x)}^f\ \cdot T^f\cdot P_x\right)}+\rho_{(i,x)}\right)}\leq {{\mu}_L }^'\cdot {C}^{\max}_i,\nonumber\\
                &\forall i\in \mathcal{N},\label{eq6} \\
            &\sum_{f=1}^F{R_{(i,j)}^f\cdot T^f}+L_{(i,j)}\leq {\mu}_L \cdot B^{\max}_{(i,j)},\  \forall i,j\in \mathcal{N},\label{eq7} \\  
            &\sum^N_{j=1}{R_{(i,j)}^f}-\sum^N_{j=1}{R^f_{(j,i)}}=
                    \begin{cases}
                    1 & i=s^f\\
                    -1 & i=d^f\\
                    0 & i\neq s^f,d^f\\
                    \end{cases},\nonumber\\
                &\forall f\in \mathcal{F},\ \forall i\in\mathcal{N},\label{eq10} \\
            &\sum^N_{j=1}{R_{(i,j)}^f}\leq 1,\ \forall i\in \mathcal{N},\ \forall f\in \mathcal{F},\label{eq11} \\
            &\sum^N_{i=1}{\sum^N_{j=1}{\left(R_{(i,j)}^f\ \cdot\ D_{(i,j)}\right)}}\leq D^{\max}_f,\ \forall f\in \mathcal{F},\label{eq12} \\  
            &U^f_{(i,x)},\ R_{(i,j)}^f\in \left\{0,\ 1\right\},\ \forall i,j\in \mathcal{N},\ \forall f \in \mathcal{F},\ \forall x \in \mathcal{X}.\nonumber
        \end{align}
    The constraint~\eqref{eq2} indicates that each flow crosses a valid VNF chain while passing through the switches (without considering the order of VNFs). Moreover, constraint~\eqref{eq3} imposes that the VNF is delivered only on crossed servers. Constraint~\eqref{eq4} checks whether the requesting VNF is supported on the specified server. Constraint~\eqref{eq5} is used to prevent gaining a VNF more than once for each flow. Constraint~\eqref{eq6} ensures that the server processing capacity to support VNFs is not exceeded. Focusing on the link capacity, constraint~\eqref{eq7} checks the link capacity between each pair of switches.\\ 
    Focusing on the flow conservation, constraint~\eqref{eq10} presents the flow management limitations. In particular, the first inequality  prevents returning to the source or leaving the destination. The second inequality imposes leaving the source switch and entering to the destination switch for each flow. The third inequality forces the input and output of each server to be equal (except for the source and destination). In order to prevent loops for each flow, constraint~\eqref{eq11} is applied. Focusing on the propagation delay, constraint~\eqref{eq12} is used to control the propagation delay for each flow. 
            
\subsubsection{SFC Constraints with VNF Ordering}\label{section:formulationWithOrdering}
	in order to consider the Service Function chaining by considering an ordered sequence of VNFs the following constraint should be added:
	    \begin{align}
        	&Q^f_{(i,j)}\geq R_{(i,j)}^f,\ \forall f\in \mathcal{F},\forall i,j\in \mathcal{N},\label{eq13} \\	
            &Q^f_{(i,j)}=Q^f_{(i,j)}\cdot R_{(i,j)}^f,\ \forall i\in \mathcal{N}-\left\{d^f\right\},\:j\in \mathcal{N},\nonumber\\
                &\forall f\in \mathcal{F},\label{eq14}\\
            & Q^f_{(d^f,i)}=0,\ \forall f\in \mathcal{F},\forall \mathcal{N}-\left\{d^f\right\},\label{eq15}\\
            &\sum^N_{j=1}{Q^f_{(i,j)}}=\sum^N_{j=1}{Q^f_{(j,i)}}+\sum^N_{j=1}{R^f_{(j,i)}},\ \forall f\in \mathcal{F},\nonumber\\
                &\forall i\in \mathcal{N}-\{s^f,d^f\},\label{eq16}\\
            &Q^f_{(d^f,d^f)}=\sum^N_{j=1}{Q^f_{(j,d^f)}}+\sum^N_{j=1}{R^f_{(j,d^f)}},\ \forall f\in \mathcal{F},\label{eq17}\\
            &\sum^N_{j=1}{Q^f_{(s^f,j)}}=1,\ \forall f\in \mathcal{F},\label{eq18}\\
            &\sum^N_{i=1}{\sum^N_{j=1}{\left(Q^f_{(i,j)}\cdot U^f_{(i,K^f_{V^f})}\right)}}\geq \sum^N_{i=1}{\sum^N_{j=1}{\left(Q^f_{(i,j)}\cdot U^f_{(i,K^f_{Z_{V^f}})}\right)}}\nonumber\\
                &,\forall f\in \mathcal{F},\forall V^f\in \left\{1,\dots ,len\left(K^f\right)\right\},\nonumber\\
                &\forall Z_{V^f}\in \left\{1,\dots ,V^f-1\right\},\label{eq19}\\
            &Q^f_{(i,j)}\in \mathbb{Z}^{\geq 0},\ \forall i,j\in \mathcal{N},\ \forall f\in \mathcal{F}.\nonumber
        \end{align}
    where constraint~\eqref{eq13} indicates that the value of the ordering matrix should be higher or equal to the rerouting matrix. Also, constraint~\eqref{eq14} hints that if the value of matrix $R_{(i,j)}^f$ is zero, then $Q^f_{(i,j)}$ becomes zero. Due to the fact that for the destination switch all output links are zeros, constraint~\eqref{eq15} is added. Constraint~\eqref{eq16} enforces that if a flow enters to a switch in its $a^{th}$ step, then it would leave that switch in the ${(a+1)}^{th}$ step, however, source and destination switches are exceptions.\\
    Since the value of $Q^f_{(d^f,d^f)}$ should be the number of crossed switches, constraint~\eqref{eq17} is considered. Focusing on the integrity of the ordering matrix, constraint~\eqref{eq18} assures that flows leave the source switch. Constraint~\eqref{eq19} guarantees the sequence of VNF chaining. To this end, for each VNF, constraint~\eqref{eq19} checks whether the VNFs with higher ordering (lower index in $K^f$) are delivered to the flow in one of the crossed servers or not. In this way, constraint \eqref{eq19} exploits i) variable $V^f$ and ii) set $Z_{V^f}$. Variable $V^f$ states the index of each required VNF for flow $f$. The set $Z_{V^f}$ contains all required VNFs with a higher order (lower index) than $V^f$. For example if $K^f= \left[\begin{smallmatrix}3\  & 2\  & 1\  & 6\end{smallmatrix}\right]$ then $V^f \in \{1,2,3,4\}$. If we consider $V^f=3$ then $Z_{V^f}$ is a member of $\{1,2\}$.
            
\subsubsection{Energy consumption constraints}\label{section:formulationWithEnergy}
    we extend the above formulations with the objective to consider the number of engaged servers and their overall energy consumption. The new constraints are as following:
    \begin{align}
            & O^t_i\leq \sum^F_{f=1}{\sum^X_{x=1}{U_{(i,x)}^f}}\,\ \forall i\in \mathcal{N}, \label{eq21}\\
            &O^t_i\cdot \sum^F_{f=1}{\sum^X_{x=1}{U_{(i,x)}^f}}=\sum^F_{f=1}{\sum^X_{x=1}{U_{(i,x)}^f}},\ \forall i\in \mathcal{N}. \label{eq22}
    \end{align}
    
    Constraint~\eqref{eq21} states that if no VNFs are used for any flow in a server, the server will not be used. On the other hand, Constraint~\eqref{eq22} specifies the essential servers for active flows (i.e., those servers that deliver at least one VNF to a flow should be in ON mode). 
            
\subsubsection{Converting Nonlinear Constraints to Linear Forms} \label{section:ConvertToLinear}
    in the above proposed formulations, constraints \eqref{eq14}, \eqref{eq19}, and \eqref{eq22} are nonlinear (i.e., only the first problem ''QoS and SFC Constraints without VNF Ordering'' is linear). In order to make the formulation suitable for solving large-scale problems (i.e with large number of flows and complex network configurations),  these constraints are converted into linear form. 
    Constraint~\eqref{eq14} is aimed to ensure that if $R^f_{i,j}$ is zero, then $Q^f_{(i,j)}$ becomes zero. Since a flow at most traverses all servers, the value of $Q^f_{(i,j)}$ is always less than or equal to $N$. Therefore, constraint~\eqref{eq14} can be substituted with constraint~\eqref{eq23}.
    \begin{align}
        &Q^f_{(i,j)}\leq N\cdot R_{(i,j)}^f,\ \forall i\in \mathcal{N}-\left\{d^f\right\},\ j\in \mathcal{N},\forall f\in \mathcal{F}.\label{eq23}
    \end{align}
            
    In order to replace the constraint~\eqref{eq19} with a linear one, we take a different approach to express the ordering constraints of VNFs belonging to a flow. The new proposed formulation is reported in constraint~\eqref{eq24}. If the server $i$ hosts the VNF $K_{V^f}$, i.e., $U^f_{(i,K_{V^f})}=1$, then the left hand side of~\eqref{eq24} considers the step of the server $i$ and it must be greater than the step of all servers $I$ hosting a VNF with index less than the index of VNF $K_{V^f}$ in $K^f$. Consider $Z_{V^f}$ as the index of any VNF in $K^f$ with an index less than VNF $K^f_{V^f}$, i.e., flow $f$ must pass VNF $K^f_{Z_{V^f}}$ before $K^f_{V^f}$. If the server $I$ hosts the VNF $K^f_{Z_{V^f}}$, i.e., $U^f_{(i,K^f_{Z_{V^f}})}=1$, then the right hand side of~\eqref{eq24} considers the step of the server $I$ and it must be greater than the step of all servers $i$ hosting a VNF with index greater than the index of VNF $K^f_{Z_{V^f}}$ in $K^f$. Since the value of $\sum^N_{j=1}{Q^f_{(i,j)}}$ is always less than $(2n-1)$, if one of $U^f_{(i,K_{V^f})}$ or $U_{(I,K_{Z^f})}^f$ is zero, then the equation is true \footnote{The value of $\sum^N_{j=1}{Q^f_{(i,j)}}$ and $\sum^N_{i=1}{Q^f_{(i,j)}}$ are always less than $2N-1$ because in the worst case, the flow crosses all switches which means that the value of $\sum^N_{j=1}{Q^f_{(i,j)}}$ is at most $(N-1)+N$. This can be visualized considering the representation of the $Q$ matrix in equation \eqref{formDec:Q}, a column can have at most two elements and they can be at most N and (N-1)}. As we mentioned before, the destination has a flow to itself with a step of at most $N+1$, therefore, in cases that both $U^f_{(i,K_{V^f})}$ and $U_{(I,K_{Z^f})}^f$ are equal to one the constraint is met \textit{if and only if} the value of $\sum^N_{j=1}{Q^f_{(i,j)}}$ is greater than $\sum^N_{j=1}{Q^f_{(I,j)}}$. This means that a server which delivers the lower index VNF is crossed before servers that deliver higher index VNFs. 
    \begin{align}
        &\left(1-U^f_{(i,K^f_{V^f})}\right)\cdot \left(2N-1\right)+\sum^N_{j=1}{Q^f_{(i,j)}}\geq \nonumber\\
            &\left(U^f_{(I,K^f_{Z_{V^f}})}-1\right)\cdot \left(2N-1\right)+\sum^N_{j=1}{Q^f_{(I,j)}\ },\nonumber\\
            & \forall f\in \mathcal{F},\forall V^f\in \left\{1,\dots ,len\left(K^f\right)\right\},\nonumber\\
            & \forall Z_{V^f}\in \left\{1,\dots ,V^f-1\right\},\ \forall I, i\in \mathcal{N}.\label{eq24}
    \end{align}
            
    Finally, a proper linear replacement for constraint~\eqref{eq22} is stated in the following. In~\eqref{eq22}, the value of $\sum^F_{f=1}{\sum^X_{X=1}{U_{(i,x)}^f}}$ is always less than $(1+F\cdot X)$. Considering the aforementioned inequality, the constraint~\eqref{eq25} is satisfied \textit{if and only if} the value of $O^t_i$ is one for servers that deliver VNFs to the flows (i.e., servers that have $\sum^F_{f=1}{\sum^X_{X=1}{U_{(i,x)}^f}}>0$). Otherwise, (i.e., if $\sum^F_{f=1}{\sum^X_{x=1}{U_{(i,x)}^f}}$ be equal to zero) the value of $O^t_i$ is zero.
    \begin{align}
            &\left(1+F\cdot X\right)O^t_i\geq \sum^F_{f=1}{\sum^X_{X=1}{U_{(i,x)}^f}},\ \forall i\in \mathcal{N}. \label{eq25}
    \end{align}
            
\subsection{Single Flow Resource Allocation (SFRA)}\label{sect:SFRA}
    {We use the constraints that are proposed in subsection \ref{section:formulation} to formulate the problem of resource allocation to a newly arrived flow. The formulation is as following:}
    \begin{align}
        &{\min_{R} \ \ }\sum^N_{i=1}{\sum^N_{j=1}{R_{(i,j)}^f}} ,\label{eq1} \\	
        &\textit{Subject to:}\nonumber \\
        &Eq.\ \eqref{eq2}-\eqref{eq13},\ \eqref{eq15}-\eqref{eq18},\ \eqref{eq23}-\eqref{eq24}. \nonumber
    \end{align}
    
    {The objective function \eqref{eq1} minimizes the number of elements that should be imported into forwarding tables of switches. In particular the objective function \eqref{eq1} minimizes the number of hops that the flow will cross from source to destination. Note that the formulation of the problem considers a number $F$ of flows to be allocated, but when dealing with Single Flow Resource Allocation only one flow is considered ($F=1$). We consider that there is no information about the rate of newly arrived flow $f$, therefore, i) the average rate of flows until now is considered as the rate of flow $f$; ii) \textit{SFRA} does not focus on minimizing the energy consumption, otherwise, it may cause congestion for big flows.}
    
    \begin{thm7}{\textit{SFRA} falls in the class of NP-Hard problems.}\label{LMA:SFRANP}\end{thm7}
    \begin{proof}
    suppose that the energy consumption of all servers are zero, $F=1$, $X=0$, $\alpha=1$, and $V^1=\emptyset$. For any types of network, we define $B'$ and $M'$ as  follows:\\\\$B'_{(i,j)}=\begin{cases}0 & B^{\max}_{(i,j)}=0\\ 1 & B^{\max}_{(i,j)}\neq 0\\ \end{cases}$\ \ \ \ \ \ \ \ \  and\ \ \ \ \ \ \ \ \  $M'=0\times M$\\\\
    consider a network with the following input parameters $B'$, $M'$, $X$, $F$, and $V$. The problem is to find a path where end-to-end delay should be less than a predefined threshold with respect to the length of the path (number of forwarding table elements). As can be seen, the problem is mapped to the \textit{weight constrained shortest path problem (WCSPP)} which is an NP-hard problem \cite{dumitrescu2001algorithms}.
    \end{proof}
 
\subsection{Global Resource Reallocation (GRR)}\label{sec:GRR}
    In this subsection, the problem of resource reallocation is presented to optimize jointly the energy consumption and the network reconfiguration overhead. Network reconfiguration overhead (or, precisely, the flow rerouting overhead) depends on the number of rerouted flows. Increasing the number of rerouted flows not only may result in the network instability, but also it may increase the packet loss and end-to-end delay. The $GRR$ formulation is as follows:
        \begin{align}
            &{\min}\ \ \  \left[\alpha\left(\sum^N_{i=1}{\sum^N_{j=1}{\sum^F_{f=1}{\left|R_{(i,j)}^f-M^f_{(i,j)}\right|}}}\right)/\left(\left(N-1\right)\cdot F\right)\right.\nonumber\\
            &\left.+(1-\alpha)\left(\sum^N_{i=1} O^{t}_i\cdot \mathcal{E}'_i\right)/\sum^N_{i=1} \mathcal{E}'_i\right], \label{eq26}\\
            &\textit{Subject to:},\nonumber\\
            &Eqs.\ \eqref{eq2}-\eqref{eq5},\ \eqref{eq10}-\eqref{eq13},\ \eqref{eq15}-\eqref{eq18},\ \eqref{eq21},\ \eqref{eq23}-\eqref{eq25},\nonumber\\
            &\sum^F_{f=1}{\sum^X_{x=1}{(U_{(i,x)}^f\ \cdot \  P_x\ \cdot \ T^f)}}\leq {{\mu}_L }^'\cdot {C}^{\max}_{i},\ \forall i\in \mathcal{N},\label{eq27}\\
            &\sum^F_{f=1}{\left(R_{(i,j)}^f \ \cdot \ T^f\right)}\leq {\mu}_L \cdot B^{\max}_{(i,j)},\forall i,j\in \mathcal{N}.\label{eq28}
        \end{align}
    
    In Eq.~\eqref{eq26}, we define the optimization function as the summation of the reconfiguration overhead and the energy consumption. We assign a relative weight to the two components by means of the $\alpha$ parameter, therefore, the two parts of the summation need to be normalized.
    The network reconfiguration overhead is defined as $\sum^N_{i=1}{\sum^N_{j=1}{\sum^F_{f=1}{\left|R_{(i,j)}^f-M^f_{(i,j)}\right|}}}$ which shows the difference of the current routing matrix with respect to the previous routing matrix. The maximum value of $\sum^N_{i=1}{\sum^N_{j=1}{R_{(i,j)}^f}}, \forall f\in \mathcal{F}$ is $N$. Therefore, $\sum^N_{i=1}{\sum^N_{j=1}{\sum^F_{f=1}{R_{(i,j)}^f}}}$ is at most $N\cdot F$. Similarly, $\sum^N_{i=1}{\sum^N_{j=1}{\sum^F_{f=1}{M^f_{(i,j)}}}}$ is at least $1\cdot F$. Therefore, $\sum^N_{i=1}{\sum^N_{j=1}{\sum^F_{f=1}{\left|R_{(i,j)}^f-M^f_{(i,j)}\right|}}}$ is always less than $\left(N-1\right)\cdot F$. In order to normalize the first part of equation~\eqref{eq26} we divide it by $\left(N-1\right)\cdot F$.
    In the second part of the objective function~\eqref{eq26}, the energy consumption of the network is $\sum^N_{i=1}{O^{t}_i\cdot \mathcal{E}'_i}$. The maximum energy consumption is $\sum^N_{i=1}{\mathcal{E}'_i}$ (i.e., when all elements of $O^t_i$ are one). Hence, we consider $\sum^N_{i=1} O^{t}_i\cdot \mathcal{E}'_i/\sum^N_{i=1} \mathcal{E}'_i$ as the normalized  value for energy consumption.
    
    Constraint~\eqref{eq27} controls the processing capacity of servers on providing a VNF. Focusing on the link capacity, constraint~\eqref{eq28} checks the link capacity between each pair of the switches. It should be mentioned that in resource reallocation algorithms, these constraints can be added because we assume to have an estimation of the rates of the flows based on the current status of the network (as explained in section \ref{sec:outline}).
    
    \begin{thm7}{\textit{GRR} falls in the class of NP-Hard problems.}\end{thm7}
    \begin{proof}
    See the proof of Lemma~\ref{LMA:SFRANP}
    \end{proof}
	
\section{Heuristics} \label{sec:heuristic}
    In this section, for each component of the Fig. \ref{fig:Architecture}, a fast heuristic algorithm is proposed.
    
\subsection{Nearest Service Function first (NSF)}\label{sect:NSF}
    Since the process of assigning resources to the newly arrived flows is a real-time one, a heuristic algorithm is proposed to solve the SFRA problem (section~\ref{sect:SFRA}) in a real-time manner, called Nearest Service Function first (NSF). Algorithm \ref{alg:NSF} presents an outline of NSF. The algorithm finds the nearest server which supports the first VNF in the chain ($K^f$) for the flow $f$ and sends the flow to that server. Thereafter, NSF removes that VNF from the chain and finds the nearest server that supports the next VNF of the chain and so on.
    	\begin{algorithm}
        	\caption{Nearest\_Service\_Function\_first (NSF)}
        	\label{alg:NSF}
        	\normalsize
        	\allowdisplaybreaks
        	\begin{algorithmic}[1]
            	\break
            	\INPUT{$K,s,d,\mathcal{N}$}
            	\OUTPUT{$SP$}
            	\Comment{$SP$ is the selected path}
            	\For{each flow $f$ in $\mathcal{F}$}
                	\State{$SP^f$=empty;}
                	\State{$CN=s$;}
                	\Comment{$CN$ is the current server}
                	\For{each VNF $k$ in $K$}
                	    \State{$[v,p]=Find\_Nearest\_Providers(CN,k,\mathcal{N})$;}
                	    \State{add $p$ to $SP^f$}
                	    \State{$CN=v$;}
                	    \Comment{$MFS$: median of flows size}
                	    \State{$B^{\max}=Reduce\_Capacity(B^{\max},MFS,p)$;}
                	\EndFor
                	\State{$p=Shortest\_Path(CN,d)$;}
            	    \State{add $p$ to $SP^f$}
            	    \State{$B^{\max}=Reduce\_Capacity(B^{\max},MFS,p)$;}
            	\EndFor\\
        	    \Return{$SP$}
        	\end{algorithmic}
    	\end{algorithm}
        	
	In Algorithm \ref{alg:NSF}, for each flow $f$ (line 1) a path $SP^f$ is selected. To this end, in the line 4, for each required VNF $k\in K$ the following procedure is done: 
    	\begin{itemize}
    	\item NSF finds the nearest server that supports that VNF (line 5; function '$Find\_Nearest\_Providers$' is precisely described in Algorithm \ref{alg:FNP}). 
    	\item The shortest path to the selected server is added to the $SP^f$ and the current status is changed to the selected server (lines 6 and 7).
    	\item Function $Reduce\_Capacity$ decreases the network capacity by the value of MFS (line 8). MFS is the median of recently communicated flows, calculated using the data stored by the network monitoring component in the knowledge base (we assume that we do not know the rate of the flow to be allocated). 
    	\item After considering all the required VNFs for the flow, the algorithm finds the shortest path to the destination and updates the links capacity matrix (lines 10-12).
    	\end{itemize}
    	
    	\begin{algorithm}
        	\caption{Find\_Nearest\_Providers}
        	\label{alg:FNP}
        	\normalsize
        	\allowdisplaybreaks
        	\begin{algorithmic}[1]
            	\break
            	\INPUT{$CN,k,OP$}
            	\Comment{$CN$ is the current server}
            	\OUTPUT{$v, p$}
            	\Comment{$k$ is requested VNF}
    	            \Statex{}
            	    \Comment{$v$: nearest server to $CN$ supporting VNF $k$}
            	    \Statex{}
            	    \Comment{$p$: nearest path from $CN$ to $v$}
            	    \Statex{}
            	    \Comment{$OP$: other required parameters}
            	    \State{$[P,Cost]=Dijkstra(CN,k,Prune(B^{\max}))$;}
            	    \Statex{}
            	    \Comment{$P$: set of nearest paths from $CN$ to other servers}
            	    \Statex{}
            	    \Comment{$Cost$: costs of nearest paths from $CN$ to other servers}
            	    \For{each server $v$ in $\mathcal{N}$}
            	        \Statex{}
        	            \Comment{$APP$: Available Processing Capacity}
        	            \Statex{}
        	            \Comment{$RPP$: Required Processing Capacity}
            	        \If{$S_{(k,v)}==0$ \textbf{OR} $APP(v)<RPP(MFS)$}
            	            \State{$Cost_v=\infty$;}
            	        \EndIf 
        	        \EndFor
            	    \State{$v$=Minimum\_Cost($Cost$);}
            	    \State{$p=P_v$;}\\
            	\Return{[$v,p$]}
    	\end{algorithmic}
    	\end{algorithm}
    	
	In the first line of Algorithm \ref{alg:FNP}, using Dijkstra, the shortest paths to all servers are calculated. It should be mentioned that just those links that have a capacity greater than the median of recently communicated flows are considered as valid links (i.e., function $Prune$ removes links that have less than MFS capacity). Thereafter, servers that meet one of the following conditions are eliminated (lines 2-8):
	i) Do not support the requested VNF; or,  ii) have a processing capacity less than the processing capacity required for processing the MFS. At this point, the nearest server to the current server is selected as destination (lines 9,10).
    	
\subsection{Offline Resource Reallocation}
    As shown in Fig. \ref{fig:Architecture}, a knowledge base components contains the traffic measurements. Based on this information, it is possible to calculate a long-term estimation of the traffic pattern (but this estimation is out of the scope of the paper). The long-term estimation of the traffic pattern can be used to turn \textit{off} or \textit{on} \footnote{Servers in the ACTIVE or IDLE modes are considered as ON.} servers at predefined time intervals. The GRR formulation (section~\ref{sec:GRR}) can be used in an ILP solver to find the optimal solution in this scenario, but this turns out to be computationally complex and in our experiment it is not applicable in medium and big sized networks. Therefore, in this subsection, a near optimal solution called 3R is proposed. Before defining 3R, we need to define \textit{Energy-aware SFRA} which is a extended version of \textit{SFRA} where there is some extra information about the rate flows.

\subsubsection{Energy-aware SFRA}\label{sec:ESFRA}
    we consider that \textit{SFRA} is unaware of the rate of flows, however, in resource reallocation we have estimation of flows rate. Therefore, we can extend \textit{SFRA} to reallocate the resources to one of the existing flows $f$ with the objective of optimizing the energy consumption of the network. Like \textit{SFRA}, \textit{Energy-aware SFRA} considers one flow at a time. The formulation is as following:
    \begin{align}
        &{\min_{O^t,R,U} \ \ }\left(\sum^N_{i=1}(1-O^{t-1}_i)\ \cdot \:O^t_i\:\cdot \:\mathcal{E}'_i\right) ,\label{eq20} \\
        &\textit{Subject to:}\nonumber \\
        &Eq.\ \eqref{eq2}-\eqref{eq13},\ \eqref{eq15}-\eqref{eq18},\ \eqref{eq21},\ \eqref{eq23}-\eqref{eq25}. \nonumber
    \end{align}
    {where the objective function \eqref{eq20} minimizes the number of servers that are required to be turned \textit{on} to support the flow $f$. The vector $ON$ specifies the servers that are required just for the flow $f$ while $O^{t-1}$ specifies the servers that are already used for all flows that are in the network. The summation $\sum_{i=1}^N{\left[(1-O^{t-1}_i).O^{t}_i\right]}$ represents with a linear expression the number of servers that have been turned on to support the flow $f$. Therefore, $\sum^N_{i=1}{\left[(1-O^{t-1}_i).O^t_i.\mathcal{E}_i\right]}$ calculates the additional amount of energy needed to support the flow $f$ (i.e., the amount of energy that is required for servers that are turned \textit{on} just for the current flow).}

\subsubsection{Relaxed Resource Reallocation (3R)}\label{sec:3R}
    since GRR considers the impact of flows on other flows and tries to find an optimal solution considering all flows simultaneously, the computational complexity increases dramatically for medium and large scale networks with large number of flows. Therefore, a relaxed version of GRR called \textit{3R} is proposed to find a near-optimal solution and provide a trade-off between the optimality gap and the computational complexity. To this end, instead of considering the impact of all flows on each other, the impact of each flow on all flows that are already rerouted is considered, i.e., $3R$ reallocates the resources to the flows one-by-one using the above defined \textit{Energy-aware SFRA}. Algorithm~\ref{alg:3r} reports the $3R$.
	\begin{algorithm}
	\caption{Relaxed Resource Reallocation (3R)}
	\label{alg:3r}
	\normalsize
	\allowdisplaybreaks
	\begin{algorithmic}[1]
    	\break
    	\INPUT{$OP$}
    	\Comment{$OP$: required parameters (Table \ref{tab:notation})}
    	\OUTPUT{$SP$}
    	\Comment{$SP^f$: selected path for flow $f$}
    	\For{each flow $f$ in $\mathcal{F}$}
    	    \State{$SP^f = Energy-aware-SFRA(B^{\max},f,T^f)$;}
    	    \For{each link $(i\rightarrow j)$ in $SP^f$}
    	        \State{$B^{\max}_{(i,j)}=B^{\max}_{(i,j)}-T^f$;}
    	    \EndFor
    	\EndFor\\
    	\Return{$SP$}
	\end{algorithmic}
	\end{algorithm}   
    	
	In lines 1 and 2 of Algorithm \ref{alg:3r}, for each flow, 3R reallocates the resources using the gathered information about the rates of the flows. Thereafter, the network status is updated by decreasing the available capacity of the links that are placed in the selected path. At this point, 3R reroutes the next flow using the same approach and so on.
    
\subsection{Online Resource Reallocation}
    \begin{table*}[!htbp]
        \caption{Traffic Generator Notation and Inputs.}\label{tab:TrafficGeneratorNotation}
        \centering
        \rowcolors{2}{gray!25}{white}
        \begin{tabular}[t]{|p{0.06\textwidth}|p{0.35\textwidth}||p{0.07\textwidth}|p{0.07\textwidth}|p{0.07\textwidth}|p{0.07\textwidth}|p{0.07\textwidth}|}\hline
        	\textbf{Symbol} & \textbf{Definition} & \textbf{Scenario 1} & \textbf{Scenario 2} & \textbf{Scenario 3} & \textbf{Scenario 4} & \textbf{Scenario 5}\\
        	\hline\hline
        	    $B^f$ & Ratio of flow size to link capacity & 0.3 & 0.3 & 0.3 & 0.3 & 0.2\\\hline 
        	    $\gamma$ & Ratio of servers that can host VNFs & 1 & 0.7 & 0.7 & 0.7 & 0.5\\\hline 
        	    $V^f$ & Average number of requested VNFs that a flow needs & 2 & 2.5 & 2 & 2.5 & 2.5\\\hline 
        	    $X_\gamma$ & Ratio of VNFs hosted by a server & 0.7 & 0.7 & 0.7 & 0.7 & 0.7\\\hline            		    
        	    $V^f_{min}$ & Minimum number of requested VNFs per flow & 2 & 2 & 2 & 2 & 2\\\hline 
        	    $V^f_{max}$ & Maximum number of requested VNFs per flow & 5 & 5 & 5 & 5 & 5\\\hline            		    
        	    $F$ & Number of flows & 25 & 22 & 26 & 28 & 45\\\hline
        	    $\tau$ & Edge switches ratio & 1 & 1 & 1 & 1 & 1\\\hline 
        	    $\tau_s$ & Ratio of edge switches that are source of a flow & 1 & 1 & 1 & 1 & 1\\\hline
        	    $\tau_d$ & Ratio of edge switches that are destination of a flow & 1 & 1 & 1 & 1 & 1\\\hline 
                $\beta$ & Coefficient of number of generated flows per source & 0.4 & 0.4 & 0.4 & 0.4 & 0.4\\\hline 
                $F_m$ & Maximum number of generated flows per source & 10 & 10 & 10 & 10 & 10\\\hline
        	    $X$ & Number of different VNFs & 10 & 10 & 10 & 10 & 10\\\hline
        \end{tabular}
    \end{table*}
    This component is designed to react to network traffic behavior in real-time. Therefore, the algorithms which are used in this subsection should have a low computational complexity. To this end, two heuristic approaches are proposed: i) ST-ENSF: it reconfigures the network to reduce the energy consumption in predefined time intervals, and ii) LT-ENSF that reconfigures the network in case of network congestion that require to switch on new servers. In the following, these two heuristic approaches are discussed precisely.
\subsubsection{Short Term Energy-aware NSF (ST-ENSF)} \label{sec:ST-ENSF}
    although the computational complexity of 3R is sufficiently lower than GRR, it is still time consuming for real-time network reconfiguration. Therefore, a faster heuristic algorithm is proposed to reallocate the resources in real-time. To this end, a greedy approach is exploited. For each flow, in each step, a server which minimizes the energy consumption of the network for the next required VNF is selected. If there are multiple servers that have equal energy consumption, then the server which is closer to the node considered in the current step is selected.
	\begin{algorithm}
	\caption{Short Term Energy-aware Nearest Service Function first (ST-ENSF)}
	\label{alg:ST-ENSF}
	\normalsize
	\allowdisplaybreaks
	\begin{algorithmic}[1]
    	\break
    	\INPUT{$k, OP$}
    	\Comment{$OP$: other required parameters}
    	\OUTPUT{$SP$}
    	\Comment{$k$: requested VNF}
    	\State{$SP$=empty;}
    	\Comment{$SP$: selected path}
    	\State{$CN=s$;}
    	\Comment{$CN$ is the current server}
    	\For{each flow $f$ in $\mathcal{F}$}
        	\For{each VNF $k$ in $K$}
        	    \State{$[costs,paths]=Dijkstra(CN,k,Prune(B^{\max}))$;}
        	    \Statex{~~~~~~$\%$\textit{Remove Redundant servers}}
        	    \For{each server $i$ in $\mathcal{N}$}
        	        \If{$S_{(k,i)}==0$}
        	            \State{$cost_i=\infty$;}
    	            \EndIf
    	        \EndFor
        	    \State{$energy=Energy\_Consumption(\mathcal{N})$;}
        	    \State{$[v,p]=ENS(paths,costs,energy)$;}
        	    \State{add $p$ to $SP^f$}
        	    \State{$CN=v$;}
        	    \State{$B^{\max}=Reduce\_Capacity(B^{\max},T^f,p)$;}
        	\EndFor
        	\State{$p=Shortest\_Path(CN,k,Prune(B^{\max}))$;}
    	    \State{add $p$ to $SP^f$}
    	    \State{$B^{\max}=Reduce\_Capacity(B^{\max},T^f,p)$;}
	    \EndFor\\
	    \Return{$SP$}
	\end{algorithmic}
	\end{algorithm}
        
    For each flow and for each VNF required by the flow, the shortest path from the current server to all other servers are calculated (line~5). It should be mentioned that just those links that have a capacity greater than the size of the flow are considered as valid links (i.e., function $Prune$ removes links that have a capacity less than $T^f$). In lines 6-10, all servers that do not support the requested VNF are omitted from the list by setting the cost of reaching them infinity (i.e., the cost is the propagation delay of paths). Additionally, it removes all links to nodes that are met previously. Afterward, the amount of extra energy that would be imposed by each server is calculated in line 11. Accordingly, if server $i$ is currently ON, then the variable $energy_i$ is zero. Otherwise, it is the energy consumption of the server. In line 12, ENS finds the nearest ON server which supports the required VNF (in the sake of cost). If there is not such a server, ENS seeks for servers that are in IDLE status. In this way, it finds the server with the minimum energy consumption.
        
    At this point, the selected path $p$ is added to $SP^f$ (line 13) and the current status is updated to the selected server (line 14). Besides, the available capacity of the links used in the selected path is reduced by the size of the flow. After meeting all required VNFs, the algorithm uses the shortest path to directly move to the destination (lines 17-19).
    
\subsubsection{VNF Placement \& ENSF (LT-ENSF)}\label{section:LT-ENSF}
    ST-ENSF is supposed to change servers' status from ACTIVE to IDLE and vice versa, however, when the network is congested it may be required to switch a server from OFF status to ON (ACTIVE/IDLE) status. This will happen based on the fact the amount of traffic load on the network will increase more than the capacity of ON servers. As a result, LT-ENSF is proposed in which the servers can switch between three different possible states. The outline of the algorithm is similar to ST-ENSF, however, the function $ENS$ in line 12 of the Algorithm \ref{alg:ST-ENSF} is completely different. $ENS$ in this algorithm seeks for a server which is already in ACTIVE status, if it is not possible then a server with minimum energy consumption would be selected. if neither possible, a nearest OFF server would be activated to handle the request. It should be mentioned that ST-ENSF selects the VNFs that are used for each flow for a set of active VNFs running in given servers (i.e. it only performs a \textit{VNF assignment}) while LT-ENSF performs the \textit{VNF Placement} and \textit{VNF assignment} simultaneously. In VNF placement, the algorithm specifies the set of supported VNFs for each server. 
    	
\subsection{Mathematical Computational Complexity}
    \begin{figure}[!htbp]
	    \centering
		\includegraphics[width=1\columnwidth]{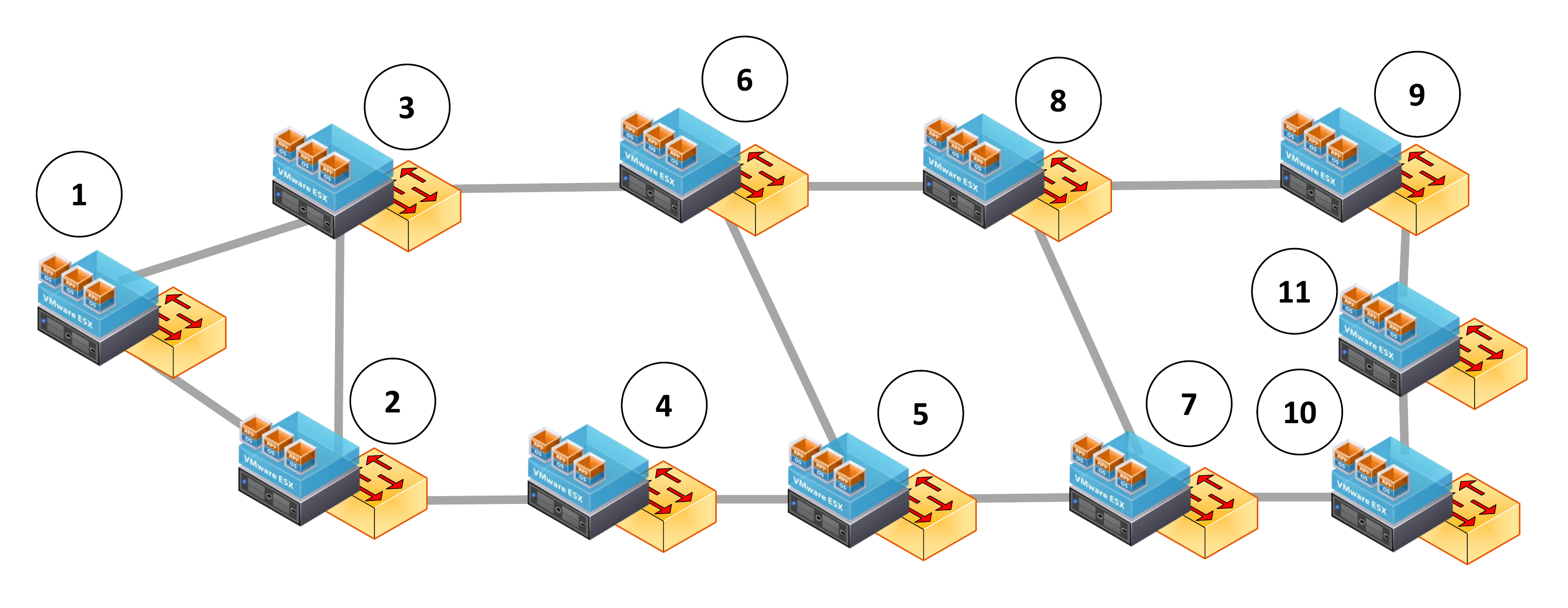}
    	\caption{Abilene Network Topology.}
    	\label{fig:abileneTopology}
    \end{figure}
    
    \begin{figure*}[!htbp]
    \begin{subfigure}{0.333\textwidth}
      \centering
      \includegraphics[width=1\linewidth]{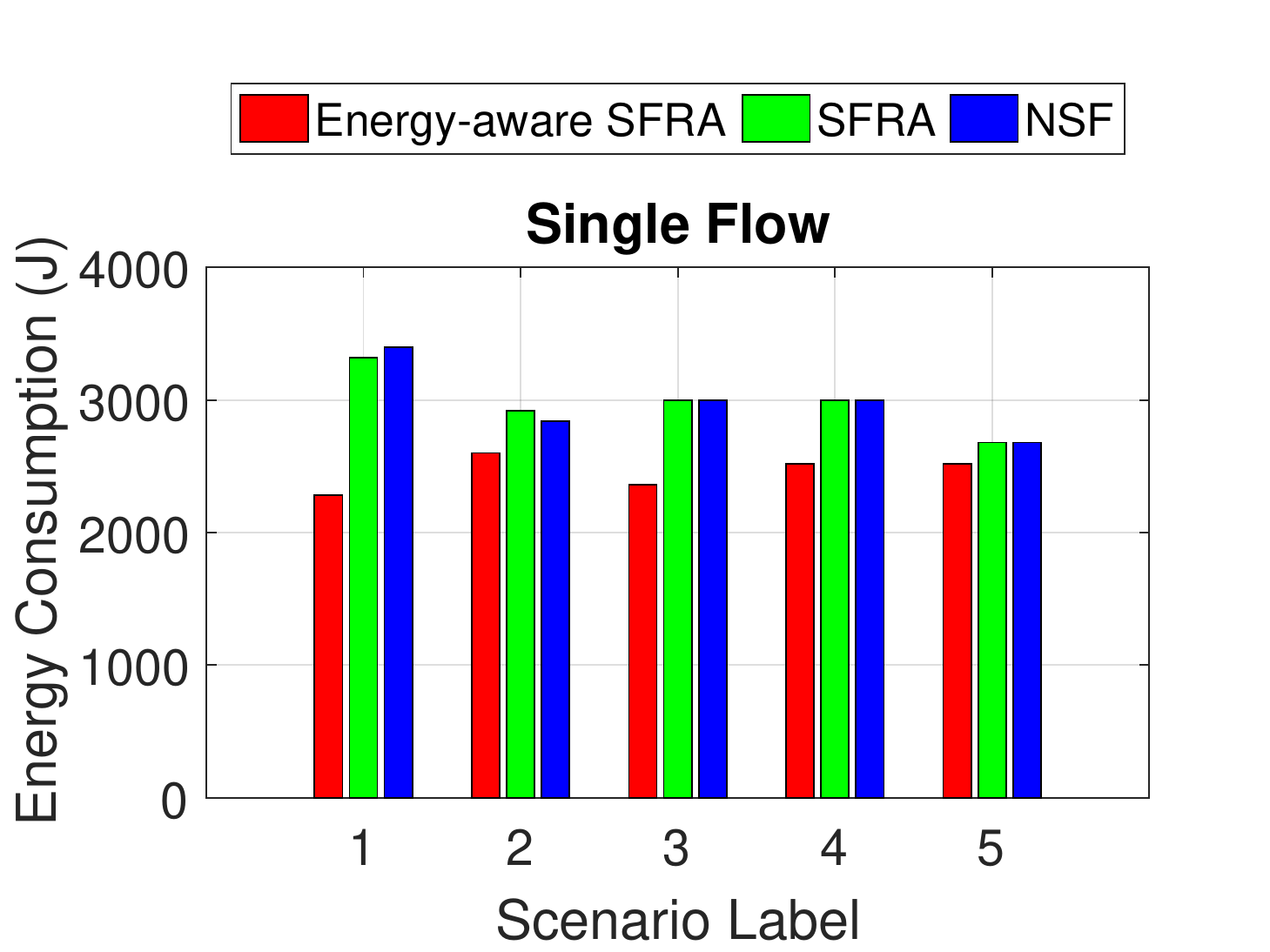}
      \caption{Energy Consumption}
      \label{fig:barpwRA}
    \end{subfigure}%
    \begin{subfigure}{0.333\textwidth}
      \centering
      \includegraphics[width=1\linewidth]{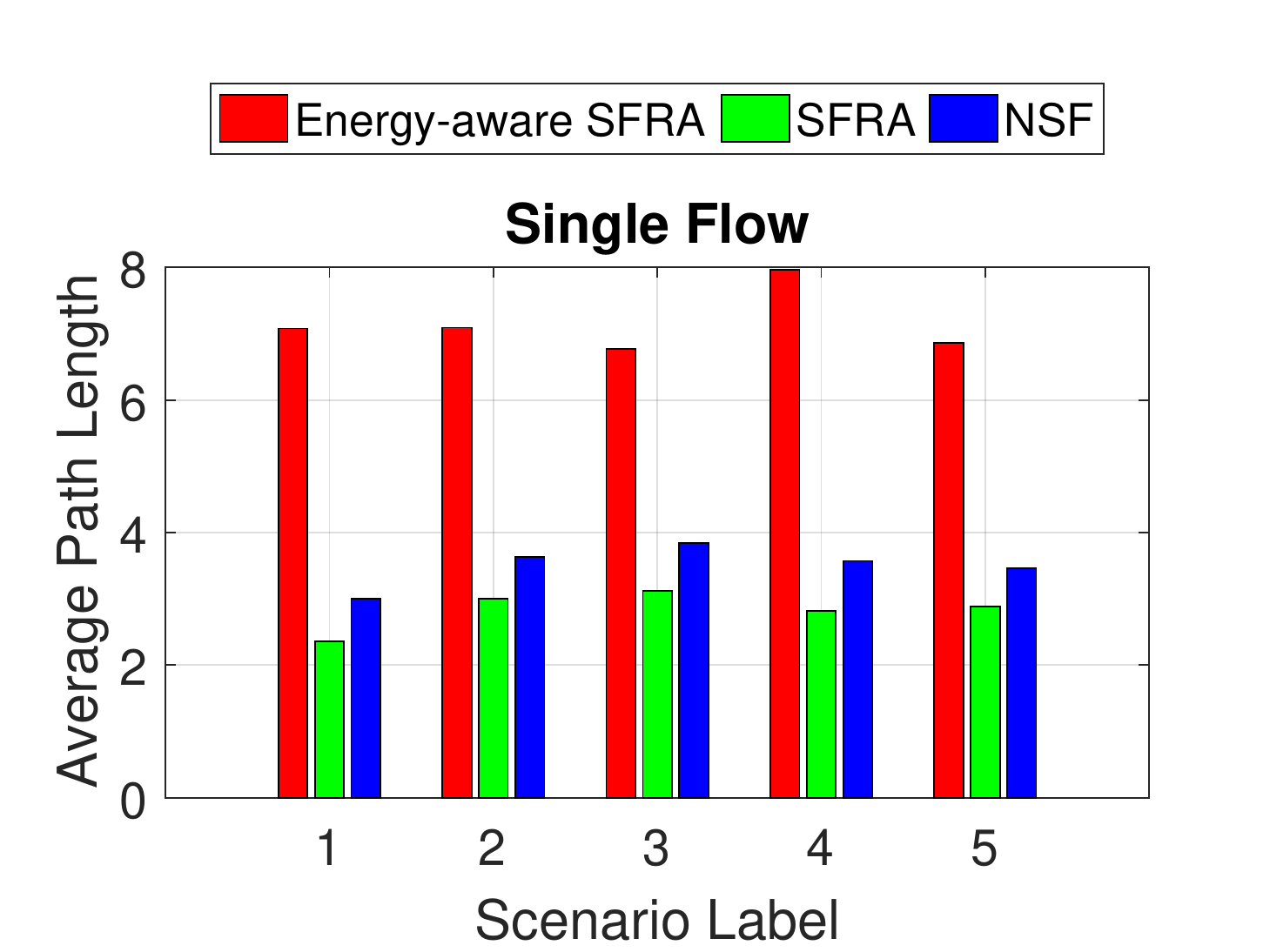}
      \caption{Path Length}
      \label{fig:barptRA}
    \end{subfigure}
    \begin{subfigure}{0.333\textwidth}
      \centering
      \includegraphics[width=1\linewidth]{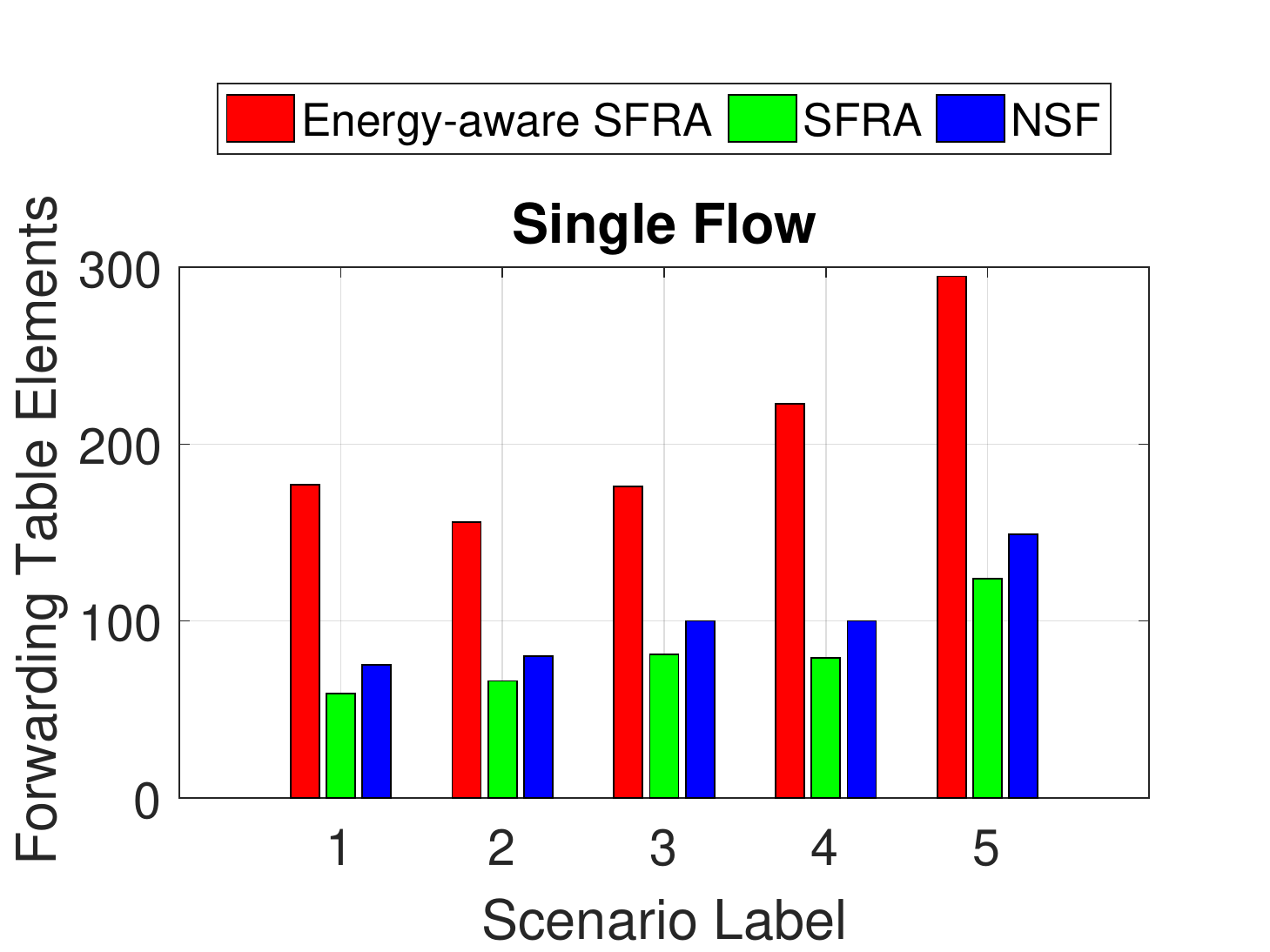}
      \caption{Reconfiguration Overhead}
      \label{fig:barnsRA}
    \end{subfigure}
    \caption{Resource Allocation}
    \label{fig:barPwrPthRA}
    \end{figure*}
    
    \begin{figure*}[!htbp]
    \begin{subfigure}{0.333\textwidth}
      \centering
      \includegraphics[width=1\linewidth]{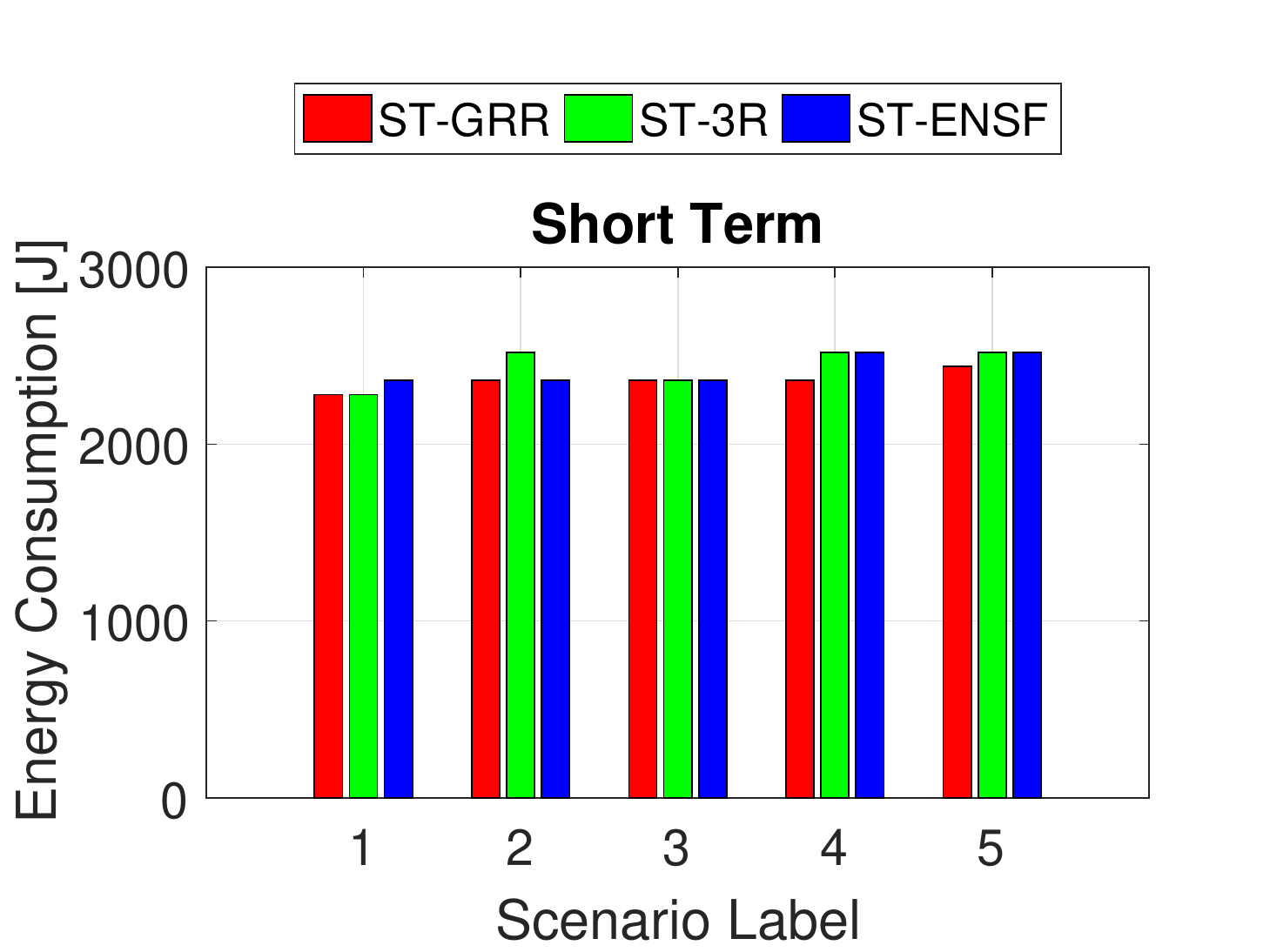}
      \caption{Energy Consumption}
      \label{fig:barpwST}
    \end{subfigure}%
    \begin{subfigure}{0.333\textwidth}
      \centering
      \includegraphics[width=1\linewidth]{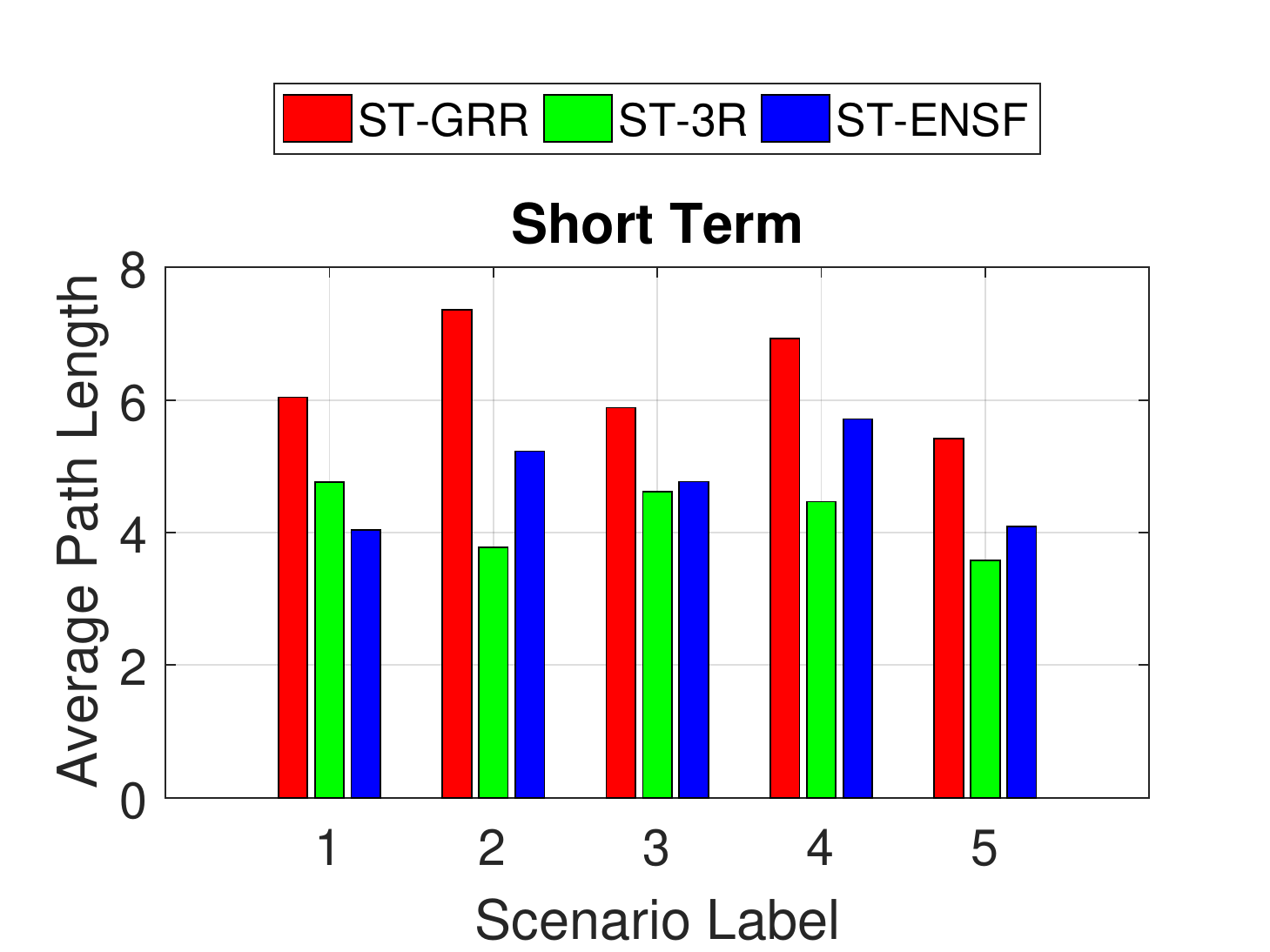}
      \caption{Path Length}
      \label{fig:barptST}
    \end{subfigure}
    \begin{subfigure}{0.333\textwidth}
      \centering
      \includegraphics[width=1\linewidth]{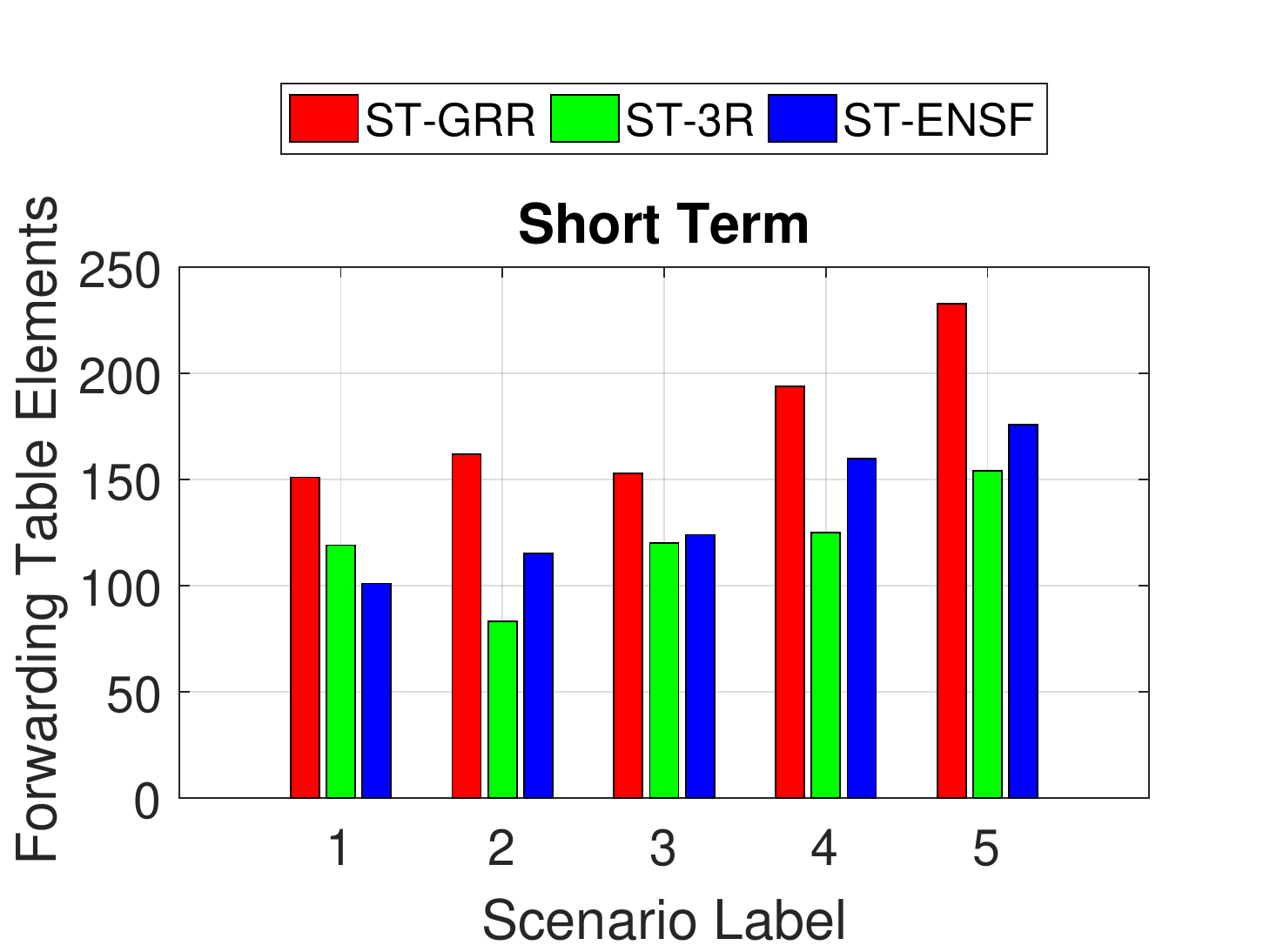}
      \caption{Configuration Overhead}
      \label{fig:barnsST}
    \end{subfigure}
    \caption{Short Term Resource Reallocation}
    \label{fig:barPwrPthST}
    \end{figure*}
    
    \begin{figure*}[!htbp]
    \begin{subfigure}{0.333\textwidth}
      \centering
      \includegraphics[width=1\linewidth]{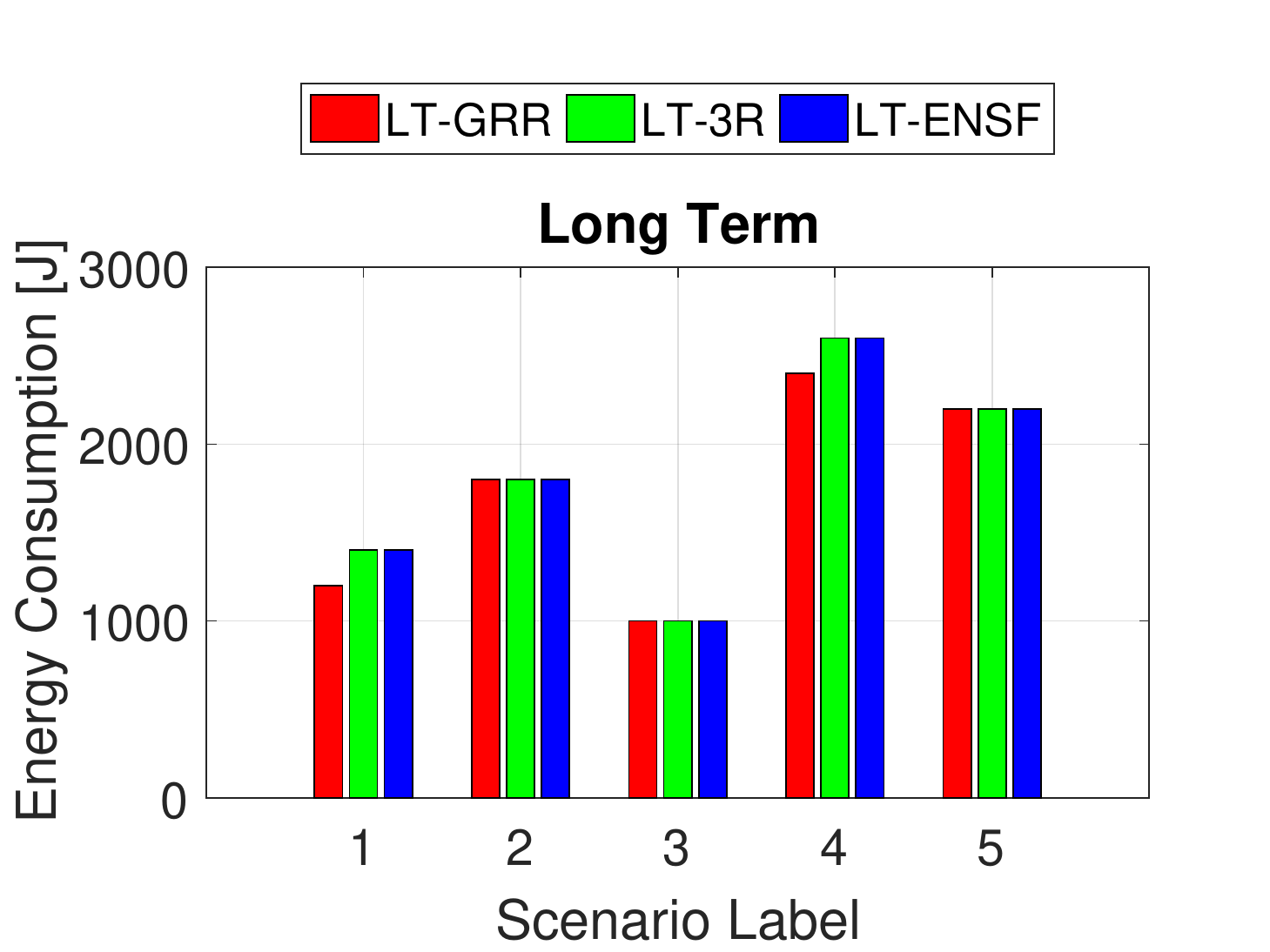}
      \caption{Energy Consumption}
      \label{fig:barpwLT}
    \end{subfigure}%
    \begin{subfigure}{0.333\textwidth}
      \centering
      \includegraphics[width=1\linewidth]{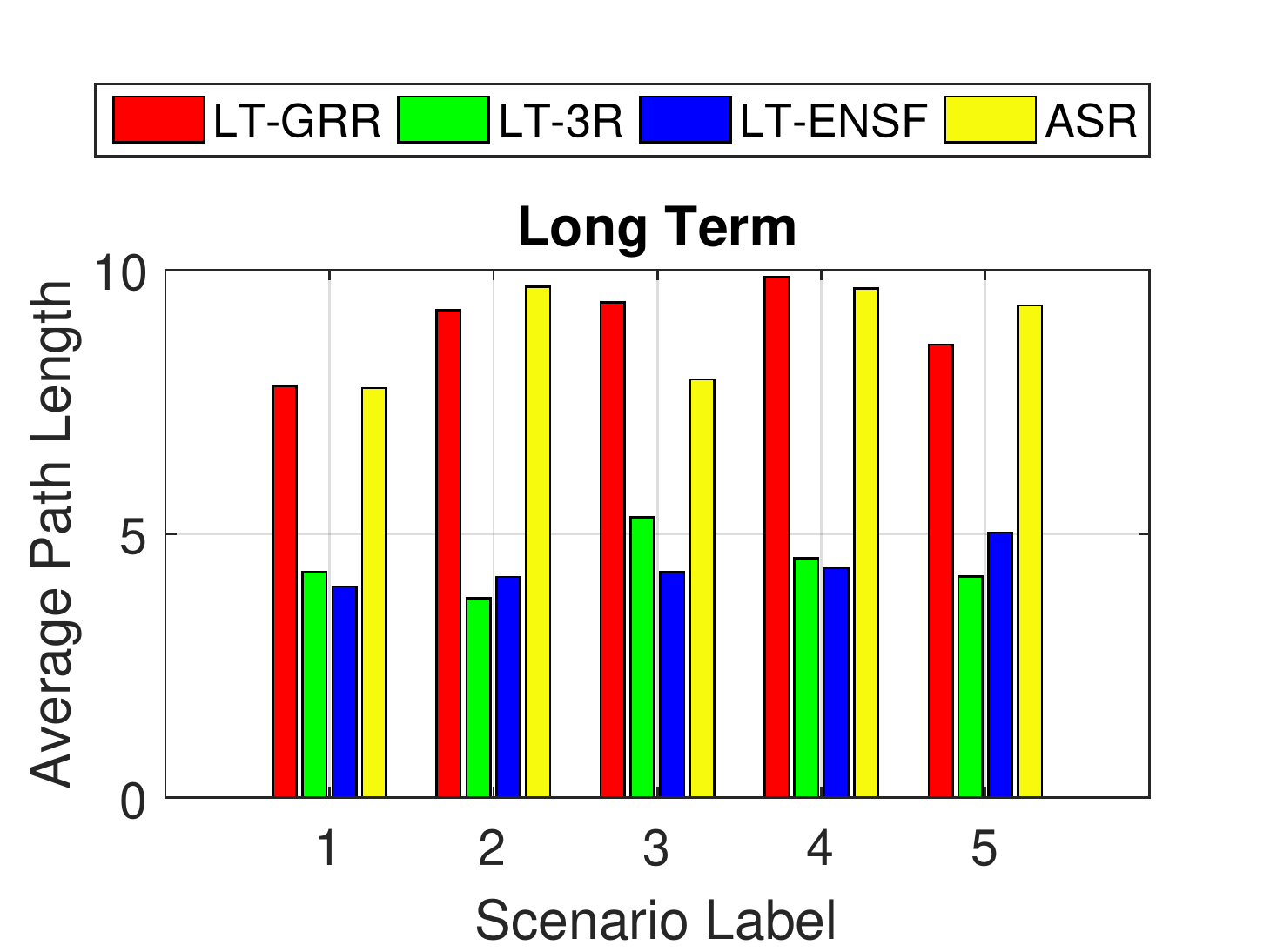}
      \caption{Path Length}
      \label{fig:barptLT}
    \end{subfigure}
    \begin{subfigure}{0.333\textwidth}
      \centering
      \includegraphics[width=1\linewidth]{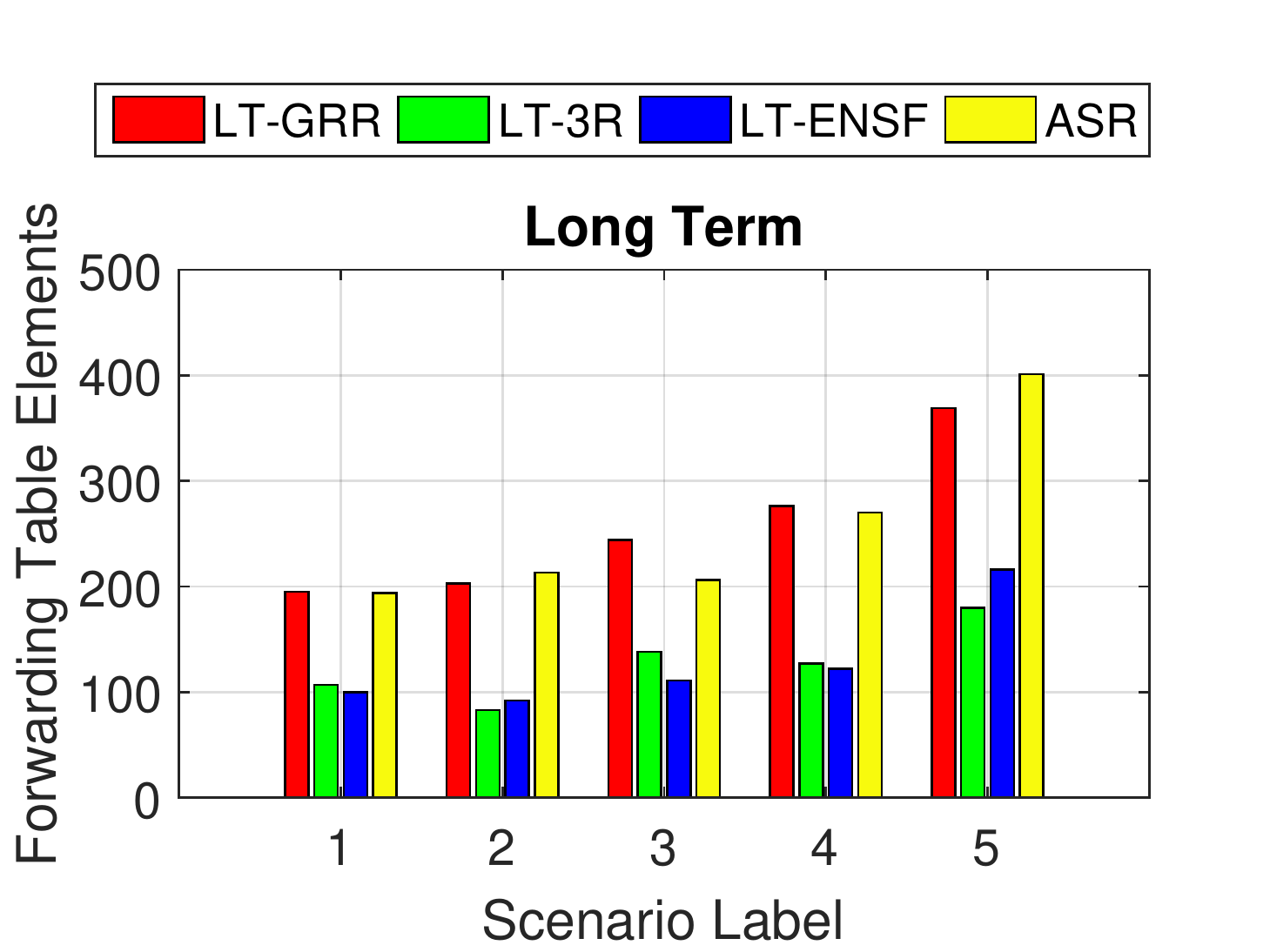}
      \caption{Reconfiguration Overhead}
      \label{fig:barnsLT}
    \end{subfigure}
    \caption{Long Term Resource Reallocation}
    \label{fig:barPwrPthLT}
    \end{figure*}
            
    \textbf{NSF:} The order of line 1 of Algorithm \ref{alg:FNP} is $\mathcal{O}(N\cdot\log{N}+|E|)$ while it is $\mathcal{O}(N)$ for lines 2 and 9. Therefore, the order of the Algorithm \ref{alg:FNP} is $\mathcal{O}(2N+N\cdot\log{N}+|E|)\approx \mathcal{O}(N\cdot\log{N}+|E|)$.
    Similarly, in Algorithm \ref{alg:NSF}, the order of lines 3, 4, and 6 are $\mathcal{O}(\Psi)$, $\mathcal{O}(N\cdot\log{N}+|E|)$, and $\mathcal{O}(N)$, respectively. Therefore, the total computational complexity of NSF is in order of $\mathcal{O}(\Psi\cdot\left[\left(N\cdot\log{N}+|E|\right)+N\right])\approx \mathcal{O}(\Psi\cdot(N\cdot\log{N}+|E|))$.\\
        
    \noindent\textbf{ENSF:} The computational complexity of lines 3 and 4 of Algorithm \ref{alg:ST-ENSF} are in order of $\mathcal{O}(F)$ and $\mathcal{O}(\Psi)$, respectively. The order of each of lines 5 and 17 is $\mathcal{O}(N\cdot\log{N}+|E|)$ while it is $\mathcal{O}(N)$ for each of lines 6, 11-15, 18, and 19. Therefore, the total computational complexity of $ENSF$ is $\mathcal{O}(F\cdot\Psi\cdot(N\cdot\log{N}+|E|))$.
	
\section{Numerical Results}\label{resultAnalysis}
    In this section, the proposed schemes are evaluated under different traffic patterns over a real-world network topology called Abilene \cite{AbilineNetwork}. The traffic-demand generator and the simulation setup are discussed in the upcoming subsections.
	    
	\subsection{Traffic-Demand Generator}\label{sec:trafficGenerator}
	In order to investigate the performance of the proposed resource allocator/reallocator algorithms, a traffic-demand generator is proposed. It takes multiple input parameters and generates network traffic flows with different specifications: rate, source and destination, VNF requirements, and end-to-end tolerable delay. Table~\ref{tab:TrafficGeneratorNotation} presents the input parameters of the traffic generator. It is important to note that the described algorithm is meant to generates the traffic pattern (and not the traffic packets).  
	    
    We consider the flows as unidirectional. The variable, $\tau$ specifies the percentage of switches that acts as edge switches (i.e. they can be source or destination of a flow, $\tau=1$ means that all switches can be considered as edge switches).  Therefore, the number of edge switches is $\tau\times N$. Similarly, the variables $\tau_s$ and $\tau_d$ are the percentage of edge switches that can act as the source or the destination of a flow, respectively, i.e., the number of source switches is $N_s = \tau\times\tau_s\times N$ and the number of destination switches is $N_d = \tau\times\tau_d\times N$. The number of potential source-destination pairs is $N_s\times N_d$.
        
    We want $\beta\times N_d$ to be the average number of flows that are generated by a source node. We assume that the number of generated flows for each source switch follows a geometric distribution \cite{zoucomputer} with $1/(\beta\times N_d)$ as the success probability. For practical reasons, we also include in the model the maximum number of flows $F_m$ that can be generated by a source node, therefore the flows will be generated with a truncated geometric distribution and the average number of flows from each source node is actually smaller than $\beta\times N_d$. 
    
    $\gamma$ is the fraction of servers that can host VNFs, therefore, $\gamma\times N$ is the number of servers that can host a VNF. On the other hand, $X_\gamma$ is the fraction of VNF types hosted by a server, meaning that a server can host at most $X_\gamma\times X$ different VNFs, where $X$ is the number of different types of VNFs. The number of VNFs that are needed by a flow is generated according to geometric distribution with average $R^f$. In order to make realistic scenarios, we consider $R^f_{max}$ and $R^f_{min}$ as the maximum and minimum number of VNFs needed by a flow, respectively. If the generated number is greater than $R^f_{max}$, the number is set to $R^f_{max}$ and similarly for the minimum (note that this approach changes the average number with respect to the average of the initial geometric distribution). The average traffic rate demand of a flow is a fraction $B^f$ of the capacity of the link, i.e., it is $B^f\times link\_capacity$. In particular, the rate of generated flows follows a uniform distribution between 0 and $2\times B^f\times link\_capacity$. 
    
    
    \subsection{Simulation Setup}
    \label{SimulationSetup}
    In this subsection, the system configuration and the network topology used in our simulations is described. Table~\ref{tab:SystemConfiguration} reports the configuration of the PC on which we run the simulations. We used CVX \cite{grant2008cvx} to solve the ILP optimization problems.
    \begin{table}[!htbp]
	\caption{Hardware Configuration.}
	\label{tab:SystemConfiguration}
	\rowcolors{2}{gray!25}{white}
	\resizebox{\columnwidth}{!}{
	\begin{tabular}[t]{|l|l|} 
		\hline
		\textbf{Name} & \textbf{Description}\\
		\hline\hline
        Processor & Intel(R) Core(TM) i5-2410M CPU @ 2.30GHz\\\hline
        IDE & Standard SATA AHCI Controller \\\hline
        RAM & 4.00 GB\\\hline
        System Type & 64-bit Operating System, Windows 10\\\hline
	\end{tabular}}
    \end{table}
    The network topology considered in our simulation is called Abilene \cite{AbilineNetwork} and is illustrated in Figure \ref{fig:abileneTopology}. We set the link capacity $B^{\max}(i,j)=1$~[Gbps] for all the links that interconnect the switches. In addition, the links that connect a server to a switch in a node have a capacity equal to the sum of the other links of the switch, so that there is no bottleneck in the local communication between a switch and a server. Each server has three modes: ACTIVE (i.e., the energy consumption is full rate energy consumption), IDLE (i.e., the energy consumption is a constant and low value), and OFF (i.e., it has no energy consumption). We assume that the processing capacity of a server $C^{\max}_i$ is equal to the sum of the capacity of all incoming links multiplied by a factor $\Theta \leq 1$. If $\Theta = 1$, the processing capacity of the server is enough to process flows at the maximum possible rate on all incoming links. By properly reducing $\Theta$ we can let the processing power of some servers become a system bottleneck. To investigate the impact of server processing capacity on energy consumption, in section \ref{sec:pwrPthRecnAnalysis}, the $\Theta$ factor has been set to 0.1, after some empirical tuning. 
    
    We assume that the energy consumption of a server $\mathcal{E}_i$ is related to its maximum processing capacity. In particular, we define as parameters the maximum and minimum energy consumption ($\mathcal{E}_{max}$, $\mathcal{E}_{min}$) of a server. We assign $\mathcal{E}_{max}$ to the server(s) that have the maximum processing capacity, $\mathcal{E}_{min}$ to the server(s) that have the minimum processing capacity, and an intermediate value between $\mathcal{E}_{min}$ and $\mathcal{E}_{max}$ to the other servers (scaled linearly). In our simulation we set $\mathcal{E}_{min}$=$200J$ and $\mathcal{E}_{max}$=$400J$.
    
    Moreover, the algorithms are examined in a sequence of five different iterations. We generate a set of traffic demands corresponding to the first iteration. In particular, the set of flows (i.e. the number of flows and their source and destination node) and the set of required VNFs for each flow are chosen according to the model described in Section~\ref{sec:trafficGenerator}. Then, in each following iteration, the flow rate is increased using an uniform distribution with an average of $10\%$ (between 0 and $20\%$) of the flow rate. In the simulations discussed here we consider $\alpha$ to be 0, i.e. we are only optimizing the energy consumption (i.e., the second term of the objective function in Eq.~\eqref{eq26}).
    
    The five considered traffic scenarios are presented in Table \ref{tab:TrafficGeneratorNotation}. In the first three lines we report the parameters that are variable in the scenarios: i) the ratio of flow size to link capacity (denoted as $B^f$), two different values for $B^f$ are considered ($0.2$ and $0.3$); ii) the ratio of servers that can host VNFs $\gamma$ (which allows to investigate the impact of servers processing capacity); iii) the average number of requested VNFs per flows $R^f$.
    In addition, we set a number of common parameters across the scenarios as reported in the other rows. Finally, the number of flows $F$ reported in the table is the output of the random generation process for the five scenarios in the considered experiments.

    


	    
    \subsection{Energy, Path Length, and Reconfiguration Side-effect}\label{sec:pwrPthRecnAnalysis}
    In Figures \ref{fig:barPwrPthRA}-\ref{fig:barPwrPthLT}, the x-axis elements are the labels of different traffic scenarios stated in Table \ref{tab:TrafficGeneratorNotation}. The energy consumption of servers in the IDLE mode is 60 percent of its energy consumption in the ACTIVE mode. We divide the algorithms into three different classes i. Resource Allocation (fig. \ref{fig:barPwrPthRA}), ii. Short Term Resource Reallocation (fig. \ref{fig:barPwrPthST}), and iii. Long Term Resource Reallocation (fig. \ref{fig:barPwrPthLT}). 
    We compare the Energy Consumption, Path Length and Reconfiguration Overhead that is the number of entries that should be changed in the switches forwarding tables to enforce the new resource allocation decision. Since for the first two classes it is not possible to change the state of a server from \textit{ON} to \textit{OFF} (and vice-versa), all servers are supposed to be in \textit{ON} mode in these cases. 
    
    Since for i. and ii. it is not possible to change the state of a server from \textit{ON} to \textit{OFF} or vice versa, all servers are supposed to be in \textit{ON} mode for these classes.
    
    {In order to provide a more realistic analyze of our work, we exploit the state-of-the-art algorithm ASR~\cite{eramo2017approach} which performs the VNF placement and the SFC routing. Since ASR considers that all VNFs are supported on each server, therefore, ASR is a long-term algorithm and we compare it with our long-term algorithms.}
    
    \begin{figure}[!htbp]
    \begin{subfigure}{0.48\columnwidth}
      \centering
     \includegraphics[width=\columnwidth]{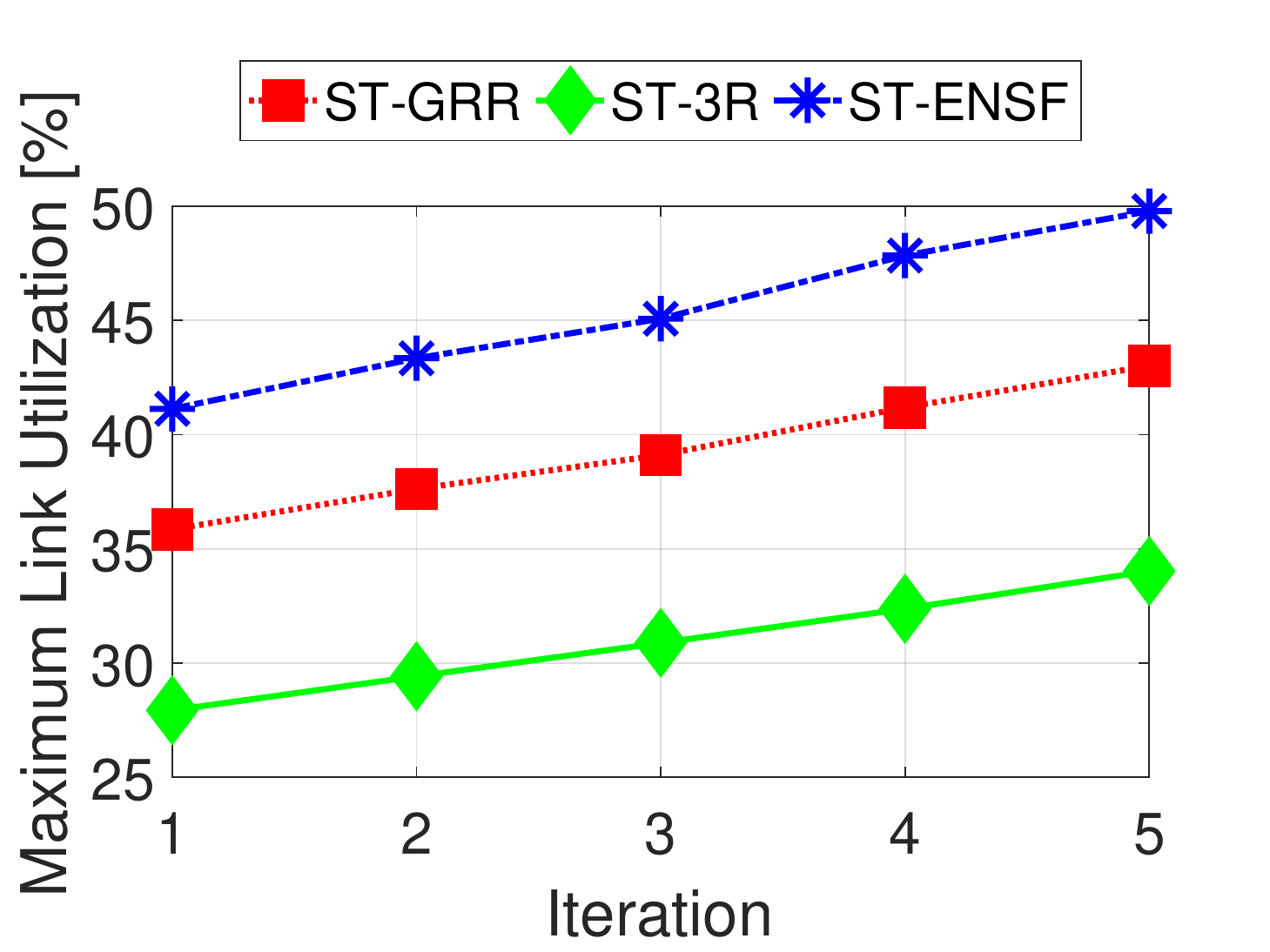}
      \caption{Maximum Link Utilization}
      \label{fig:p25ml}
    \end{subfigure}
    \begin{subfigure}{0.48\columnwidth}
      \centering
      \includegraphics[width=\columnwidth]{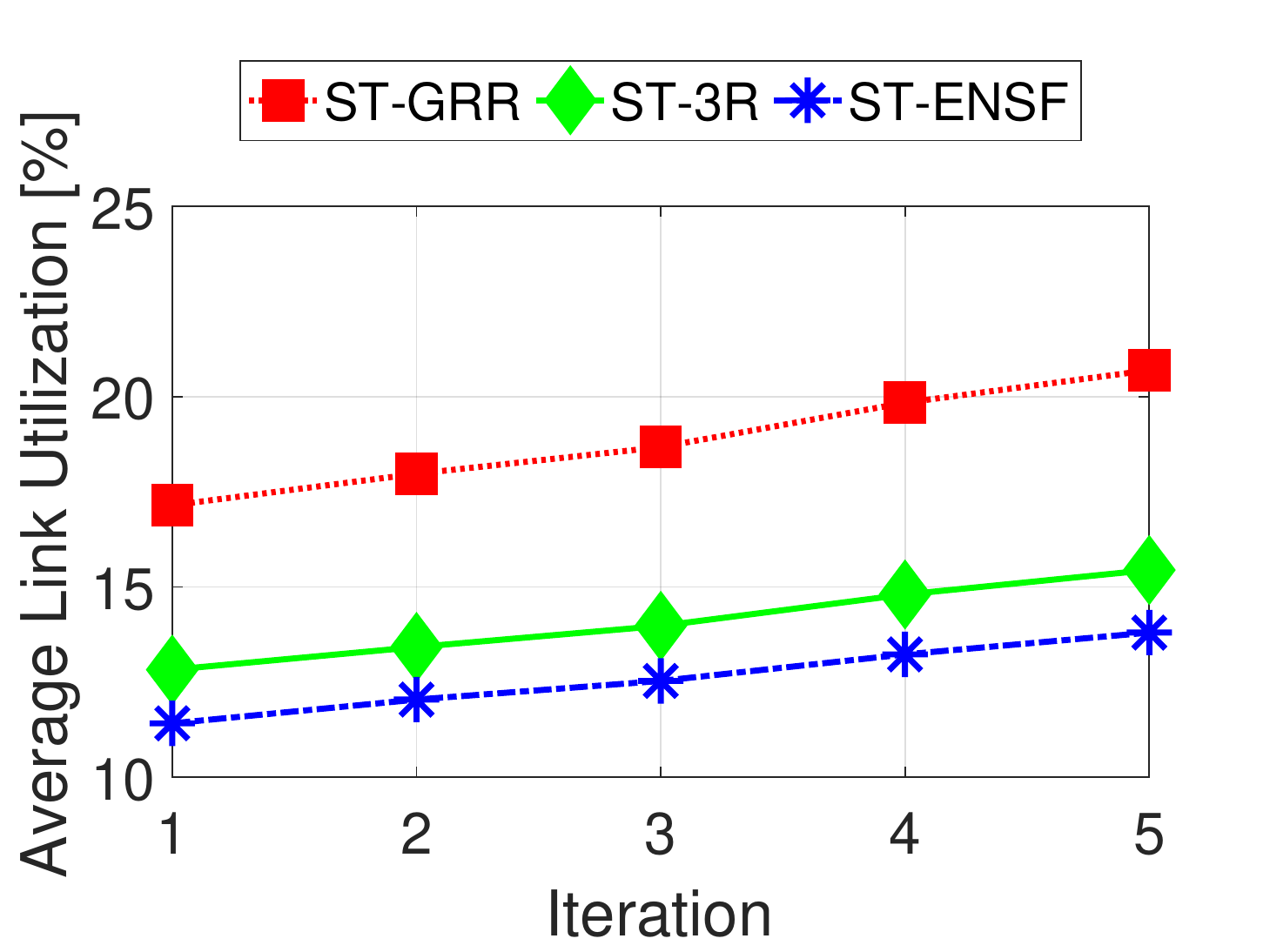}
      \caption{Average Link Utilization}
      \label{fig:p25ln}
    \end{subfigure}\hfill
    \begin{subfigure}{0.48\columnwidth}
      \centering
      \includegraphics[width=\columnwidth]{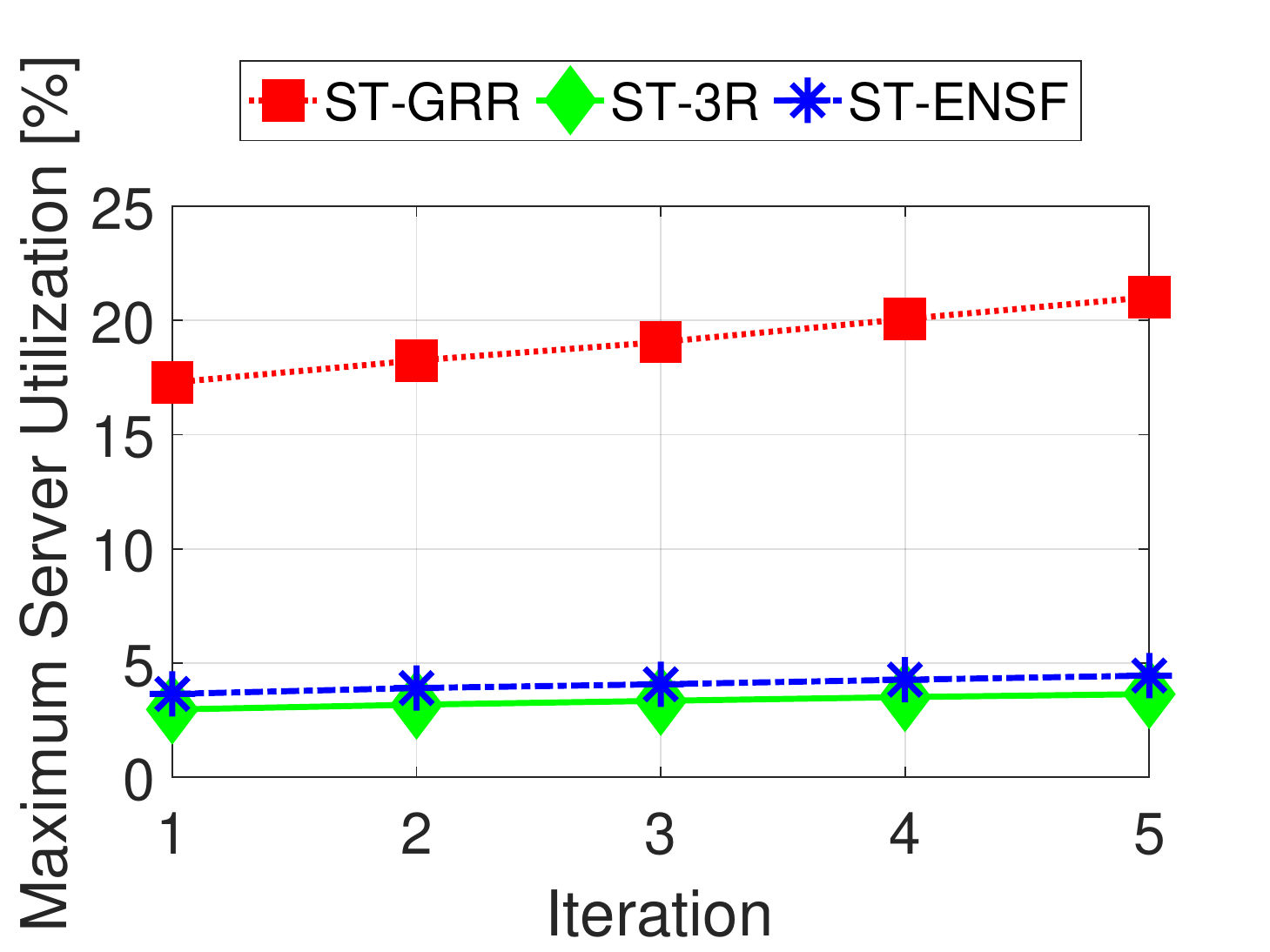}
      \caption{Maximum Server Utilization}
      \label{fig:p25mn}
    \end{subfigure}
    \begin{subfigure}{0.48\columnwidth}
      \centering
      \includegraphics[width=\columnwidth]{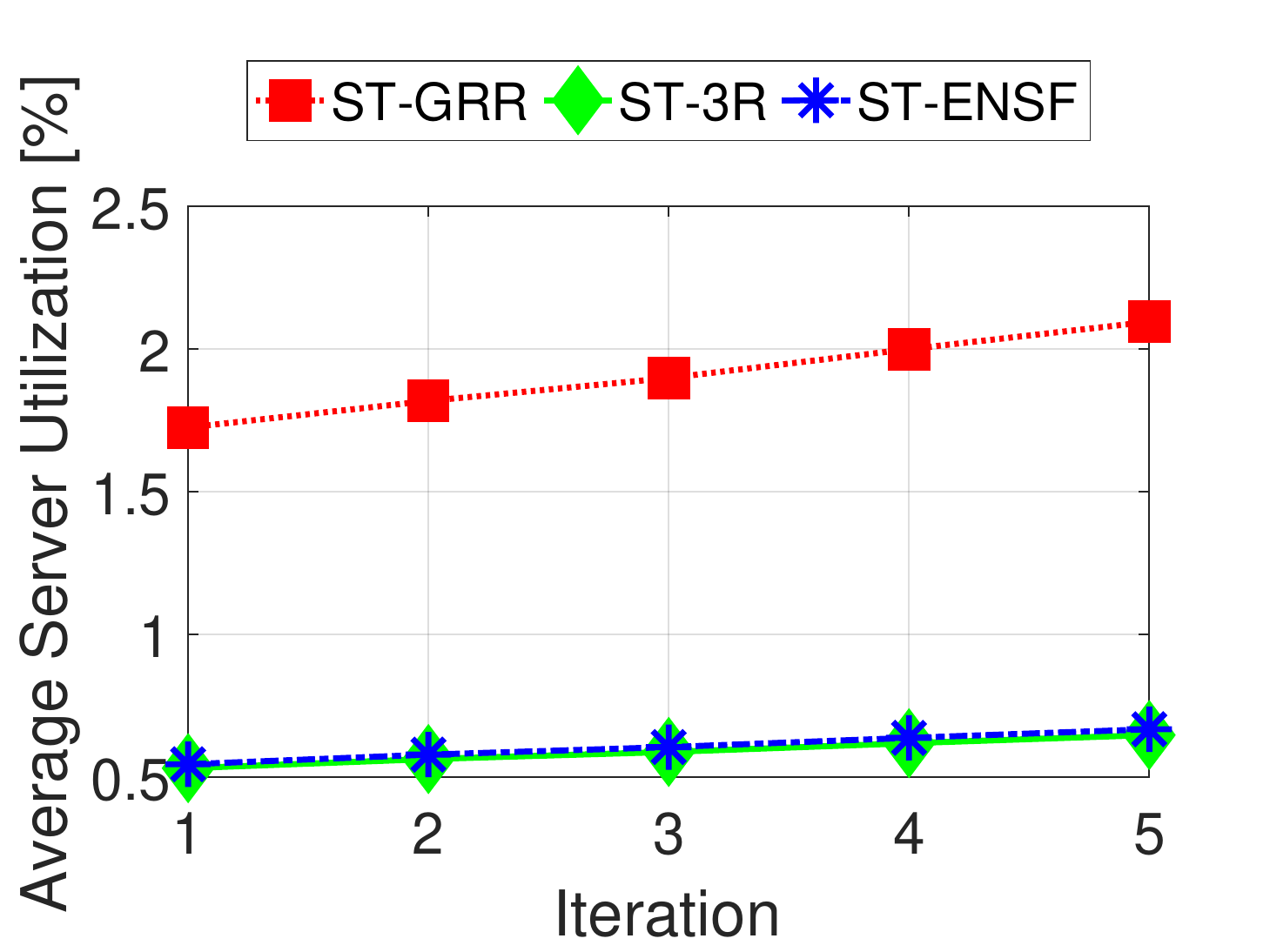}
      \caption{Average Server Utilization}
      \label{fig:p25nd}
    \end{subfigure}
    \caption{Traffic Scenario 1}
    \label{fig:p25ut}
    \end{figure}
    
    \begin{figure}[!htbp]
    \begin{subfigure}{0.48\columnwidth}
      \centering
     \includegraphics[width=\columnwidth]{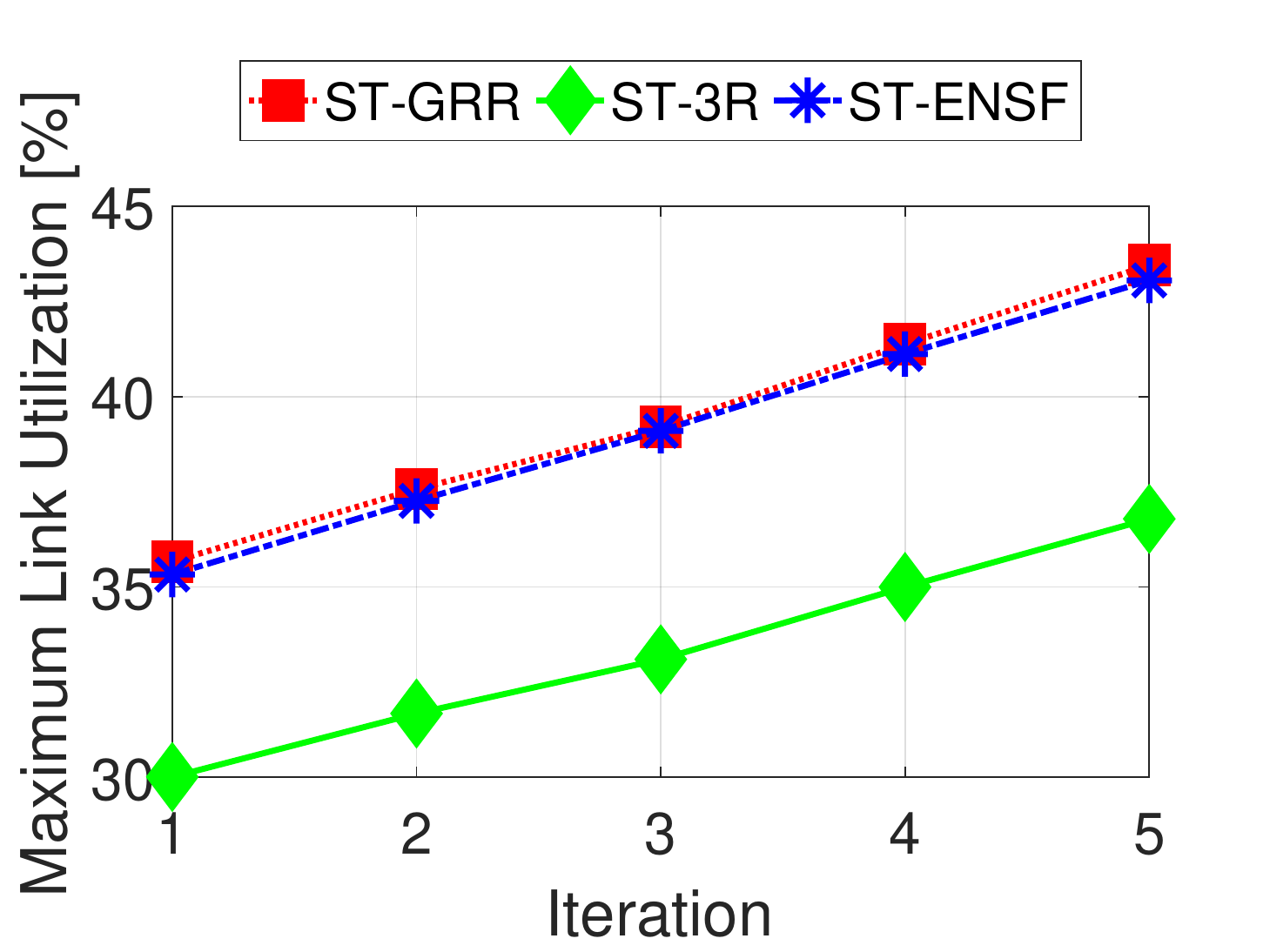}
      \caption{Maximum Link Utilization}
      \label{fig:p22ml}
    \end{subfigure}
    \begin{subfigure}{0.48\columnwidth}
      \centering
      \includegraphics[width=\columnwidth]{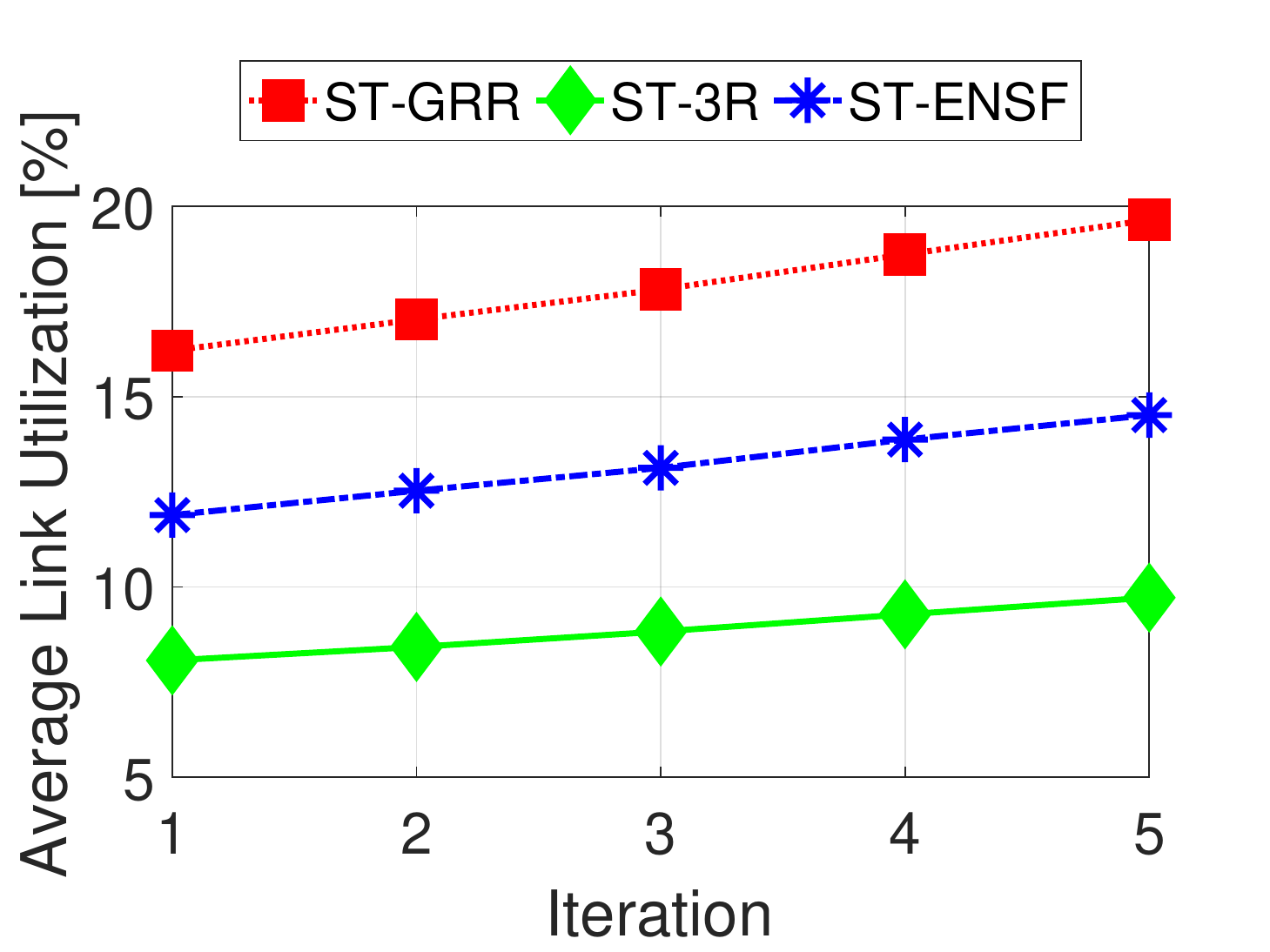}
      \caption{Average Link Utilization}
      \label{fig:p22ln}
    \end{subfigure}\hfill
    \begin{subfigure}{0.48\columnwidth}
      \centering
      \includegraphics[width=\columnwidth]{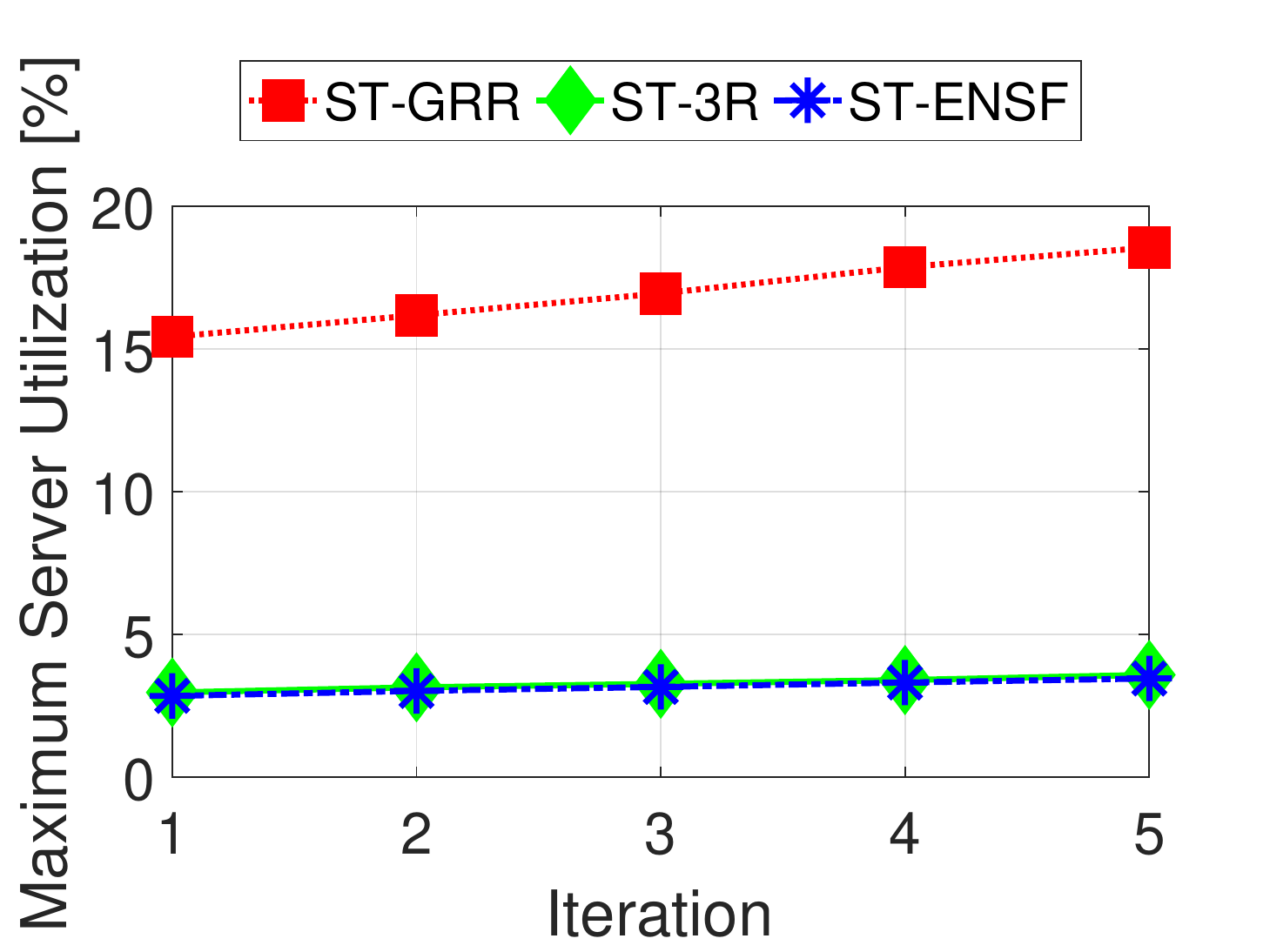}
      \caption{Maximum Server Utilization}
      \label{fig:p22mn}
    \end{subfigure}
    \begin{subfigure}{0.48\columnwidth}
      \centering
      \includegraphics[width=\columnwidth]{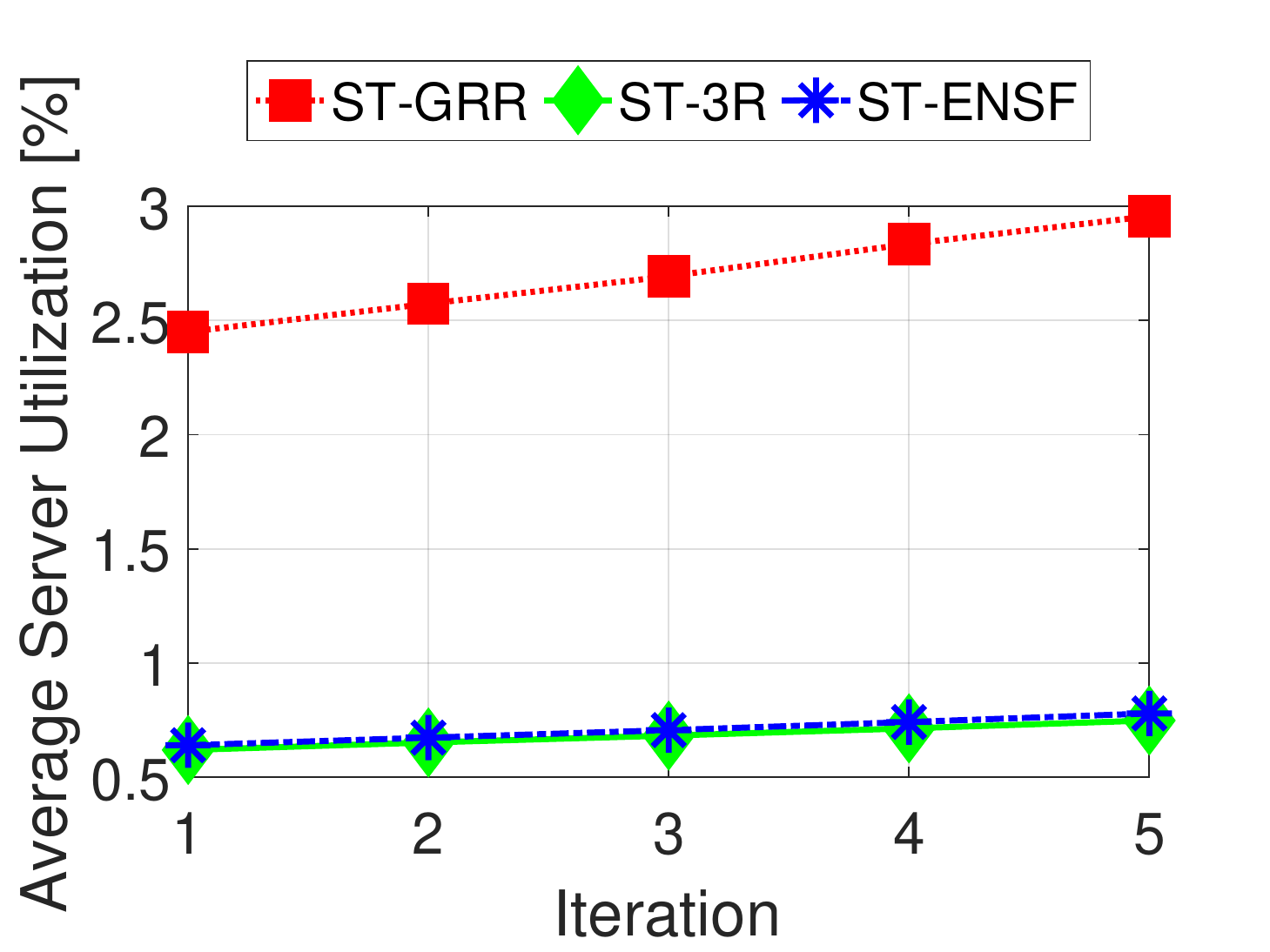}
      \caption{Average Server Utilization}
      \label{fig:p22nd}
    \end{subfigure}
    \caption{Traffic Scenario 2}
    \label{fig:p22ut}
    \end{figure}
    
    \begin{figure}[!htbp]
    \begin{subfigure}{0.48\columnwidth}
      \centering
     \includegraphics[width=\columnwidth]{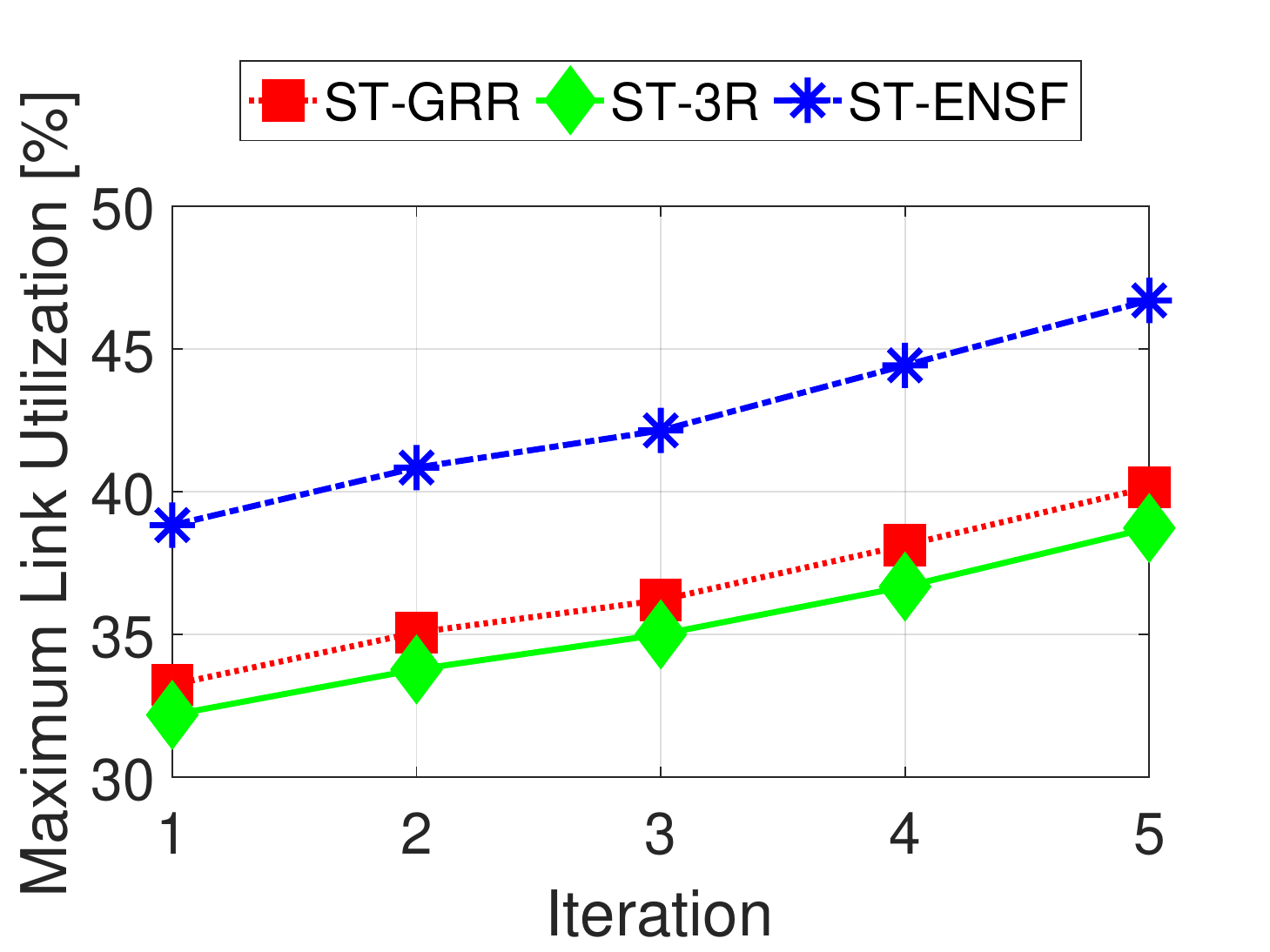}
      \caption{Maximum Link Utilization}
      \label{fig:p26ml}
    \end{subfigure}
    \begin{subfigure}{0.48\columnwidth}
      \centering
      \includegraphics[width=\columnwidth]{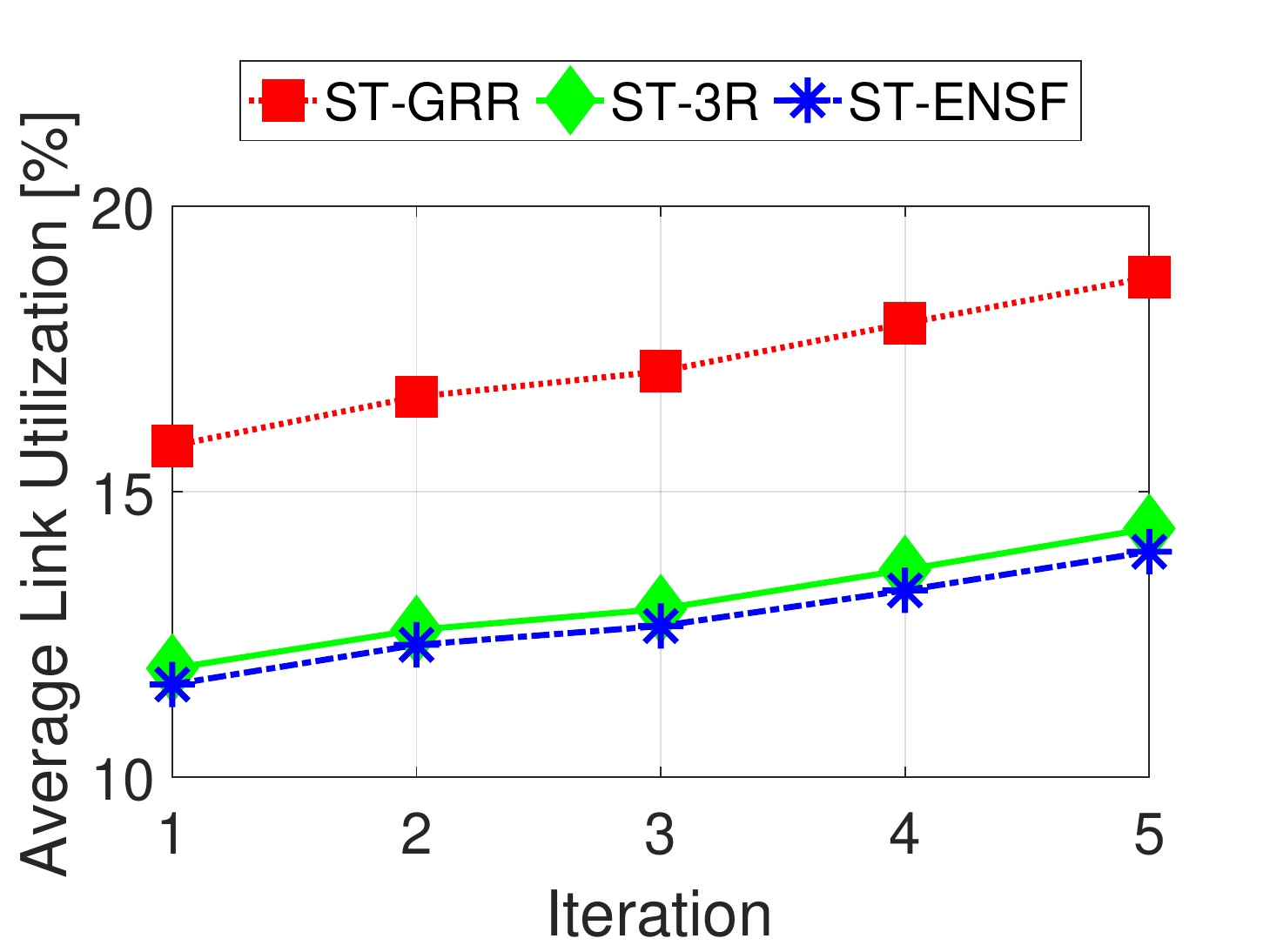}
      \caption{Average Link Utilization}
      \label{fig:p26ln}
    \end{subfigure}\hfill
    \begin{subfigure}{0.48\columnwidth}
      \centering
      \includegraphics[width=\columnwidth]{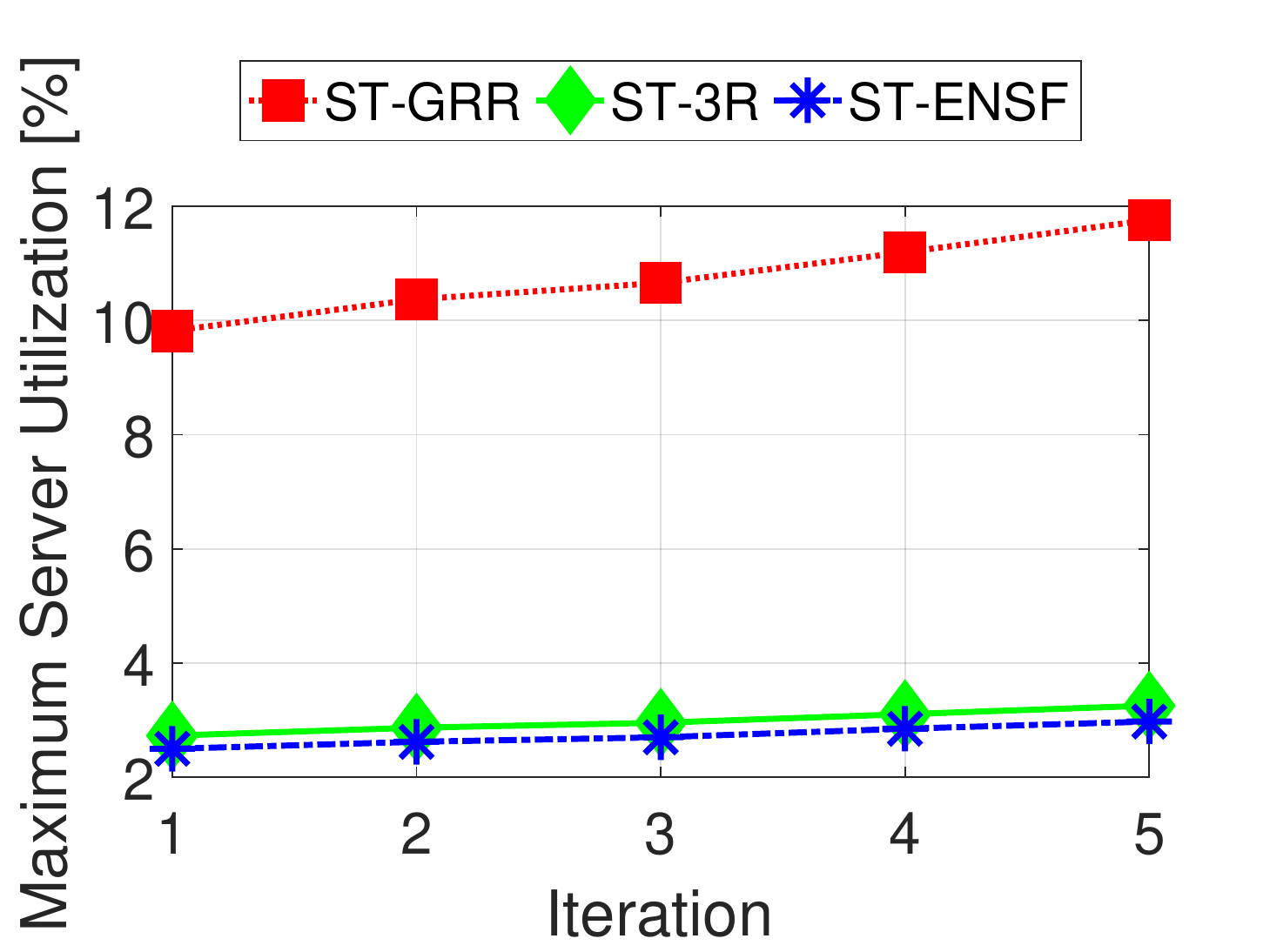}
      \caption{Maximum Server Utilization}
      \label{fig:p26mn}
    \end{subfigure}
    \begin{subfigure}{0.48\columnwidth}
      \centering
      \includegraphics[width=\columnwidth]{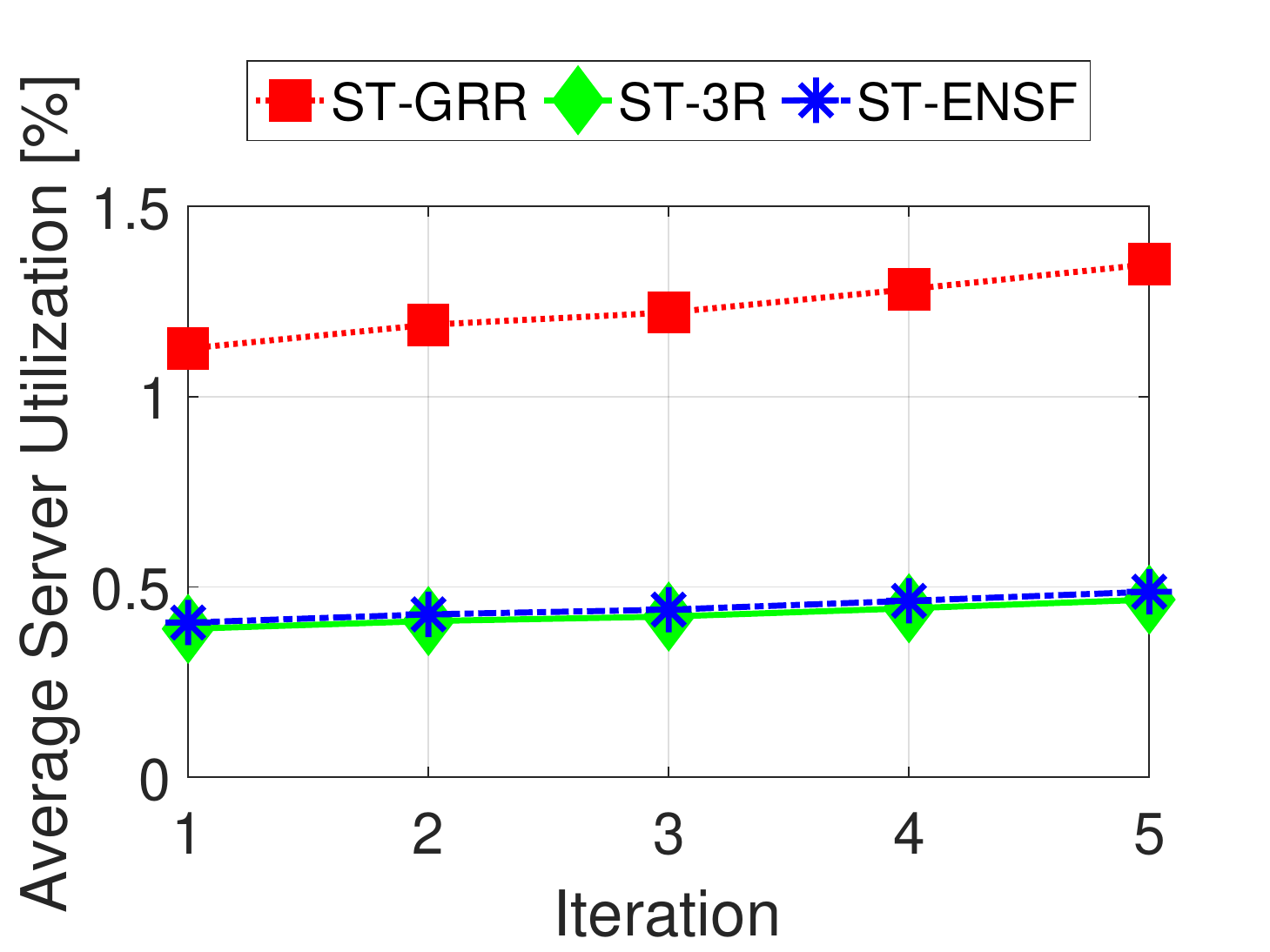}
      \caption{Average Server Utilization}
      \label{fig:p26nd}
    \end{subfigure}
    \caption{Traffic Scenario 3}
    \label{fig:p26ut}
    \end{figure}
    
    \begin{figure}[!htbp]
    \begin{subfigure}{0.48\columnwidth}
      \centering
     \includegraphics[width=\columnwidth]{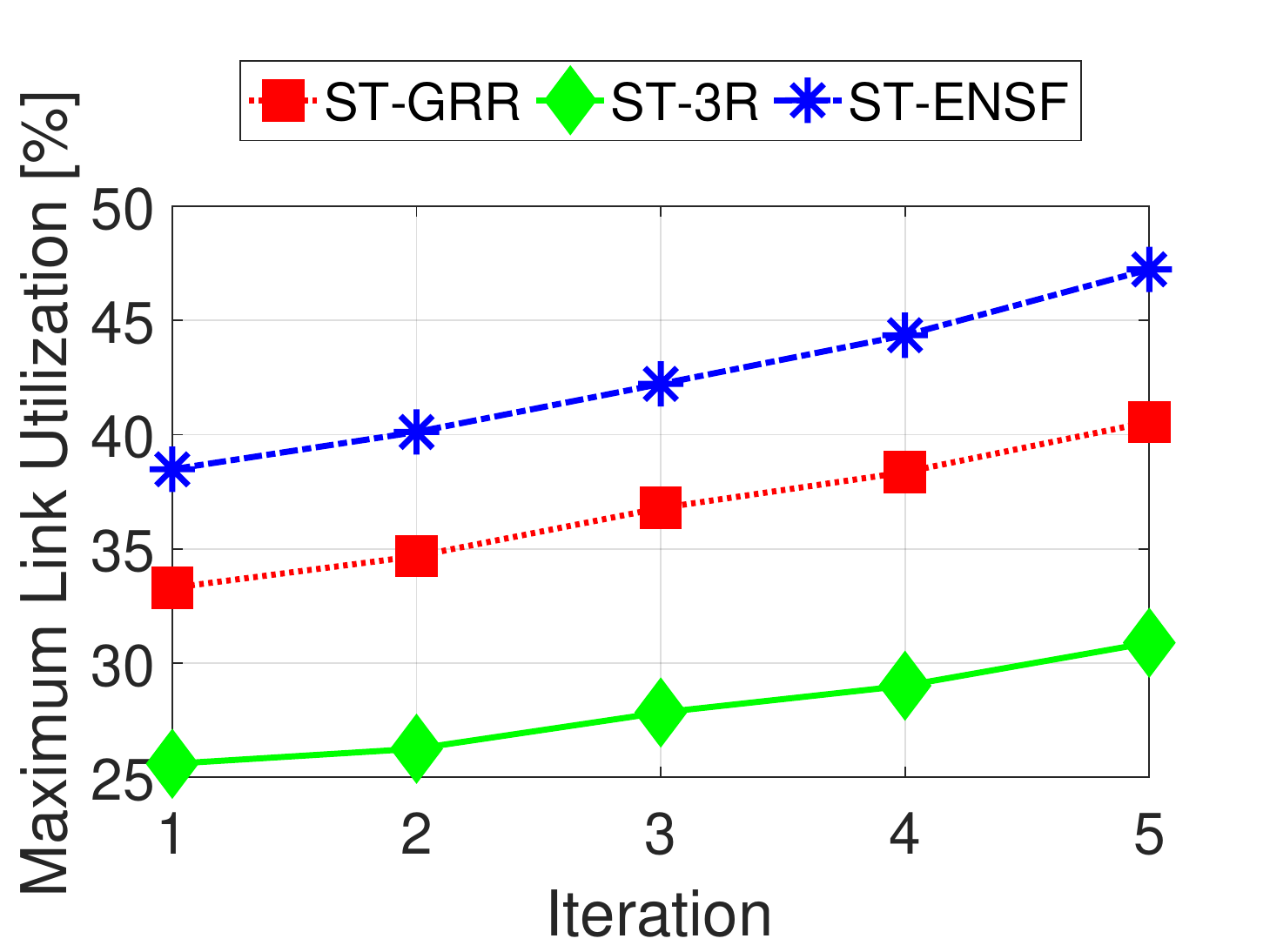}
      \caption{Maximum Link Utilization}
      \label{fig:p28ml}
    \end{subfigure}
    \begin{subfigure}{0.48\columnwidth}
      \centering
      \includegraphics[width=\columnwidth]{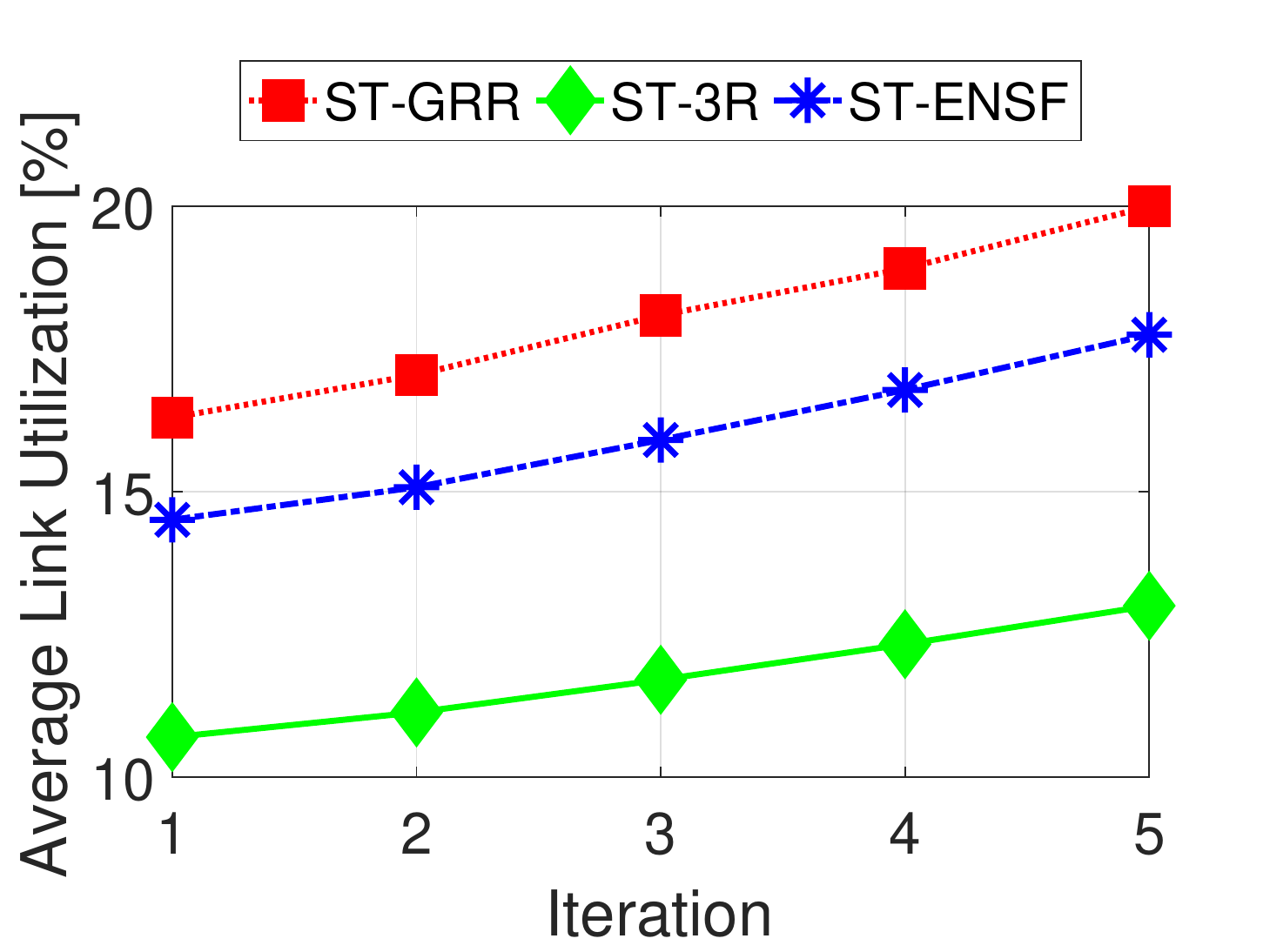}
      \caption{Average Link Utilization}
      \label{fig:p28ln}
    \end{subfigure}\hfill
    \begin{subfigure}{0.48\columnwidth}
      \centering
      \includegraphics[width=\columnwidth]{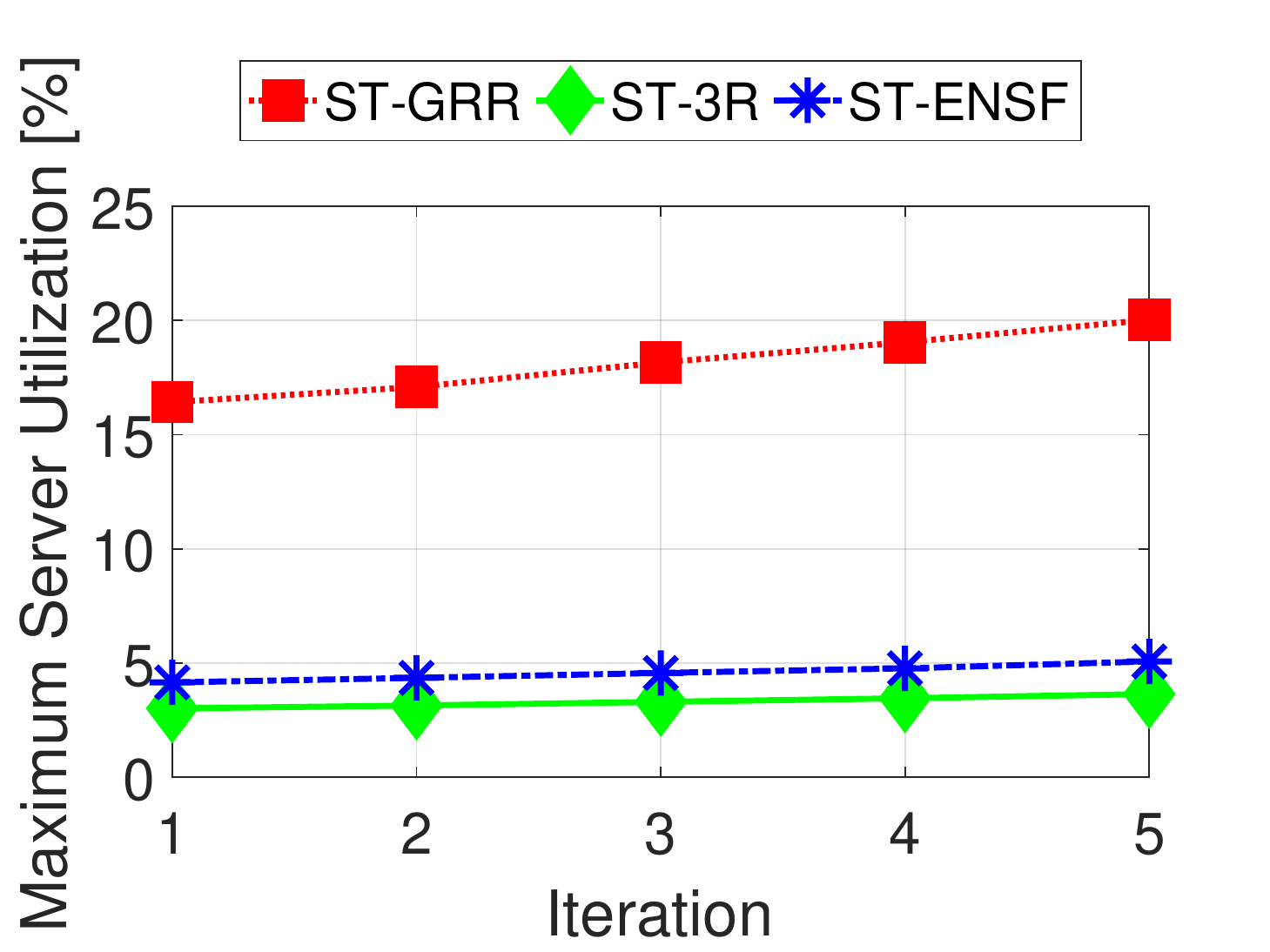}
      \caption{Maximum Server Utilization}
      \label{fig:p28mn}
    \end{subfigure}
    \begin{subfigure}{0.48\columnwidth}
      \centering
      \includegraphics[width=\columnwidth]{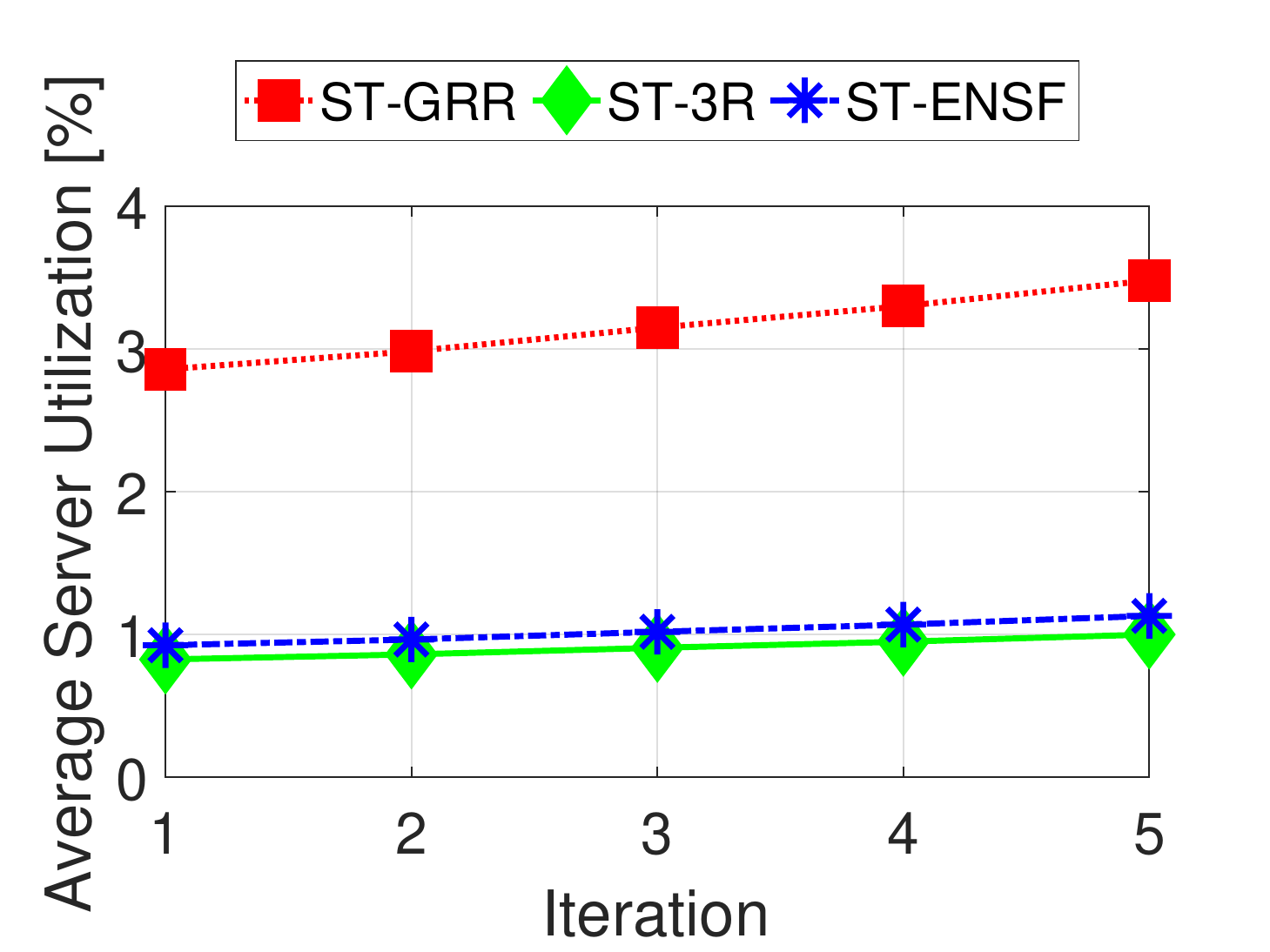}
      \caption{Average Server Utilization}
      \label{fig:p28nd}
    \end{subfigure}
    \caption{Traffic Scenario 4}
    \label{fig:p28ut}
    \end{figure}
     
    As for the resource allocation algorithms, we compare \textit{SFRA}, \textit{NSF}, and \textit{Energy-aware SFRA} which are defined in sections \ref{sect:SFRA}, \ref{sect:NSF}, and \ref{sec:ESFRA}, respectively. SFRA (formulation of SFC constraint with VNF ordering) and NSF take as input flow demands whose size is not specified, while \textit{Energy-aware SFRA} takes as input flow demands with a specified size. As can be seen, \textit{SFRA} and \textit{NSF} have a similiar ''energy consumption'' while the ''path length'' and ''configuration overhead'' of \textit{SFRA} is lower than \textit{NSF}. All of these algorithms (\textit{NSF}, \textit{SFRA}, and \textit{Energy-aware SFRA}) are sub-optimal because they consider flows one-by-one.
    
    Fig. \ref{fig:barPwrPthST}, compares the short term reallocation algorithms \textit{ST-GRR} \ref{sec:GRR}, \textit{ST-3R} \ref{sec:3R}, and \textit{ST-ENSF} \ref{sec:ST-ENSF} considering the three different metrics. In all of the test cases, the energy consumption of both \textit{ST-ENSF} and \textit{ST-3R} are near optimal. In all scenarios except scenario 1, the path length and reconfiguration overhead of \textit{ST-3R} is lower than \textit{ST-ENSF}.
    
    Finally, fig. \ref{fig:barPwrPthLT} analyzes our performance metrics of the Long-Term Resource Reallocation algorithms \textit{LT-GRR} (section~\ref{sec:GRR}), \textit{LT-3R} (section~\ref{sec:3R}), \textit{LT-ENSF} (section~\ref{section:LT-ENSF}), and \textit{ASR}~\cite{eramo2017approach}. In all test cases, the energy consumption obtained by running \textit{LT-3R} and \textit{LT-ENSF} are close the optimal one and far away from \textit{ASR}. Moreover, both the path length and the reconfiguration overhead are dramatically lower than \textit{ASR} and optimal solution. This happens because the focus of the optimal solution is on the energy consumption but not the length of the paths.
    
	    
\subsection{Server and Link Utilization}
\begin{figure}[!htbp]
    \begin{subfigure}{0.48\columnwidth}
      \centering
      \includegraphics[width=1\columnwidth]{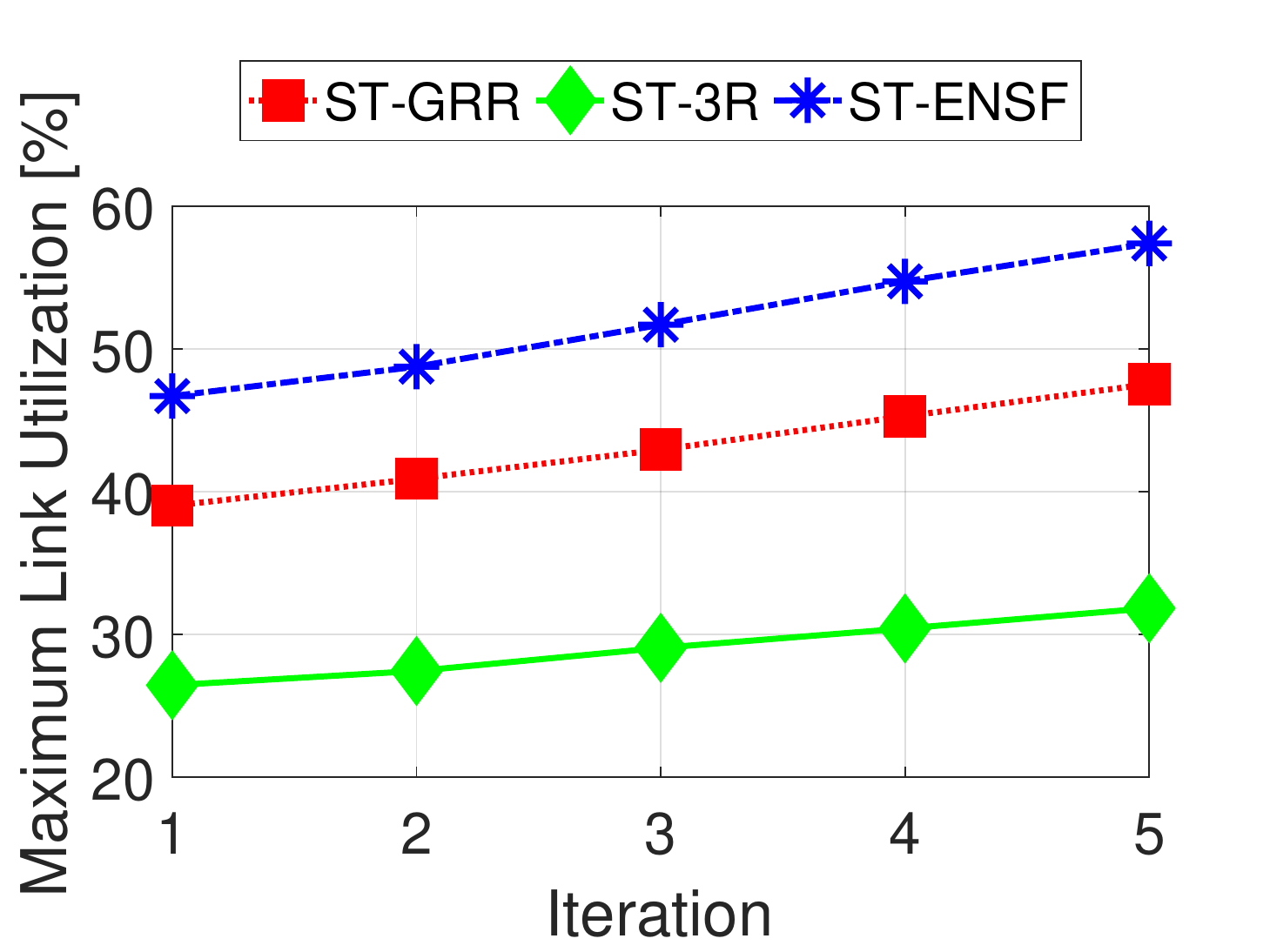}
      \caption{Maximum Link Utilization}
      \label{fig:p43ml}
    \end{subfigure}
    \begin{subfigure}{0.48\columnwidth}
      \centering
      \includegraphics[width=1\columnwidth]{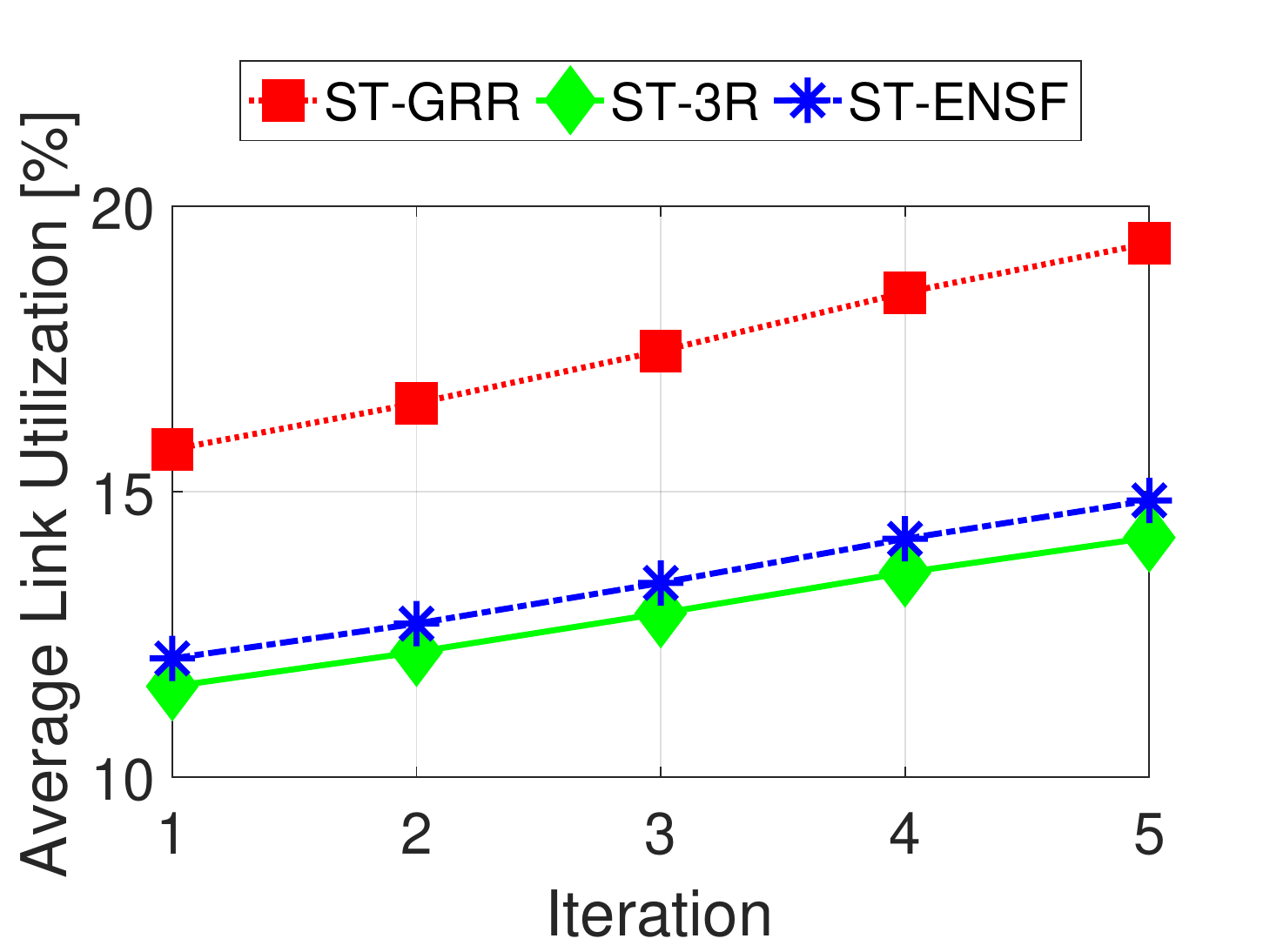}
      \caption{Average Link Utilization}
      \label{fig:p43ln}
    \end{subfigure}\hfill
    \begin{subfigure}{0.48\columnwidth}
      \centering
      \includegraphics[width=1\columnwidth]{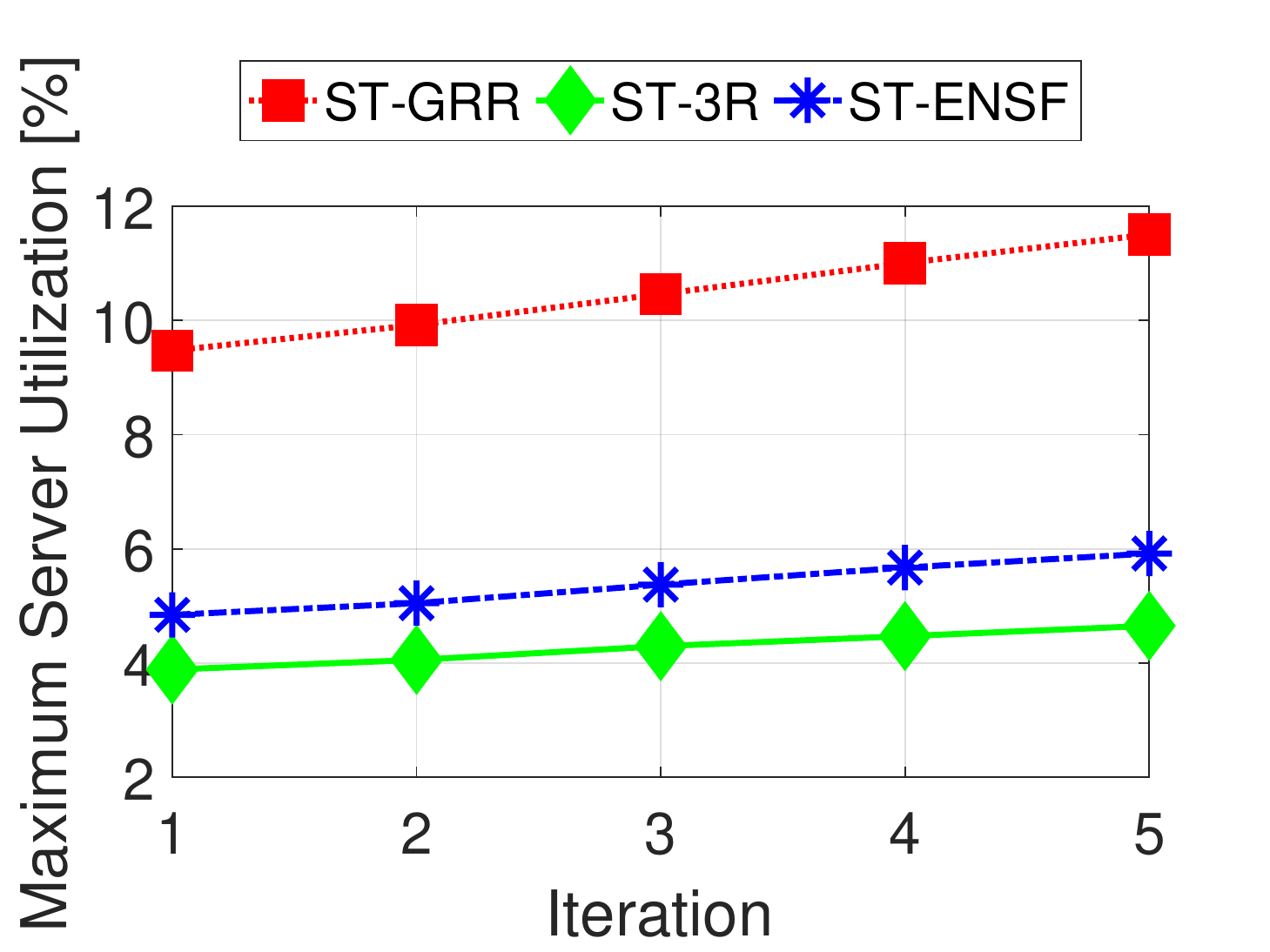}
      \caption{Maximum Server Utilization}
      \label{fig:p43mn}
    \end{subfigure}
    \begin{subfigure}{0.48\columnwidth}
      \centering
      \includegraphics[width=1\columnwidth]{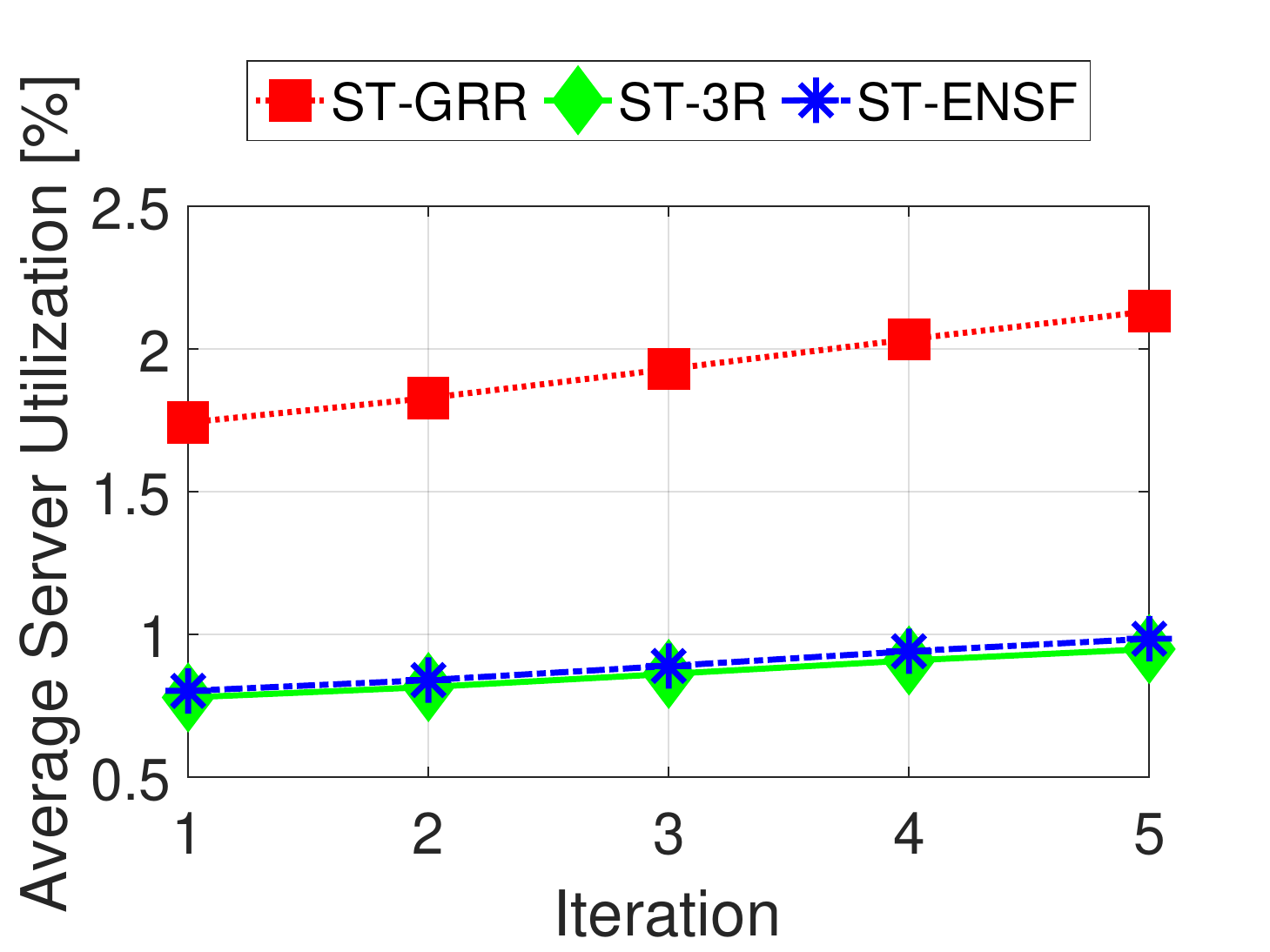}
      \caption{Average Server Utilization}
      \label{fig:p43nd}
    \end{subfigure}
    \caption{Traffic Scenario 5}
    \label{fig:p43ut}
    \end{figure}

    In this subsection, the proposed schemes are compared considering the link and server utilization metrics. To this end, the average utilization and maximum utilization of both links and servers are evaluated for the five iterations described in \ref{SimulationSetup}. We select two different scenarios and compare these measures among them in Fig.~\ref{fig:p25ut} and Fig.~ \ref{fig:p43ut}, respectively.

    As can be seen, in all test cases the average and the maximum server utilization of \textit{ST-ENSF} and \textit{ST-3R} are lower than \textit{ST-GRR} while the total throughputs are similar. This happens because the focus of \textit{ST-GRR} is only on the energy consumption, therefore, it tries to minimize the number of active servers. As a result, the maximum and the average server utilization of \textit{ST-GRR} is higher than the heuristic approaches in all test cases. On the other hand, \textit{ST-ENSF} tries to find the nearest VNFs in the sake of delay and path length, therefore, the maximum link utilization of \textit{ST-ENSF} is higher than the other approaches. 
            
\subsection{Computational Complexity}
    In this subsection, the execution time of the heuristic algorithms and the optimal solutions are compared. Table \ref{tab:SystemConfiguration} presents the configuration of the system used to execute the algorithms. The results are stated in Table \ref{tab:ComputationalComplexity}. As shown, the execution time of \textit{ENSF} is very low, hence, they can be used as real-time heuristics to reconfigures the network for online resource reallocation. On the other hand, \textit{3R} can be used for offline resource reallocation.
    	    
    \begin{table}[!htbp]
    \caption{Computational Complexity.}
    \label{tab:ComputationalComplexity}
	\rowcolors{2}{gray!25}{white}
	\resizebox{\columnwidth}{!}{
    \begin{tabular}[t]{|c|c|c|c|c|}
		\hline
		\textbf{Flow} & \textbf{GRR} & \textbf{RRR [s]} & \textbf{NSF [s]} & \textbf{ENSF [s]}\\
		\hline\hline
        10 & 58 s & 36 & 0.005 & 0.005\\\hline
        20 & 16 min & 77 & 0.005 & 0.005\\\hline 
        30 & 49 min & 154 & 0.005 & 0.005\\\hline 
        40 & 4 hour & 183 & 0.005 & 0.005\\\hline 
        250 & $\geq$ 1 Month & 821 & 0.097 & 0.101\\\hline
        500 & $\geq$ 1 Month & 1641 & 0.196 & 0.204\\\hline
        750 & $\geq$ 1 Month & 2443 & 0.298 & 0.312\\\hline
        1000 & $\geq$ 1 Month & 3252 & 0.395 & 0.426\\\hline
        1500 & $\geq$ 1 Month & 4871 & 0.675 & 0.608\\\hline
        2000 & $\geq$ 1 Month & 6494 & 0.871 & 0.812\\\hline
	\end{tabular}}
    \end{table}
	\begin{figure}[!htbp]
	\centering
	\begin{subfigure}{.49\columnwidth}
	    \includegraphics[width=\columnwidth]{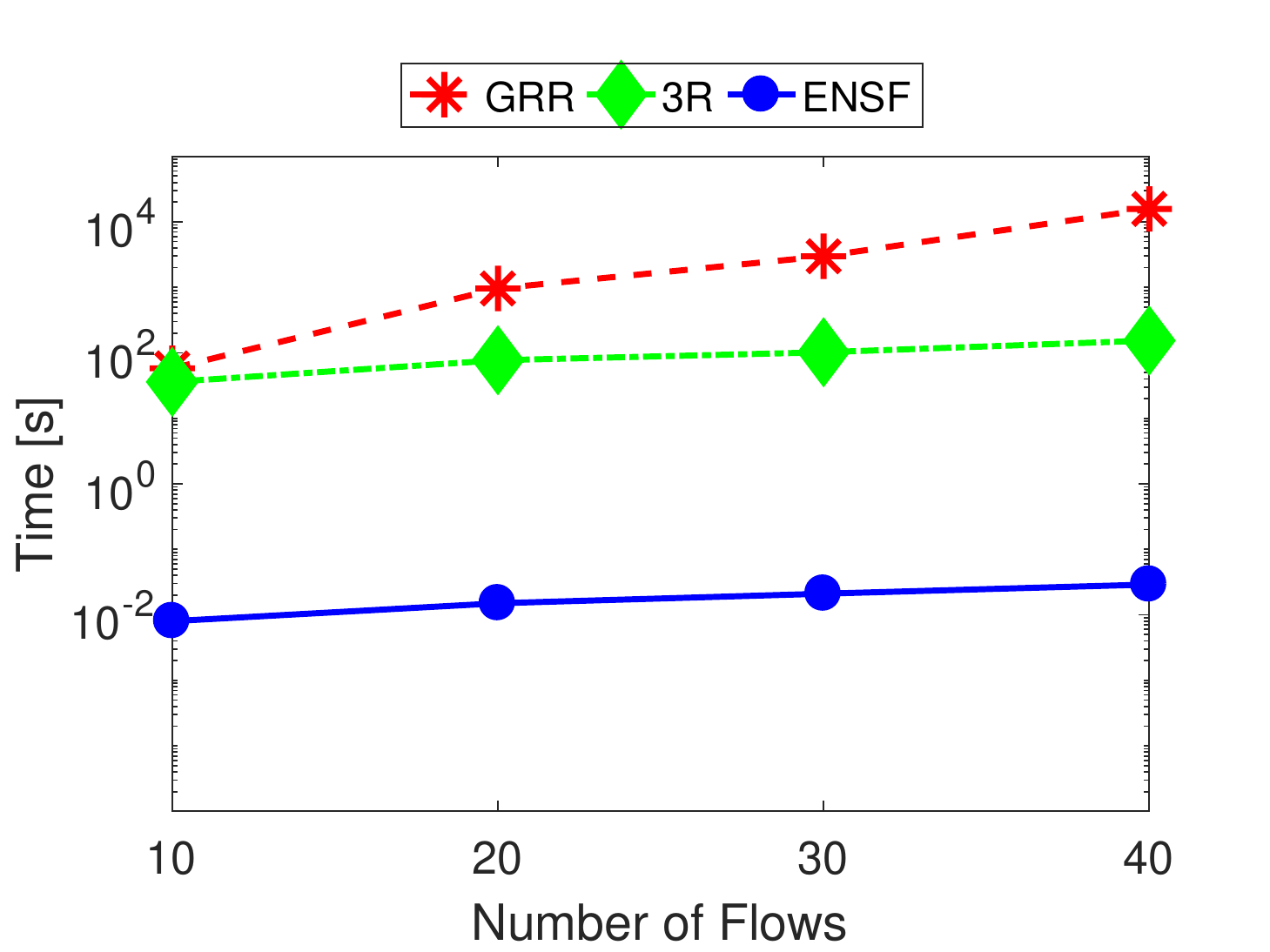}
	\end{subfigure}
	\begin{subfigure}{.49\columnwidth}
	    \includegraphics[width=\columnwidth]{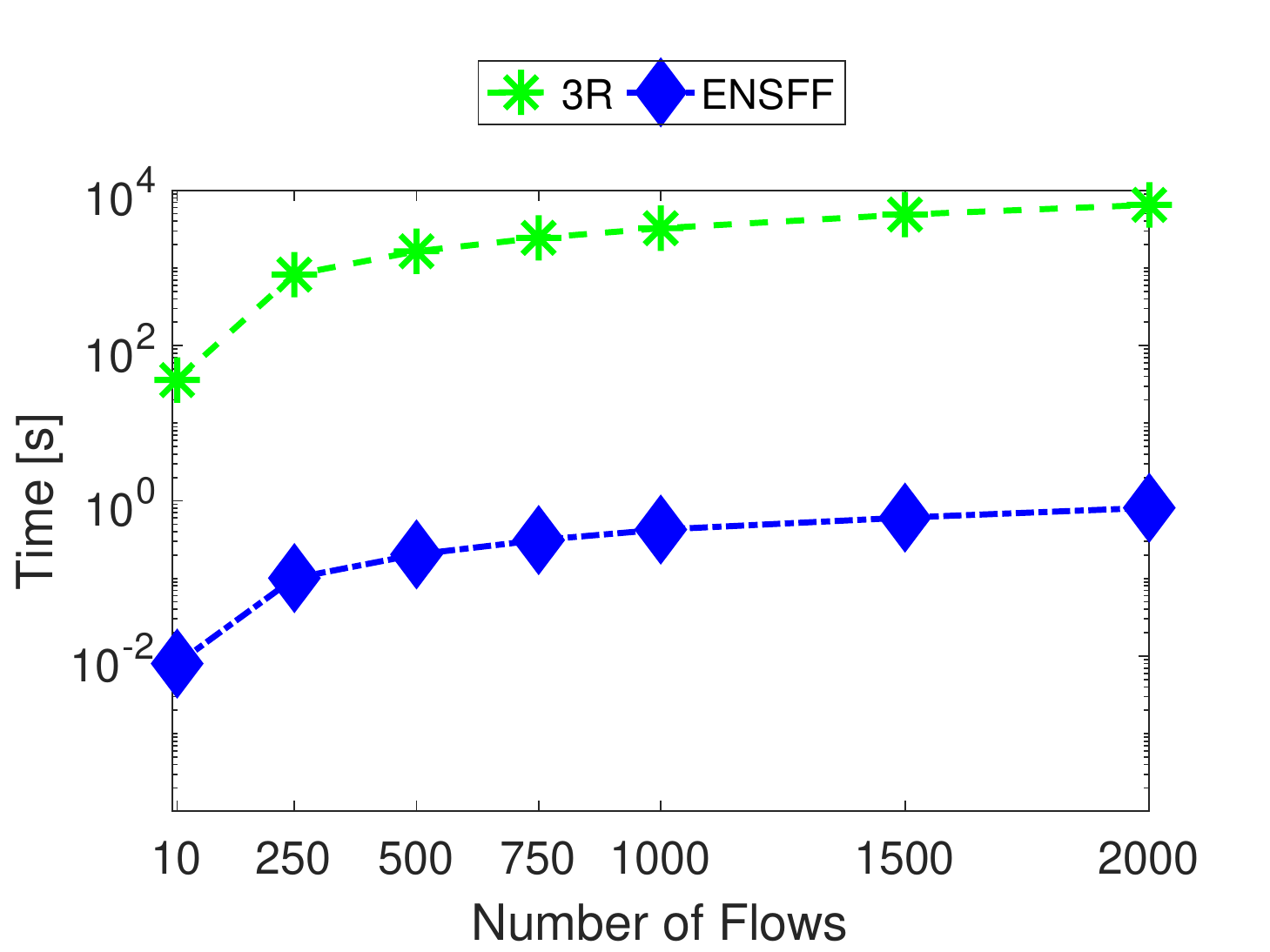}
	\end{subfigure}
	\caption{Computational Complexity.}
	\label{fig:complxity}
        	
    \end{figure}
 
\section{Conclusion and Future Works}\label{conclusion}
    In this paper, we proposed a novel resource allocation architecture for joint VNF placement and routing in the SDN-based networks. Besides, the problem of service function chaining with the goal of minimizing the energy consumption and network reconfiguration overhead is addressed. To this end, we mathematically formulated the problem and proposed several heuristics to solve them. The objective is to minimize the energy consumption and the side-effect of network reconfiguration while the flow requirements are met. Simulation results showed that the proposed schemes allocate the network resource in a way that the energy consumption is near to the optimal solution. Future work will be dedicated to make the algorithms robust against burst traffics. Another field of future interest would be considering the queuing delay of the switches and the processing delay of the server to improve the QoSs.
    
\section*{Acknowledgment}
    This work has received funding from the Horizon 2020 EU project SUPERFLUIDITY (grant agreement No. 671566). The authors would like to thank Francesco Giacinto Lavacca and Vincenzo Eramo for their support on the evaluation of our algorithms in comparison with the ASR algorithm.
    \bibliographystyle{IEEEtran}
    \bibliography{references.bib}

\begin{thebibliography}{10}
\providecommand{\url}[1]{#1}
\csname url@samestyle\endcsname
\providecommand{\newblock}{\relax}
\providecommand{\bibinfo}[2]{#2}
\providecommand{\BIBentrySTDinterwordspacing}{\spaceskip=0pt\relax}
\providecommand{\BIBentryALTinterwordstretchfactor}{4}
\providecommand{\BIBentryALTinterwordspacing}{\spaceskip=\fontdimen2\font plus
\BIBentryALTinterwordstretchfactor\fontdimen3\font minus
  \fontdimen4\font\relax}
\providecommand{\BIBforeignlanguage}[2]{{%
\expandafter\ifx\csname l@#1\endcsname\relax
\typeout{** WARNING: IEEEtran.bst: No hyphenation pattern has been}%
\typeout{** loaded for the language `#1'. Using the pattern for}%
\typeout{** the default language instead.}%
\else
\language=\csname l@#1\endcsname
\fi
#2}}
\providecommand{\BIBdecl}{\relax}
\BIBdecl

\bibitem{fischer2013virtual}
A.~Fischer, J.~F. Botero, M.~T. Beck, H.~De~Meer, and X.~Hesselbach, ``Virtual
  network embedding: A survey,'' \emph{IEEE Communications Surveys \&
  Tutorials}, vol.~15, no.~4, pp. 1888--1906, 2013.

\bibitem{halpern2015service}
\BIBentryALTinterwordspacing
J.~Halpern and C.~Pignataro, ``{Service Function Chaining (SFC)
  architecture},'' Tech. Rep., 2015. [Online]. Available:
  \url{https://www.rfc-editor.org/info/rfc7665}
\BIBentrySTDinterwordspacing

\bibitem{medhat2017service}
A.~M. Medhat, T.~Taleb, A.~Elmangoush, G.~A. Carella, S.~Covaci, and
  T.~Magedanz, ``Service function chaining in next generation networks: State
  of the art and research challenges,'' \emph{IEEE Communications Magazine},
  vol.~55, no.~2, pp. 216--223, 2017.

\bibitem{hepburn2006regulating}
C.~Hepburn, ``Regulating by prices, quantities or both: an update and an
  overview,'' \emph{Oxford Review of Economic Policy}, vol.~22, no.~2, pp.
  226--247, 2006.

\bibitem{khosravi2017dynamic}
A.~Khosravi, L.~Andrew, and R.~Buyya, ``Dynamic vm placement method for
  minimizing energy and carbon cost in geographically distributed cloud data
  centers,'' \emph{IEEE Transactions on Sustainable Computing}, 2017.

\bibitem{etsi-mano}
``{ETSI GS NFV-MAN 001: Network Functions Virtualisation (NFV); Management and
  Orchestration, V 1.1.1},'' 2014.

\bibitem{mills2011comparing}
K.~Mills, J.~Filliben, and C.~Dabrowski, ``Comparing vm-placement algorithms
  for on-demand clouds,'' in \emph{Cloud Computing Technology and Science
  (CloudCom), 2011 IEEE Third International Conference on}.\hskip 1em plus
  0.5em minus 0.4em\relax IEEE, 2011, pp. 91--98.

\bibitem{filiposka2016balancing}
S.~Filiposka, A.~Mishev, and C.~Juiz, ``Balancing performances in online vm
  placement,'' in \emph{ICT Innovations 2015}.\hskip 1em plus 0.5em minus
  0.4em\relax Springer, 2016, pp. 153--162.

\bibitem{bhamare2017optimal}
D.~Bhamare, M.~Samaka, A.~Erbad, R.~Jain, L.~Gupta, and H.~A. Chan, ``Optimal
  virtual network function placement in multi-cloud service function chaining
  architecture,'' \emph{Computer Communications}, vol. 102, pp. 1--16, 2017.

\bibitem{bari2015orchestrating}
M.~F. Bari, S.~R. Chowdhury, R.~Ahmed, and R.~Boutaba, ``On orchestrating
  virtual network functions,'' in \emph{Network and Service Management (CNSM),
  2015 11th International Conference on}.\hskip 1em plus 0.5em minus
  0.4em\relax IEEE, 2015, pp. 50--56.

\bibitem{even2016approximation}
G.~Even, M.~Rost, and S.~Schmid, ``An approximation algorithm for path
  computation and function placement in sdns,'' in \emph{International
  Colloquium on Structural Information and Communication Complexity}.\hskip 1em
  plus 0.5em minus 0.4em\relax Springer, 2016, pp. 374--390.

\bibitem{xu2015effective}
S.~Xu, B.~Fu, M.~Chen, and L.~Zhang, ``An effective correlation-aware vm
  placement scheme for sla violation reduction in data centers,'' in
  \emph{International Conference on Algorithms and Architectures for Parallel
  Processing}, vol. 125, no.~1.\hskip 1em plus 0.5em minus 0.4em\relax
  Springer, 2015, pp. 64--75.

\bibitem{rocha2015network}
L.~Rocha and F.~Verdi, ``A network-aware optimization for vm placement,'' in
  \emph{Advanced Information Networking and Applications (AINA), 2015 IEEE 29th
  International Conference on}.\hskip 1em plus 0.5em minus 0.4em\relax IEEE,
  2015, pp. 619--625.

\bibitem{zhao2015joint}
Y.~Zhao, Y.~Huang, K.~Chen, M.~Yu, S.~Wang, and D.~Li, ``Joint vm placement and
  topology optimization for traffic scalability in dynamic datacenter
  networks,'' \emph{Computer Networks}, vol.~80, pp. 109--123, 2015.

\bibitem{calcavecchia2012vm}
N.~M. Calcavecchia, O.~Biran, E.~Hadad, and Y.~Moatti, ``Vm placement
  strategies for cloud scenarios,'' in \emph{Cloud Computing (CLOUD), 2012 IEEE
  5th International Conference on}.\hskip 1em plus 0.5em minus 0.4em\relax
  IEEE, 2012, pp. 852--859.

\bibitem{meng2010improving}
X.~Meng, V.~Pappas, and L.~Zhang, ``Improving the scalability of data center
  networks with traffic-aware virtual machine placement,'' in \emph{INFOCOM,
  2010 Proceedings IEEE}.\hskip 1em plus 0.5em minus 0.4em\relax IEEE, 2010,
  pp. 1--9.

\bibitem{eramo2017migration}
V.~Eramo, M.~Ammar, and F.~G. Lavacca, ``Migration energy aware
  reconfigurations of virtual network function instances in nfv
  architectures,'' \emph{IEEE Access}, vol.~5, pp. 4927--4938, 2017.

\bibitem{zhang2016co}
B.~Zhang, P.~Zhang, Y.~Zhao, Y.~Wang, X.~Luo, and Y.~Jin, ``Co-scaler:
  Cooperative scaling of software-defined nfv service function chain,'' in
  \emph{Network Function Virtualization and Software Defined Networks
  (NFV-SDN), IEEE Conference on}.\hskip 1em plus 0.5em minus 0.4em\relax IEEE,
  2016, pp. 33--38.

\bibitem{abdelsalam2017implementation}
A.~AbdelSalam, F.~Clad, C.~Filsfils, S.~Salsano, G.~Siracusano, and L.~Veltri,
  ``Implementation of virtual network function chaining through segment routing
  in a linux-based nfv infrastructure,'' \emph{arXiv preprint
  arXiv:1702.05157}, 2017.

\bibitem{khoshbakht2016sdte}
M.~Khoshbakht, M.~M. Tajiki, and B.~Akbari, ``Sdte: Software defined traffic
  engineering for improving data center network utilization,'' \emph{Int. J.
  Inf. Commun. Technol. Res}, vol.~8, no.~4, pp. 15--24, 2016.

\bibitem{kulkarni2017neo}
S.~Kulkarni, M.~Arumaithurai, K.~Ramakrishnan, and X.~Fu, ``Neo-nsh: Towards
  scalable and efficient dynamic service function chaining of elastic network
  functions,'' in \emph{Innovations in Clouds, Internet and Networks (ICIN),
  2017 20th Conference on}.\hskip 1em plus 0.5em minus 0.4em\relax IEEE, 2017,
  pp. 308--312.

\bibitem{soares2015toward}
J.~Soares, C.~Gon{\c{c}}alves, B.~Parreira, P.~Tavares, J.~Carapinha, J.~P.
  Barraca, R.~L. Aguiar, and S.~Sargento, ``Toward a telco cloud environment
  for service functions,'' \emph{IEEE Communications Magazine}, vol.~53, no.~2,
  pp. 98--106, 2015.

\bibitem{tajiki2017optimal}
M.~M. Tajiki, B.~Akbari, and N.~Mokari, ``Optimal qos-aware network
  reconfiguration in software defined cloud data centers,'' \emph{Computer
  Networks}, vol. 120, pp. 71--86, 2017.

\bibitem{reddy2016robust}
V.~S. Reddy, A.~Baumgartner, and T.~Bauschert, ``Robust embedding of
  vnf/service chains with delay bounds,'' in \emph{Network Function
  Virtualization and Software Defined Networks (NFV-SDN), IEEE Conference
  on}.\hskip 1em plus 0.5em minus 0.4em\relax IEEE, 2016, pp. 93--99.

\bibitem{ghaznavi2016service}
M.~Ghaznavi, N.~Shahriar, R.~Ahmed, and R.~Boutaba, ``Service function chaining
  simplified,'' \emph{arXiv preprint arXiv:1601.00751}, 2016.

\bibitem{jiang2012joint}
J.~W. Jiang, T.~Lan, S.~Ha, M.~Chen, and M.~Chiang, ``Joint vm placement and
  routing for data center traffic engineering,'' in \emph{INFOCOM, 2012
  Proceedings IEEE}.\hskip 1em plus 0.5em minus 0.4em\relax IEEE, 2012, pp.
  2876--2880.

\bibitem{wang2016joint}
L.~Wang, Z.~Lu, X.~Wen, R.~Knopp, and R.~Gupta, ``Joint optimization of service
  function chaining and resource allocation in network function
  virtualization,'' \emph{IEEE Access}, vol.~4, pp. 8084--8094, 2016.

\bibitem{Akbari2016qrtp}
M.~M. Tajiki, B.~Akbari, and N.~Mokari, ``Qrtp: Qos-aware resource reallocation
  based on traffic prediction in software defined cloud networks,'' in
  \emph{Telecommunications (IST), 2016 8th International Symposium on}.\hskip
  1em plus 0.5em minus 0.4em\relax IEEE, 2016, pp. 527--532.

\bibitem{marotta2017energy}
A.~Marotta, F.~D’Andreagiovanni, A.~Kassler, and E.~Zola, ``On the energy
  cost of robustness for green virtual network function placement in 5g
  virtualized infrastructures,'' \emph{Computer Networks}, 2017.

\bibitem{shojafar2016energy}
M.~Shojafar, N.~Cordeschi, and E.~Baccarelli, ``Energy-efficient adaptive
  resource management for real-time vehicular cloud services,'' \emph{IEEE
  Transactions on Cloud Computing}, vol.~PP, pp. 1--14, 2016.

\bibitem{tang2015hybrid}
M.~Tang and S.~Pan, ``A hybrid genetic algorithm for the energy-efficient
  virtual machine placement problem in data centers,'' \emph{Neural Processing
  Letters}, vol.~41, no.~2, pp. 211--221, 2015.

\bibitem{gu2015joint}
L.~Gu, D.~Zeng, S.~Guo, and B.~Ye, ``Joint optimization of vm placement and
  request distribution for electricity cost cut in geo-distributed data
  centers,'' in \emph{Computing, Networking and Communications (ICNC), 2015
  International Conference on}.\hskip 1em plus 0.5em minus 0.4em\relax IEEE,
  2015, pp. 717--721.

\bibitem{eramo2017approach}
V.~Eramo, E.~Miucci, M.~Ammar, and F.~G. Lavacca, ``An approach for service
  function chain routing and virtual function network instance migration in
  network function virtualization architectures,'' \emph{IEEE/ACM Transactions
  on Networking}, vol.~25, no.~4, pp. 2008--2025, 2017.

\bibitem{herrera2016resource}
J.~G. Herrera and J.~F. Botero, ``Resource allocation in nfv: A comprehensive
  survey,'' \emph{IEEE Transactions on Network and Service Management},
  vol.~13, no.~3, pp. 518--532, 2016.

\bibitem{rost2016service}
M.~Rost and S.~Schmid, ``Service chain and virtual network embeddings:
  Approximations using randomized rounding,'' \emph{arXiv preprint
  arXiv:1604.02180}, 2016.

\bibitem{HajMeity}
M.~M. Tajiki, S.~Salsano, M.~Shojafar, L.~Chiaraviglio, and B.~Akbari,
  ``Energy-efficient path allocation heuristic for service function chaining,''
  in \emph{Innovations in Clouds, Internet and Networks (ICIN), 2018 21th
  Conference on}.\hskip 1em plus 0.5em minus 0.4em\relax IEEE, 2018, pp. 1--8.

\bibitem{haleplidis2015software}
E.~Haleplidis, K.~Pentikousis, S.~Denazis, J.~H. Salim, D.~Meyer, and
  O.~Koufopavlou, ``{Software-Defined Networking (SDN): Layers and Architecture
  Terminology},'' IETF RFC 7426, 2015.

\bibitem{chiaraviglio2010polisave}
L.~Chiaraviglio and M.~Mellia, ``Polisave: efficient power management of campus
  pcs,'' in \emph{Software, Telecommunications and Computer Networks (SoftCOM),
  2010 International Conference on}.\hskip 1em plus 0.5em minus 0.4em\relax
  IEEE, 2010, pp. 82--87.

\bibitem{chiaraviglio2017measurement}
L.~Chiaraviglio, N.~Blefari-Melazzi, C.~Canali, F.~Cuomo, R.~Lancellotti, and
  M.~Shojafar, ``A measurement-based analysis of temperature variations
  introduced by power management on commodity hardware,'' in \emph{Transparent
  Optical Networks (ICTON), 2017 19th International Conference on}.\hskip 1em
  plus 0.5em minus 0.4em\relax IEEE, 2017, pp. 1--4.

\bibitem{dumitrescu2001algorithms}
I.~Dumitrescu and N.~Boland, ``Algorithms for the weight constrained shortest
  path problem,'' \emph{International Transactions in Operational Research},
  vol.~8, no.~1, pp. 15--29, 2001.

\bibitem{AbilineNetwork}
\BIBentryALTinterwordspacing
``Abilene network,'' Jun 2017, [Online; posted 24-March-2012]. [Online].
  Available: \url{https://uit.stanford.edu/service/network/internet2/abilene}
\BIBentrySTDinterwordspacing

\bibitem{zoucomputer}
X.~Zou, ``Computer communication networks cs 6/75202 g1 project source traffic
  modeling and generation,'' \emph{Mohammad M. Tajiki received his BS degree in
  Computer Engineering from Shahid Bahonar University, Kerman, Iran, in}, 2011.

\bibitem{grant2008cvx}
M.~Grant, S.~Boyd, and Y.~Ye, ``Cvx: Matlab software for disciplined convex
  programming,'' 2008.

\end{thebibliography}
    
\section*{Biography}
\vspace{-5em}
\begin{IEEEbiography}[{\includegraphics[width=1in,height=1.25in,clip,keepaspectratio]{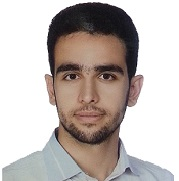}}]{Mohammad M. Tajiki} received his B.S. degree in Computer Engineering from Shahid Bahonar University, Kerman, Iran, in 2011. In 2013, he graduated from Electrical and Computer Engineering School of Tehran University, Tehran, Iran. Currently, he is a PhD candidate in Tarbiat Modares University, Tehran, Iran. His main research interests are Network QoS, media streaming over the Internet, data center networking, traffic engineering, and Software Defined Networking (SDN).
\end{IEEEbiography}
\vspace{-5em}
\begin{IEEEbiography}[{\includegraphics[width=1in,height=1.25in,clip,keepaspectratio]{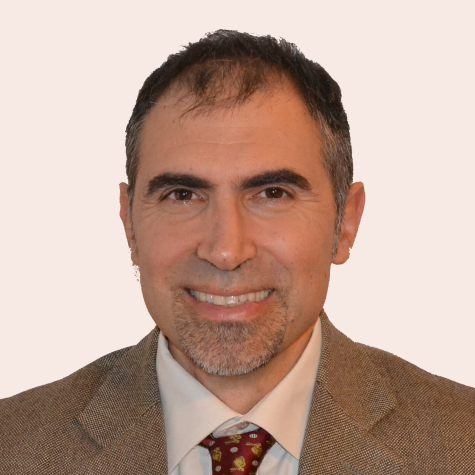}}]{Stefano Salsano}
(M'98-SM'13) received his PhD from University of Rome “La Sapienza” in 1998. He is Associate Professor at the University of Rome Tor Vergata. He participated in 15 research projects founded by the EU, being project coordinator in one of them and technical coordinator in two of them. He has been principal investigator in several research and technology transfer contracts funded by industries. His current research interests include Software Defined Networking, Network Virtualization, Cybersecurity, Information Centric Networking. He is co-author of an IETF RFC and of more than 140 peer-reviewed papers and book chapters.
\end{IEEEbiography}
\vspace{-5em}
\begin{IEEEbiography}[{\includegraphics[width=1in,height=1.25in,clip,keepaspectratio]{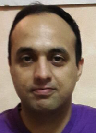}}]{Mohammad Shojafar} is currently is a CNIT senior researcher and Intel innovator at the University of Rome Tor Vergata to work on European H2020 “Superfluity” project. Recently, he completed an Italian project named ``SAMMClouds'' by the University of Modena and Reggio Emilia, Modena, Italy. He received the Ph.D. degree from Sapienza University of Rome, Rome, Italy, in 2016. He received the Msc and Bsc in QIAU and IUST, Iran in 2010 and 2006, respectively. He was a programmer/analyzer at National Iranian Oil Company (NIOC) and Tidewater ltd. in Iran from 2008-2013, respectively.
\end{IEEEbiography}
\vspace{-5em}
\begin{IEEEbiography}[{\includegraphics[width=1in,height=1.25in,clip,keepaspectratio]{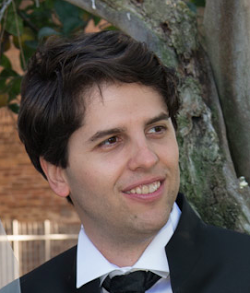}}]{Luca Chiaraviglio} is Tenure-Track Assistant Professor in the Department of Electronic Engineering, University of Rome Tor Vergata, Italy. He holds a Ph.D. in Telecommunication and Electronics engineering obtained at Politecnico di Torino, Italy. Previously to join the University of Rome Tor Vergata, Luca has worked in different institutions, including: Senior Researcher at CNIT, Assistant Professor at University of Rome Sapienza, ERCIM Fellow at INRIA Sophia Antipolis, and Post-Doc at Politecnico di Torino.
\end{IEEEbiography}
\vspace{-5em}
\begin{IEEEbiography}[{\includegraphics[width=1in,height=1.25in,clip,keepaspectratio]{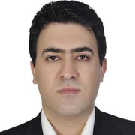}}]{Behzad Akbari} received the B.S., M.S., and PhD degree in computer engineering from the Sharif University of Technology, Tehran, Iran, in 1999, 2002, and 2008 respectively. His research interest includes Computer Networks, Multimedia Networking Overlay and Peer-to-Peer Networking, Peer-to-Peer Video Streaming, Network QOS, Network Performance Analysis, Network Security, Network Security Events Analysis and Correlation, Network Management, Cloud Computing and Networking, Software Defined Networks.
\end{IEEEbiography}


\end{document}